%% file: paper.tex
\documentclass[letterpaper,twocolumn,10pt]{article}
\usepackage{usenix-2020-09}
\usepackage{etoolbox}
\providetoggle{techreport}
\settoggle{techreport}{true}%{false}%
\providetoggle{journal}
\settoggle{journal}{false}%{true}%
\providetoggle{nsdi22}
\settoggle{nsdi22}{false}%{true}%

\usepackage{booktabs} % For formal tables
\usepackage{xspace}
\usepackage{multirow}
\usepackage{makecell}
\usepackage{graphicx,amssymb,amsmath,endnotes}
\usepackage{caption}
\usepackage{ulem}
\usepackage[tight]{subfigure}
\usepackage{fancyhdr}
\usepackage{textcomp, libertine}
\usepackage{xcolor}
\usepackage{varwidth}
\usepackage[noend]{algpseudocode}
\iftoggle{journal}{\usepackage{hyperref}[backref,pdfpagelabels,bookmarks,hyperindex,hyperfigures]}{}
\iftoggle{journal}{}{\pagenumbering{gobble}}
\usepackage{etoolbox}
\makeatletter
\patchcmd{\maketitle}
	{\@maketitle}
	{\vspace{-6em}\@maketitle\vspace{-4em}}% change the value as needed
	{}
	{}
\renewcommand\subsection{\@startsection{subsection}{2}{\z@}%
                                     {-2.25ex\@plus -1ex \@minus -.2ex}%
                                     {0.5ex \@plus .2ex}%
                                     {\normalfont\bf}}% 
\renewcommand\subsubsection{\@startsection{subsubsection}{3}{\z@}%
                                     {-1.25ex\@plus -1ex \@minus -.2ex}%
                                     {0.5ex \@plus .2ex}%
                                     {\normalfont\bf}}% 
\makeatother

\begin{document}
\title{A Case for Task Sampling based Learning for Cluster Job Scheduling}
\iftoggle{journal}{\author{Akshay Jajoo, Y. Charlie Hu,~\IEEEmembership{Fellow,~IEEE,} Xiaojun Lin,~\IEEEmembership{Fellow,~IEEE}, Nan Deng}
}{\author{
{\rm Akshay Jajoo\thanks{The work was done while the author was pursuing his Ph.D. at Purdue University.}}\\
%Purdue University
\small{akshay.jajoo@nokia-bell-labs.com}
%\small{akshayjajoo.research@gmail.com}
%ajajoo@purdue.edu
\and
{\rm Y. Charlie Hu}\\
\small{ychu@purdue.edu}%\\
%\{ ajajoo, ychu, linx \} @purdue.edu\\
%Purdue University
\and
{\rm Xiaojun Lin}\\
\small{linx@purdue.edu}
\and
{\rm Nan Deng}\\
\small{dengnan@google.com}
}}
%\author{
\xspace
%% If anonymous submission change id in the acmart.cls line num 1117
%Paper ID: 119, pages = 12+2\\
%\date{\today}
%\date{}
%} %% end author
\maketitle
\newcommand{\oldstuff}[1]{}
\newcommand{\yes}{$\checkmark$}
\newcommand{\limited}{Limited}
\newcommand{\no}{$\times$}
%\definecolor{gray}{rgb}{0.5,0.5,0.5}
\newcommand{\etc}{\emph{etc.}\xspace}
\newcommand{\etcc}{\emph{etc.}}
\newcommand{\ie}{\emph{i.e.,}\xspace}
\newcommand{\eg}{\emph{e.g.,}\xspace}
\newcommand{\etal}{\emph{et al.}\xspace}
\newcommand{\SmallCrunch}{\vspace{-0cm}}
\newcommand{\smallcrunch}{\vspace{-0cm}}

\renewcommand{\paragraph}[1]{\smallskip\noindent{\bf{#1}}}

\definecolor{red}{rgb}{1,0,0}
\renewcommand{\rm}[1]{{}}
%\definecolor{purple}{RGB}{127,0,127}

%%%%%%%%%%%%%%%%% below for USENIX sty no use of color %%%%%%%%%%%%%%%%

%\newcommand{\commentlinx}[1]{{ \textit{linx: #1}}}
%\newcommand{\commentaj}[1]{{\textit{aj: #1}}}
%\newcommand{\questionaj}[1]{{\textbf{questionAJ: #1}}}
%\newcommand{\updated}[1]{{{}}}
%\newcommand{\todoaj}[1]{{\textit{TODO(AJ): #1}}}
%\newcommand{\editaj}[2]{{\sout{#1}{#2}}}
%\newcommand{\addaj}[1]{{ {ADD?: #1}}}
%\newcommand{\updated}[1]{{{updated: #1}}}
%%%%%%%%%%%%%%%%% above for USENIX sty no use of color %%%%%%%%%%%%%%%%

\newcommand{\comment}[1]{{\it #1}}
\newcommand{\deadlineCS}[1]{{\em #1}}
\newcommand{\soccReviewEdit}[2]{{\em \textit{socc19 (Reviewer - #1): }#2}}
\newcommand{\commentlinx}[1]{{\em \textit{linx: #1}}}
\newcommand{\commentaj}[1]{{\em \textit{aj: #1}}}
\newcommand{\questionaj}[1]{{\em \textbf{questionAJ: #1}}}
\newcommand{\updated}[1]{{{}}}
\newcommand{\editaj}[2]{{\sout{#1} {#2}}}
\newcommand{\removesl}[2]{{{#2}}}%New word
\newcommand{\removeajbcoz}[2]{{\em{\textbf{Remove following:}}}{\sout{#1}}{\em {\textbf{ - Because} #2}}}
\newcommand{\removeaj}[1]{{\em{\sout{#1}}}}
\newcommand{\addaj}[1]{{\em {ADD?: #1}}}
\newcommand{\addajbcoz}[2]{{\em {ADD?: #1} \textit{\textbf{ - } #2}}}
\newcommand{\todoaj}[1]{{\em \textit{TODO(AJ): #1}}}
\iftoggle{techreport}{\newcommand{\editnsdiSHP}[3]{#1}}{\newcommand{\editnsdiSHP}[3]{{#2}}}
\iftoggle{techreport}{\newcommand{\addnsdiSHP}[2]{#1}}{\newcommand{\addnsdiSHP}[2]{#1}}
\newcommand{\removensdiSHP}[2]{{\sout{#1} \em{#2}}}%Suggest Mode

\newcommand{\insitu}{{\em in situ}\xspace}

\newcommand{\numTraces}{{3}\xspace}
\newcommand{\thinLimit}{{3}\xspace}
\newcommand{\namepredict}{{\sc SLearn}\xspace}
\newcommand{\las}{{\sc LAS}\xspace}
\newcommand{\pointestimator}{{\sc Point-Est}\xspace}
\newcommand{\oracle}{{\sc Oracle}\xspace}
\newcommand{\fifo}{{\sc {FIFO}}\xspace}

\newcommand{\slearndag}{{\sc SLearn-DAG}\xspace}
\newcommand{\primarybasedag}{{\sc 3Sigma-DAG}\xspace}
\newcommand{\lasdag}{{\sc LAS-DAG}\xspace}
\newcommand{\pointestimatordag}{{\sc Point-Est-DAG}\xspace}
\newcommand{\oracledag}{{\sc Oracle-DAG}\xspace}
\newcommand{\fifodag}{{\sc {FIFO-DAG}}\xspace}

\newcommand{\name}{{\sc SLearn}\xspace}
\newcommand{\namee}{{\sc SLearn}}
\newcommand{\lTechnique}{{\sc SLearn}\xspace}
\newcommand{\slearn}{{\sc SLearn}\xspace}

\newcommand{\gs}{{\sc GS}\xspace}
\newcommand{\gsdl}{{\sc GS-DL}\xspace}
\newcommand{\resistance}{\textit{resistance}\xspace}
\newcommand{\ychurm}[1]{{\hspace{-0.2cm}}}
\newcommand{\primarybase}{{3Sigma}\xspace}
\newcommand{\primarybasepredict}{{3Sigma}\xspace}
\newcommand{\primarybasepredictTL}{{3SigmaTL}\xspace}

\newcommand{\dagtrace}{{GTrace19-DAG}\xspace}

\newcommand{\panic}[1]{\vspace{-#1 plus 1pt minus 1pt}}
\newcommand{\panictwo}[1]{\vspace{-#1 plus 2pt minus 2pt}}

\newcommand{\nsection}[1]{\panictwo{0pt}\section{#1}\panictwo{0pt}}
\newcommand{\nsubsection}[1]{\panictwo{0pt}\subsection{#1}\panictwo{0pt}}
\newcommand{\nsubsubsection}[1]{\panictwo{0pt}\subsubsection{#1}\panictwo{0pt}}

\newcommand{\flow}{flow\xspace}

\newcommand{\linx}[1]{{\color{blue} #1}}

%%%%%%%%%%%%%%%%%%%%%%%%%%%%			%%%%%%%%%%%%%%%%%%%%%%%%%%%%
%%%%%%%%%%%%%%%%%%%%%%%%%%%% Below by prof. Lin %%%%%%%%%%%%%%%%%%%%%%%%%%%%
%%%%%%%%%%%%%%%%%%%%%%%%%%%%			%%%%%%%%%%%%%%%%%%%%%%%%%%%%
\newcommand{\nnone}{n_1}
\newcommand{\nntwo}{n_2}
\newcommand{\mone}{m_1}
\newcommand{\mtwo}{m_2}
\newcommand{\muone}{\mu_1}
\newcommand{\mutwo}{\mu_2}
\newcommand{\sigmaone}{\sigma_1}
\newcommand{\sigmatwo}{\sigma_2}
\newcommand{\Tc}{T^c}
\newcommand{\Tcprime}{\tilde{T}^c}
%%%%%%%%%%%%%%%%%%%%%%%%%%%%			%%%%%%%%%%%%%%%%%%%%%%%%%%%%
%%%%%%%%%%%%%%%%%%%%%%%%%%%% Above by prof. Lin %%%%%%%%%%%%%%%%%%%%%%%%%%%%
%%%%%%%%%%%%%%%%%%%%%%%%%%%%			%%%%%%%%%%%%%%%%%%%%%%%%%%%%
\newcommand{\sigmanot}{\sigma_o}
\newcommand{\sigmanotsqrd}{\sigmanot^2} 
\newcommand{\sigmaonesqrd}{\sigmaone^2}
\newcommand{\yvec}{\vec{y}}

\maketitle

\input abstract
\input intro
\input back
\input sampling

\input comparison

\input case-study
\input dag
\input discussion
\input conc
\input paper_acknowledgement
\iftoggle{nsdi22}{}{\input appendix}
\iftoggle{journal}{}{\pagebreak}

\iftoggle{journal}{\bibliographystyle{abbrv}}{
\bibliographystyle{plain}}
\bibliography{js}
%\bibliography{sample-bibliography}

\end{document}

%% file: abstract.tex
\section*{Abstract}
The ability to accurately estimate job runtime properties allows a
scheduler to effectively schedule jobs.
State-of-the-art online cluster job schedulers use history-based
learning, which uses past job
execution information to estimate the runtime properties of newly arrived
jobs. However, with fast-paced development in cluster technology (in both
hardware and software) and changing user inputs, job runtime properties
can change over time, which lead to inaccurate predictions.

In this paper, we explore the potential and limitation of real-time learning of
job runtime properties, by proactively sampling and scheduling a small fraction
of the tasks of each job. Such a task-sampling-based approach exploits the
similarity among runtime properties of the tasks of the same job and is
inherently immune to changing job behavior.
\if 0
Our study focuses on two key questions in comparing 
task-sampling-based learning (\textit{learning in space}) and
history-based learning (\textit{learning in time}):
(1) Can learning in space be more accurate than learning in time?
vvnvv(2) If so, can delaying scheduling the remaining
tasksaxf of a job till the completion of sampled tasks
be more than compensated by the improved accuracy
and result in improved job performance?
\fi
Our analytical and experimental analysis of \numTraces production traces with
different skew and job distribution shows that learning in space can be
substantially more accurate.
%  and led to  1.28$\times$, 1.56$\times$, and 1.32$\times$
%  reduction in the average Job Completion Time (JCT) when
%  compared to the prior-art history-based predictor
%  in our simulation and testbed evaluation on Azure
%  using \numTraces production cluster job traces
Our simulation and testbed evaluation on Azure of
the two learning approaches anchored in a generic job scheduler using \numTraces
production cluster job traces shows that despite its online overhead, learning
in space reduces the average Job Completion Time (JCT) by 1.28$\times$,
1.56$\times$, and 1.32$\times$ compared to the prior-art history-based
predictor.
{Finally, we show how the sampling-based 
learning can be extended to schedule DAG jobs and achieve
similar speedups over the prior-art history-based predictor.
}

%% file: intro.tex
\section{Introduction}
\label{sec:intro}

In big-data compute clusters, jobs arrive online and compete to share the
cluster resources. In order to best utilize the cluster and to ensure
that jobs also meet their service level objectives, efficient
scheduling is essential. However, as jobs arrive online, their runtime
characteristics are not known a priori. Due to this lack of
information, it is challenging for the cluster scheduler to determine
the right job execution order that optimizes scheduling metrics
such as maximal resource utilization or application service level objectives.

An effective way to tackle the challenges of cluster scheduling is to
learn the runtime characteristics of pending jobs, 
% as accurately
% estimating job runtime characteristics 
which
allows the scheduler to
exploit offline scheduling algorithms that are known to be optimal,
\eg Shortest Job First (SJF) for minimizing the average completion time.
Indeed, there has been
{a large amount of}
%{many}
work~\cite{tetrisched, morpheus,
3Sigma, IfYouAreLateDontBlameUs:socc14, DontCryOverSpilledRecords, corral,
perforator:socc2016, cdef:atc18} on learning job runtime characteristics to
facilitate cluster job scheduling.
%        , \eg to leverage optimal offline algorithms using the estimated job
%        runtime information.  or improve avaialbility by predicting error
%        rates~\cite{cdef:atc18}}.  error rates of what?
%{Indeed, there have been a large amount of work on learning job runtime
%characteristics online to facilitate cluster job scheduling, \eg to leverage
%optimal offline algorithms to minimize job compeltion time \cite{tetrisched,
%morpheus, 3Sigma, IfYouAreLateDontBlameUs:socc14, DontCryOverSpilledRecords}
%or better resource utilization \cite{shufflewatcher, corral}, or meeting
%deadlines \cite{morpheus, 3Sigma} or improving availability by predicting
%error rates \cite{cdef:atc18} or optimize dollar cost in virtual machine
%scheduling \cite{perforator:socc2016, stratus:socc2018}.}

In essence, all of the previous learning algorithms learn job
runtime characteristics from observing historical executions of the
same jobs, which execute the same code but process different sets of
data, or of similar jobs, which have matching features such as the same
application name, the same job name, or the same user who submitted the
job.

The effectiveness of the above {\em history-based} learning schemes
critically rely on two conditions to hold true: 
(1) The jobs are recurring; (2) The performance of the same or
similar jobs will remain consistent over time.

In practice, however, the two conditions often do not hold true.
First, many previous work have acknowledged that not all jobs are
recurrent. For example, in the traces used in Corral \cite{corral} and
Jockey~\cite{jockey:eurosys2012}, only
40\% of the jobs are recurrent, and Morpheus \cite{morpheus} shows that
only 60\% of the jobs are recurrent.
Second, even the authors of history-based prediction schemes such as
3Sigma~\cite{3Sigma} and Morpheus~\cite{morpheus} strongly argued why
runtime properties of jobs, even with the same input, will not remain
consistent and will keep evolving. The primary reason is due to
updates in cluster hardware, application software, and user scripts to execute
the cluster jobs.
% can affect job runtime characteristics.
%
%\commentaj{Why \S\ref{sec:back} mentioned below?}
Third, our own analysis of three
production cluster traces (\S\ref{sec:accuracy}) have also shown that historical job runtime
characteristics have considerable variations.

In this paper, we explore an alternative approach to learning runtime
properties of distributed jobs online to facilitate cluster
job scheduling.
The approach is motivated by the following key observations about
distributed jobs running on shared clusters: (1) a job typically has a
{\em spatial dimension}, \ie it typically consists of many tasks; and (2)
the tasks (in the same phase) of a job typically execute the same code
and process different chunks of similarly sized
data~\cite{googleClusterData2011-2Schema,
  personalCommunication:MarkAstley}.  These observations suggest that
if the scheduler first schedules a few sampled tasks of a job, known as
pilot tasks, to run till finish, it can use the observed runtime
properties of those tasks to accurately estimate those of the whole
job.  Effectively, such a {\em task-sampling-based} approach learns job
properties in the spatial dimension.  We denote the new learning
scheme as \slearn, for ``learning in space''.

Intuitively, by using the execution of pilot tasks to predict the
properties of other tasks, \lTechnique avoids the primary drawback of
history-based learning techniques, \ie relying on jobs to be recurring
and job properties to remain stationary over time.  However, learning
in space introduces two new challenges: (1) its estimation accuracy
can be affected by the variations of task runtime properties, \ie task skew; 
(2) delaying scheduling the remaining tasks of a job till the
completion of sampled tasks may potentially hurt the job's completion
time.

In this paper, we perform a comprehensive comparative study
of history-based learning (learning in time) and sampling-based learning
(learning in space), to systematically answer the following questions:
% \begin{enumerate}
% \item 
{\em
(1) Can learning in space be more accurate than learning in time?
% \item 
(2) If so, can {delaying scheduling} the remaining tasks of a job till the completion
  of sampled tasks
  be more than compensated by the improved accuracy and result
  %   negatively affect   the overall job performance,
  in improved job performance, \eg completion time?
%\end{enumerate}
}

We answer the first question via quantitative analysis, and trace and
experimental analysis based on three production job traces, including two public
cluster traces from Google released in 2011 and 
2019~\cite{googleTraceGithub,googleClusterData2019} and a private trace
from 2Sigma~\cite{2Sigma:website}.  We answer the second
question by designing a generic scheduler that schedules jobs based on
job runtime estimates to optimize a given performance metric, \eg
average job completion time (JCT), and then plug into the scheduler
different prediction schemes, in particular, learning in time and
learning in space, to compare their effectiveness.
% using three production job traces.

%\subsection{Emperical Analysis}
%\label{sec:intro:empAna}
%Using two production cluster traces, we studied variation in runtime properties
%of jobs across history and across tasks of same job. We used a publicaly
%available trace extracted from clusters of Google~\cite{googleTraceGithub} and
%a private trace from 2Sigma~\cite{2Sigma:website}.
%Figure~\ref{fig:intro:empAna} shows coeffiecient of variation()waitingTimes in
%
%\begin{figure}[tp]
%\centering
%\subfigure[GTrace]{
%\vspace{-0.2in}
%	\includegraphics[width=0.9\linewidth]{figures/simulation/average_task_waiting_time_gTrace.pdf} %done
%	\label{fig:sim:waitingTimes:gTrace}
%	\vspace{-0.1in}
%}
%\subfigure[2STrace]{
%	\includegraphics[width=0.9\linewidth]{figures/simulation/average_task_waiting_time_2STrace.pdf} %done
%	\label{fig:sim:waitingTimes:2Strace}
%	\vspace{-0.1in}
%}
%\caption{Waiting times for job.}
%\label{fig:intro:empAna}
%\vspace{-0.2in}
%\end{figure}

We summarize the major findings and contributions of this paper as follows:
\begin{itemize} %% [noitemsep,topsep=0pt,leftmargin=0.2in]
\item Based on literature survey and analysis using three production
	cluster traces,
	%, one from Google and another 2Sigma, 
	we show that history is not a stable and accurate predictor for
	runtime characteristics of distributed jobs.

%\item We propose, \namepredict, the novel idea of applying sampling in the spatial dimension
      \item We propose \lTechnique, a novel learning approach
        % for distributed jobs. % \lTechnique
        that uses sampling in the spatial dimension of jobs to learn job runtime
        properties online.  \addnsdiSHP{We also provide solutions to practical 
        issues such as dealing with thin jobs (jobs with a few tasks only) and
        work 
        conservation.}{C6} % with high accuracy.

\item Via quantitative, trace and experimental analysis, we demonstrate
  that \lTechnique can predict job runtime properties with much higher
  accuracy than history-based schemes. {For the 2Sigma,
 Google 2011, and Google 2019 cluster traces, the median prediction error are 18.98\%, 
13.68\%, and 51.84\% for \lTechnique but 36.57\%, 21.39\%, and 71.56\% for the
state-of-the-art history-based \primarybasepredict,
  respectively.}
\item We show that learning job runtime properties by sampling
  job tasks, although delays scheduling the remaining tasks
  of a job, can be more than compensated by the improved accuracy, and
  as a result reduces the average JCT.
  In particular, our extensive simulations and testbed experiments
  using a prototype on a 150-node cluster in Microsoft Azure
  show that compared to the prior-art history-based predictor, \slearn
  reduces the average JCT by 1.28$\times$, 1.56$\times$, and 1.32$\times$ for the
  extracted 2Sigma, Google 2011 and Google 2019 traces, respectively.
\item We show how the sampling-based learning can be extended to schedule
  DAG jobs. Using a DAG trace generated from the Google 2019 trace,
  we show a hybrid sampling-based and history-based scheme 
  reduces the average JCT by 1.25$\times$
over a pure history-based scheme.
\end{itemize}

%% file: back.tex
\section{Background and Related Work}
\label{sec:back}

In this section, we provide a brief background on the cluster scheduling problem,
review existing
% schedulers based on learning job runtime characteristics,
learning-based schedulers,
and discuss their weaknesses.

%\vspace{-0.1in}
\subsection{Cluster Scheduling Problem}
\label{sec:back:problem}

%Growing use of cloud for variety of applications, increase in shared usage of
%the cluster among multiple users and execution of diverse workloads in the same
%cluster due to cluster consolidation leads to challenge of appropriately
%scheduling jobs in order to best utilize the cluster. The notion of best
%utilization might vary. In some cases requirement is to ensure fairness accross
%users, in some other meeting deadlines for maximum number of jobs is most
%important. Whereas for some cases quickly completing maximum number of jobs is
%the desired goal. Also, in many cases minimizing the operational cost of the
%cluster is the primary goal and for achieving that it is important to minimize
%average job completion time. Last two situations leads to the cluster
%scheduling problem with goal of minimizing average job completion time. In this
%paper we focus on online cluster scheduling problem with goal of minimizing
%average job completion time.

%\questionaj{Why the first para here is not good? Is it not a way to tell that
%why scheduling is a problem?\\}
%Businesses ranging from Fortune-500 companies to small seed funded start-ups are
%increasingly relying on shared clusters for executing thier business critical
%jobs. The efficiency of cluster usage has a direct impact on business running cost.
%This makes efficient cluster scheduling and meeting desired scheduling
%objectives very crucial.

In both public and private clouds,
clusters are typically shared among multiple users to execute diverse jobs. Such
jobs typically arrive online and compete for shared resources. In order to best
utilize the cluster and to ensure that jobs also meet their service level
objectives (SLOs), efficient job scheduling is essential. Since jobs arrive online,
their runtime characteristics are not known a priori. This lack of information
makes it challenging for the scheduler to determine the right order for running
the jobs that maximizes resource utilization and/or meets application 
SLOs. Additionally, jobs have different SLOs. For some
meeting deadlines is important while for others faster completion or minimizing
the use of networks is more important. Such a diverse set of objectives pose
further challenges to effective job
scheduling~\cite{drf:nsdi11,jockey:eurosys2012, shufflewatcher, corral,
morpheus, delay:eurosys10, cdef:atc18}.

%\soccReviewEdit{A, B}{\paragraph{Job model.}
%\vspace{-0.1in}
\subsection{Job Model}
\label{sec:back:jobmodel}

We consider big-data compute clusters running data-parallel frameworks
such as Hadoop~\cite{hadoop:web}, Hive~\cite{hive:web},
Dryad~\cite{dryad:eurosys2007}, 
Scope~\cite{scope:2008},
and Spark~\cite{spark:web} that run simple MapReduce
jobs~\cite{mapreduce:osdi04} or more complex DAG-structured jobs, where each job
processes a large amount of data. Each job consists of one or multiple
stages, such as map or reduce, and each stage 
partitions the data into
manageable chunks and runs many parallel tasks,
each for processing one data chunk.  

\subsection{Existing Learning-based Schedulers}
\label{sec:back:existing}

\begin{table}[tp]
%\vspace{-0.05in}
\caption{Summary of selected previous work that use history-based learning techniques.}
\label{table:prevwork}
	%\questionaj{I have not mentioned all the prior works in table \ref{table:prevwork} which are in \S\ref{sec:back:existing}. The works I have mentioned in the table cover all varieties. I didn't add other things to keep the table succint. Do you think we should expand the table?\\}
%\commentaj{I am not mentioning prediction accuracy as they are differently measured in different papers and are on different traces. Comparison of those values are not very sensible.}
\centering
{\small
\vspace{-0.1in}
\begin{tabular}{|c|c|c|c|c|}
\hline
		\textbf{ Name} & \textbf{Property} & \textbf{Estimation} & \textbf{Learning} \\ %& \textbf{Cluster traces used}\\
			& \textbf{estimated} & \textbf{technique} & \textbf{frequency} \\ %& \textbf{(Private + Public)}\\
	 
%\hline
\hline	
	  %\textbf{Corral \cite{corral}} & Job runtime & Profiling models & Makes several impractical assum- &  1 + 1 \\
	  \textbf{Corral } & Job runtime & Offline model & On arrival \\ %& 1 + 1 \\
	  \textbf{\cite{corral}} &  & (not updated) & \\% &\\
\hline
	  \textbf{DCOSR} & Memory elasti- & Offline model & Scheduler \\ %& 0 + 0 \\
	  \textbf{\cite{DontCryOverSpilledRecords}}& city profile & (not updated) & dependent \\% &\\
\hline
	  \textbf{Jockey} & Job runtime & Offline & Periodic \\ %& 1 + 0 \\
	  \textbf{\cite{jockey:eurosys2012}}& & simulator  &  \\% &\\
\hline
	  \textbf{3Sigma} & Job runtime & Offline & On arrival \\ %& 3 + 1 \\
	  \textbf{\cite{3Sigma}} & history dist. & model &  \\% &\\

	%%%%% Morpheus was cut from text so we have removed it from table as well %%%%%	  
%\hline
%	  \textbf{Morpheus} & SLOs; Resource & Offline & On arrival \\ %& 1 + 0 \\
%	  \textbf{\cite{morpheus}} & requirements &  model   &  \\% &\\
%\hline
%	  \textbf{Tetrisched \cite{tetrisched}} &  &  &  & 0 + 1 \\
\hline
\end{tabular}
%\begin{tabular}{|c|c|c|c|c|c|}
%\hline
%		\textbf{ Name} & \textbf{Property} & \textbf{Estimation} & \textbf{Target} & \textbf{Learning}  & \textbf{Cluster traces used}\\
%			& \textbf{estimated} & \textbf{technique} & \textbf{job types} & \textbf{frequency} & \textbf{(Private + Public)}\\
%	 
%%\hline
%\hline	
%	  %\textbf{Corral \cite{corral}} & Job runtime & Profiling models & Makes several impractical assum- &  1 + 1 \\
%	  \textbf{Corral \cite{corral}} & Job runtime & Offline & Recurrent and & On arrival & 1 + 1 \\
%	  &  & static model & offline jobs &  &\\
%\hline
%	  \textbf{Jockey \cite{jockey:eurosys2012}} & Job runtime  & Offline & Recurrent jobs & Periodic & 1 + 0 \\
%	  &  & simulator & &  &\\
%\hline
%	  \textbf{3Sigma \cite{3Sigma}} & Job runtime & Distribution of job & All jobs & On arrival & 3 + 1 \\
%	  & & runtime history & &  &\\
%\hline
%	  \textbf{Morpheus \cite{morpheus}} & SLOs; Resource & Offline  & Recurrent jobs & On arrival & 1 + 0 \\
%	  & requirements &  static model &  &  &\\
%\hline
%	  \textbf{Don't cry \cite{DontCryOverSpilledRecords}} & Memory elasticity & Offline & Recurrent jobs & Scheduler & 0 + 0 \\
%	  & profile & static model &  & dependent &\\
%%\hline
%%	  \textbf{Tetrisched \cite{tetrisched}} &  &  &  & 0 + 1 \\
%\hline
%%\vspace{-0.2in}
%\end{tabular}
}
\vspace{-0.1in}
\end{table}

An effective way to tackle the challenges of cluster scheduling is to
learn runtime characteristics of pending jobs.
%  If we can accurately
%  estimate jobs characteristics, we can leverage offline scheduling
%  algorithms that are known to be optimal, \eg \removesl{Shortest Job First}{SJF}
%  for minimizing the average completion time.
% Indeed the problem of learning runtime characteristics of jobs has been intensively
% studied.
As such cluster schedulers using various learning methods have been proposed
\cite{corral, morpheus, shufflewatcher, 3Sigma, tetrisched,
DontCryOverSpilledRecords, perforator:socc2016, Apollo:osdi2014, wsmith:IEEE98, stratus:socc2018, roughSetEstimation:IEEE:Shonali}.  
In essence, all previous learning schemes
are {\em history-based}, \ie they learn job characteristics by observations
made from the past job executions.\footnote{\addnsdiSHP{
Some recent work use the characteristics of completed mini-batches as 
a proxy for the remaining mini-batches, to improve the scheduling of ML jobs \cite{gandiva:osdi18}.
However, such jobs are different in that the mini-batches in general experience 
significantly less (task-level) variations than what we studied in this paper.}{C4}
}
 In particular,
existing
% history-based
learning approaches can be broadly categorized into the following groups, as
summarized in Table~\ref{table:prevwork}.

%\cite{DontCryOverSpilledRecords} attempt to come up with a resource assignment
%using offline models.
%\paragraph{Offline prediction or profiling.}

\paragraph{Learning offline models.}
\if 0
\commentaj{Sampling based approach is fundamentally different from online
updates.  Learning by sampling is in-principle same as offline learning for
scheduling all the non-sampling tasks as non-sampling tasks are not scheduled
till sampling is over. Whereas online updates act as an error correction
mechanism while scheduling in the remaining tasks, if any, of the job or to
estimate remaining execution time of running jobs while scheduling new jobs.\\}
Corral~\cite{corral} uses an offline predictor for estimating job running
times, which it acknowledges may not be highly accurate.
\fi
Corral's prediction model is designed with the primary assumptions that most
jobs are recurring in nature,
% and % . It also makes several
% additional assumptions
% such as 
and the latency of each stage of a multi-stage job is proportional
to the amount of data processed by it, which do not always hold true~\cite{corral}.

DCOSR~\cite{DontCryOverSpilledRecords} predicts the memory usage for
data parallel compute jobs using an offline model built from a fixed
number of profile runs that
% .  These profile runs
are specific to the
framework and depend on the  framework's properties.
% as well as the hardware they run on.
Any software update in the existing frameworks,
addition of new framework or hardware update will require an update in
profile.

%The Yarn capacity scheduler \cite{yarnCapacity:web} also uses offline modeling
%created by profiling data gathered from execution history. \todoaj{Find
%details about following - to do scheduling? modeling what? and how is the
%model used?}

%Tetris \cite{MultiResourcePackingForClusterSchedulers} also uses history
%for predicting job running times for recurring jobs. 
For analytics jobs that perform the same computation periodically on different sets
of data, Tetris~\cite{MultiResourcePackingForClusterSchedulers} takes
measurements from past executions of a job to estimate the requirements for
the current execution.

%{how to predict? -- aj: Paper doesn't provide any more details on it.}

%\questionaj{This paper uses 4 different ways for prediction. Two are plain
%hueristics. Third is for reccuring jobs and uses history. Another is for such
%jobs whose tasks arrive at different times and so they use measurements from
%first few tasks for the later task. The second one is not exactly sampling but
%in principle it is the same.  Do you think mentioning this paper might make
%confused about our novelity?}

%Another reservation based scheduler \cite{IfYouAreLateDontBlameUs:socc14}
%uses history baesd predictor. % Commenting this as no name for the scheduler or predictor, hard to cite and not very important. 

\paragraph{Learning offline models with periodic updates.}
Jockey \cite{jockey:eurosys2012} periodically characterizes job progress at
runtime, which along with a job's current resource allocation is used
{by an offline simulator to estimate the job's completion time
% . Based on   the estimated completion time, 
and update the job's resource allocation.}
%  {as input to an offline estimated random variable function. The output of the
% function is used to update the resource allocation of the job so that its utility
% is optimized.}
\rm{The running time estimates made by the simulator are based
on performance statistics extracted from one or more previous runs of the
recurring job.}
%
% Jockey's learning module relies on the periodicity of job
Jockey relies on job
recurrences and cannot work with new jobs.

%Tetrisched \cite{tetrisched} leverages the estimated job running
%times \comment{how does it estimate?}  and deadlines provided by the
%reservation systems of scheduling frameworks like Hadoop YARN
%\comment{to plan ahead in deciding whether or not to wait for a busy preferred resource.} It
%also coordinates with the reservation system to re-evaluate the
%immediate-term scheduling plan for all pending jobs at every scheduling
%cycle. \comment{this sentence is strange. above has not talked about estimation?}
%It is based on assumption that most of the jobs will be
%reccuring in nature.

%Tetrisched \cite{tetrisched} works in conjunction with reservation systems of
%scheduling frameworks like Hadoop YARN and leverages information provided by
%them about job deadlines and estimated runtimes to plan ahead in deciding
%whether to wait for a busy preferred resource or work with less preferred
%options. Tetrisched coordinates with the reservation system to re-evaluate the
%immediate-term scheduling plan for all pending jobs every scheduling cycle.
%Tetrisched is based on assumption that most of the jobs will be reccuring in
%nature.

%Perforator \cite{perforator:socc2016} also leverages job structure and
%profiling to predict job runtimes.

\paragraph{Learning from similar jobs.}
% JVuPredict~\cite{jamiasvu} uses a black-box approach to learn job properties.
Instead of using execution history from the exact same jobs,
JVuPredict~\cite{jamiasvu}
matches jobs on the basis of some common features such as application name,
job name, the user who owns the job, and the resource requested by the job.
\if 0
Additionally, 
% instead of using just one mathematical metric for estimating
% running times from the distribution, 
it uses multiple metrics, such as rolling
average and median,
in estimating the running time of a new job from such similar jobs.
%   In principle, this elaborate design of JVuPredict 
%   revokes the reliance on recurring jobs, but it still depends on historical data.
\fi
%
%\paragraph{Distribution based learning.}
3Sigma~\cite{3Sigma} 
%  shares the two ideas of
extends JVuPredict~\cite{jamiasvu}
% , \ie matching jobs on the basis of some common features and using multiple metrics,
% and hence is also applicable to non-recurring jobs.
by introducing a new idea on prediction: instead of using point
metrics to predict runtimes, it uses full distributions of relevant
runtime histories.
\if 0
3Sigma~\cite{3Sigma} differs from previous history-based learning in three ways.
(1) Instead of using point metrics to predict runtimes, it uses full
distributions of relevant runtime histories.
(2) Instead of using execution history from the
exact same jobs, it matches jobs on the basis of some common features such as
application name, job name, the user who owns the job, and the resource requested by the
job.
%
(3) Instead of using just one mathematical metric for
estimating running times from the distribution, it uses multiple metrics, such as
rolling average and median.
\fi
%  It maintains a running metric to pick a feature
%  value and metric pair that can be best estimator.
%It maintains a running metric which tells that
%which feature value and metric pair can be best estimator and always use the
%one with the highest score.
%  {Though in principle, this elaborate design of
% 3Sigma revokes the reliance on recurring jobs, it still depends on historical data.
However, since it is impractical to maintain precise distributions for each
feature value, it resorts to approximating distributions, which
compromises the benefits of having full distributions. 

\subsection{Learning from History: Assumptions and Reality}
\label{sec:back:whatsWrong}

Predicting job runtime characteristics from history information relies on
the following two
conditions to hold,
%  : (1) The jobs are recurring; (2) The performance of
%  the same {or similar} jobs
%  remain consistent over time. Below,
which we argue
% why these conditions
may not be applicable to modern day clusters.

%One of the major assumptions history based predictors rely on is that most of
%the jobs are recurring.

%Different history-based predictors are designed in accordance with the distribution
%and behavior pattern of jobs in the cluster.
%%Though they incorporate different ways in their design to mitigate the harm
%%which will happen when such assumptions fail\cite{corral, jockey:eurosys2012,
%%morpheus, 3Sigma}.
%Such predictors will not perform in an expected way when there is a
%considerable change in the fraction of recurring jobs in the cluster.
%%Or overall the cluster property is not very similar to properties for which the
%%predictor was designed.
%Different cluster schedulers which use history-based prediction mechanisms,
%based on the requirement of their workload, incorporate different ways in
%their design to mitigate the harm which will happen when such assumptions
%fail\cite{corral, jockey:eurosys2012, morpheus, 3Sigma}. Such predictors might
%not perform in an expected way when there is a considerable change in the
%fraction of recurring jobs in the cluster or overall the cluster property is
%not very similar to properties for which the predictor was designed.

%Cluster schedulers which use these prediction mechanisms 
%appropriately build around exploiting this assumption in different ways
%\cite{jockey:eurosys2012, 3Sigma, morpheus, corral}.  As these schedulers are
%primarily exploiting the reccurance of jobs and with the 
%might not perform in expected way when there is considerable change in fraction
%of reccurring jobs in the cluster. 

%\noindent
%\vspace{-0.2in}
\paragraph{Condition 1: The jobs are recurring.}
Many previous works have acknowledged that not all jobs are
recurrent. For example,
the traces used in Corral \cite{corral} and Jockey
\cite{jockey:eurosys2012} show that only 40\% of the jobs are recurrent and
Morpheus \cite{morpheus} shows that 60\% of the jobs are recurrent.
%A trace analysis study
%\cite{workloadDiversity:atc18} has shown that there can be a huge variation in
%properties across traces from different clusters as well as among jobs in the
%same cluster.
%In 3Sigma \cite{3Sigma} performance gain measured for the same metric can vary
%upto 8$\times$ just by varying the trace.
%%In 3Sigma \cite{3Sigma},performance gain measured for the same metric can vary
%%from 1.5$\times$ to 8$\times$ just by varying the trace.
%Also, the number of jobs with >95\% error in predicted runtimes varies upto
%4$\times$ and number of jobs which have zero to negligible error varies upto
%3.5$\times$ across different traces.
%History based predictors, directly or indirectly, rely on some common
%assumptions.  However, many of these assumptions are not applicable for modern
%day's clusters.
%So, cluster schedulers have to provide mechanisms to mitigate
%effects of failure of these assumptions \cite{jockey:eurosys2012, 3Sigma,
%morpheus, corral}.  
%In this subsection, we discuss those assumption and their validity.

%\paragraph{Assumption 2: Characteristics of recurrent jobs are highly predictable.}
%Another property that history based predictors would ideally want to have is
%that characteristics of the recurrent jobs are highly predictable.
%\paragraph{Assumption 2: Runtime characteristics of newly arriving jobs can be
%predicted with high accuracy using historical data.\\}

%\noindent
%\vspace{-0.2in}
\paragraph{Condition 2: The performance of the same
{or similar} jobs will remain consistent over time.}
Previous works~\cite{3Sigma, morpheus, corral, jockey:eurosys2012} that exploited
history-based prediction have considered jobs in one of
the following two categories.
%\begin{itemize}
(1) {\em Recurring jobs}: A job is re-scheduled to run on newly arriving data;
(2) {\em Similar jobs:} A job has not been seen before but has some
attributes in common with some jobs executed in the past~\cite{jamiasvu,3Sigma}.
%  The attributes can be application name, job name, user of
%  the job, or amount of resources requested by the job.
%\end{itemize}
Many of the history-based approaches only predict for recurring jobs \cite{morpheus, corral,
jockey:eurosys2012}, while some others \cite{3Sigma, jamiasvu, stratus:socc2018, roughSetEstimation:IEEE:Shonali} work for both categories of jobs.

However, even the authors of history-based prediction schemes such as
3Sigma~\cite{3Sigma} and Morpheus~\cite{morpheus} strongly argued why
runtime properties of jobs, even with the same input, 
% will not remain consistent and 
will keep evolving.  The primary reason is that updates in cluster
hardware, application software, and user scripts to execute the cluster jobs
affect the job runtime characteristics.
{They found that}
in a large Microsoft production cluster, within a one-month period,
applications corresponding to more than 50\% of the recurring jobs were
updated. The source code changed by at least 10\% for applications
corresponding to 15-20\% of the jobs.  Additionally, over a one-year period,
the proportion of two different types of machines in the cluster
changed from 80/20 to 55/45. For a same production Spark job, there is
a 40\% difference between the running time observed on the two types of
machines~\cite{morpheus}.  

For these reasons, although the state-of-the-art history-based system 3Sigma~\cite{3Sigma} 
uses sophisticated prediction techniques,
% Machine Learningd also shown that
the predicted running time for more than 23\% of the jobs have at
least 100\% error, and for many the prediction is off by an order of
magnitude.
\if 0
 In our analysis of three production cluster traces (see
Figure~\ref{fig:accuracy:trace_analysis_window:sp3} on page~\pageref{fig:accuracy:trace_analysis_window:sp3}), we observed similar
levels of high variability in the runtime characteristics of the jobs with the same
attributes.
\fi

% \questionaj{The trace analysis will be disscused in
% \S\ref{sec:comparison:quantity}. Do you think there is any need to add more of it here?}

%Above observations make it quite evident that even in short duration runtimes
%can vary significantly for similar jobs also.\\
%The above observations make it clear that the assumption that a large fraction of
%jobs is recurrent is not very reliable.
%Additionally, characteristics of jobs
%can vary a lot from one cluster settings to others and hence the design of the
%scheduler needs to make those accomodations.
%\cite{jockey:eurosys2012} and \cite{morpheus} have been evaluated only on
%single trace.

\if 0
\paragraph{}
Although the previous history based predictors have acknowledged
the above discussed drawbacks, they did not provided a solution to tackle the
root cause of the limitation, \ie the assumption that
job runtime properties remains stationary over time \cite{jockey:eurosys2012, 3Sigma, morpheus, jamiasvu}.
%They atmost provide steps to mitigate the effects of
%the drawbacks \cite{jockey:eurosys2012, 3Sigma, morpheus, jamiasvu}.
\fi

%% file: sampling.tex
\section{\lTechnique\ -- Learning in Space}
\label{sec:sampling}

In this paper, we explore an alternative approach to learning job runtime
properties online in order to facilitate cluster job scheduling.  The approach
is motivated by the following
%  \lTechnique, a novel learning technique with a very different
%  approach for learning runtime properties of distributed jobs. \lTechnique makes
key observations about distributed jobs running in shared clusters: (1) a
distributed job has a spatial dimension, \ie it typically consists of many
tasks; (2) all the tasks in the same phase of a job typically execute 
the same code with the same settings~\cite{googleClusterData2011-2Schema,
personalCommunication:MarkAstley,googleClusterData2019Schema}, 
and  differ in that they process different
chunks of similarly sized data. 
%   \commentaj{Also, if there is a need to run different types
%   of tasks with different resource requirements, they are executed as a separate
% jobs~\cite{googleClusterData2019Schema}.}
Hence, it is likely that their runtime behavior
will be statistically similar.

The above observations suggest that if the scheduler first schedules a few
sampled tasks of a job to run till finish, it can use the observed runtime
properties of those tasks to accurately estimate those of the whole job. 
In a modular design, such an online learning scheme can be decoupled from the
cluster scheduler.  In particular, upon a job arrival, the predictor first
schedules sampled tasks of the job, called {\em pilot tasks}, till their
completion, to learn the job runtime properties. The learned job properties are
then fed into the cluster job scheduler, which can employ different scheduling
polices to meet respective SLOs.  Effectively, the new scheme learns job
properties in the spatial dimension, \ie  {\em learning in space}.  We denote
the new learning scheme  as \lTechnique.

% \lTechnique is a generic learning approach which can be used for estimating
%runtime properties of distributed jobs.
%  different schedulers can then use the estimated values to come up with a
%  schedule to meet the desired scheduling goals.
%the property of other tasks of the job.

%  Intuitively, by using the execution of pilot tasks to predict the properties of
%  other tasks, \lTechnique avoids the primary drawback of history-based learning
%  techniques, \ie relying on job properties to remain stationary over time.

\if 0
Learning in space introduces two challenges:
%
(1) its estimation can be affected by the variations of task runtime
properties, \ie task skew;
%
(2) delaying scheduling the remaining tasks of a job till the completion of
sampled tasks may potentially hurt the job's completion time.
\fi
\begin{table}[tp]
%\vspace{-0.05in}
\caption{Comparison of learning in time and learning in space of job runtime properties.}
\label{table:proscons}
\centering
{\small
\vspace{-0.1in}
\begin{tabular}{|c|c|c|c|c|c|}
\hline
                        & Applicability & Adapti- & Accuracy & Runtime \\
	                &               &         veness     &          & overhead\\
%\hline
\hline
	Time & Recurring jobs & No/Yes & Depends  & No\\
	%based& ing jobs  & &      &\\
\hline
	Space &New/Recurring jobs & Yes & Depends & Yes\\
	 %based &    &&&\\
\hline
%\vspace{-0.2in}
\end{tabular}
}
\vspace{-0.2in}
\end{table}

% \paragraph{Learning in Time vs. Learning in Space}
% \label{sec:comparison}

Table~\ref{table:proscons} summarizes the pros and cons of the two
learning approaches along four dimensions:
%
%  \begin{itemize}%[leftmargin=*]
%  \vspace{-0.05in}
%  \item 
(1) {\bf Applicability:} As discussed in \S\ref{sec:back:existing}, most history-based
predictors cannot be used for the jobs of a new category or for categories
for which the jobs are rarely executed.  In contrast, learning in space
has no such limitation; it can be applied to any new job.
% \vspace{-0.05in}
% \item 
(2) {\bf Adaptiveness to change:} Further, history-based predictors assume
job runtime properties persist over time, which often does not hold, as discussed 
in \S\ref{sec:back:whatsWrong}.
% In contrast, learning in space does not have such limitation.
% \vspace{-0.05in}
% \item 
(3) {\bf Accuracy:}
The accuracy of the two approaches are directly affected by
how they learn, \ie in space versus in time.
The accuracy of history-based approaches is affected by how stable
the job runtime properties persist over time, while that  of
sampling-based approach is affected by the variation of the task runtime properties,
\ie the extent of task skew.
%  \questionaj{Should we add a briefly add here the intuitive argument as why
%  ampling will work in task runtime skew? Similar to the one we added for 
%  the coflow paper. I have written this in Appendix \S\ref{sec:appendixSkew}.}
%
% \vspace{-0.05in}
% \item 
(4) {\bf Runtime overhead:} The history-based approach
has an inherent advantage of having very low to zero runtime
overhead. It performs offline analysis of historical data to generate
a prediction model.
\rm{Afterwards there is almost no overhead in
estimating runtime characteristics of newly arriving jobs. Variations
of history-based predictors that use runtime feedback to update the
prediction models may have some cost, but usually such systems are
optimized to have low runtime update overhead.}  In contrast,
sampling-based predictors do not have offline cost, but
need to first run a few pilot tasks till completion
before scheduling the remaining tasks. This may potentially
delay the execution of non-sampled tasks.
% job completion time.
%  Though, another point to be noted is that the
%  sampling based systems do not have any offline cost.
% \end{itemize}

The above qualitative comparison of the two learning approaches raises the
following two questions:
{\em
(1) Can learning in space be more accurate than learning in time?
(2) If so, can {delaying scheduling} the remaining tasks of a job till the completion
of sampled tasks be more than compensated by the improved accuracy, so that the
overall job performance, \eg completion time, is improved?
}
We answer the first question
via analytical, trace and experimental analysis
in \S\ref{sec:accuracy}
and
the second question
via a case study of cluster job
scheduling using the two types of predictors in \S\ref{sec:study}.

%% file: comparison.tex
\section{Accuracy Analysis}
\label{sec:accuracy}
% 1. Write about using training data as best of the history here and in \S\ref{sec:study}.\\ 
% 2. Argument for using the 2Sigma trace as the main
% evaluation trace and other answers about the trace diversity here and in
% \S\ref{sec:study}.}

In this section, we perform an in-depth study of the prediction
accuracy of the two learning approaches: {\em learning in time}
(history-based learning) and {\em learning in space}
(task-sampling-based learning).  
{
Both approaches can potentially be used
to learn  different job properties for different optimization objectives.
In this paper, we focus on job completion time 
because it is an important metric that has been intensively studied
in recent work~\cite{cora:infocom2015,DontCryOverSpilledRecords,kairos:socc2018,varys:sigcomm14,corral,3Sigma,AltruisticScheduling,aalo:sigcomm15}.
}

\if 0
We first derive analytical bounds on
their prediction errors (\S\ref{sec:accuracy:quantity}).  We then
measure and compare the bounds in real traces from two production
datacenters (\S\ref{sec:accuracy:trace}).  Finally, we
experimentally compare the prediction accuracy of learning in space
with a history-based predictor
\primarybasepredict~\cite{3Sigma} in estimating the job runtimes
(\S\ref{sec:accuracy:experiment}).
We pick \primarybasepredict because it is a state-of-the-art 
history-based predictor that can learn for non-recurrent jobs.
\fi

%  , and thus specific choices are somewhat orthogonal to the main focus of our
%  paper, which is to show that sampling-based prediction can be a viable
%  alternative to history-based prediction, explain the situations where
%  it will work well, and demonstrate its effectiveness empirically
%  through production traces. 
% Consistent with this objective, we mainly focus on estimating task running-times.

\vspace{-0.1in}
%\subsection{Quantitative Comparison}
\subsection{Analytical Comparison}
\label{sec:accuracy:quantity}

We first  present a theoretical analysis of the prediction accuracies of the two
approaches. We caution that here we use a highly-stylized model (e.g., two
jobs and normal task-length distributions), which does not capture the
possible complexity in real clusters, such as heavy parallelism across servers and highly-skewed 
task-length distributions. Nonetheless, it reveals important insights
that help us understand in which regimes history-based schemes or sampling-based
schemes will perform better. Consider a simple case of two jobs
$j_1$ and $j_2$, where each job has $n$ tasks. The size of each task of $j_1$
is known. Without loss of generality, let us assume that the task size of $j_1$
is 1. Thus, the total size of $j_1$ is $n$. The size of a task of $j_2$ is
however unknown.
% Instead, we know the following about $j_2$: 
%  \editaj{(1) The average
%  task size across all tasks of $j_2$, denoted by $x$, follows a normal
%  distribution with mean $\mu$ and variance $\sigmanotsqrd$;}{
Let $x$ denote the average task size of $j_2$, and this its total size
is $n x$. Clearly, if we knew $x$ precisely, then we should have
scheduled $j_1$ first if $x > 1$ and $j_2$ first if $x \le
1$. However, suppose that we only know the following:
(1) % The average task size of $j_2$, denoted by
(Prior distribution:) $x$ follows a normal
distribution with mean $\mu$ and variance $\sigmanotsqrd$;
(2) Given $x$, the size of a random task of the job follows a normal 
distribution with mean $x$ and variance $\sigmaonesqrd$.
Intuitively, $\sigmanotsqrd$ captures the variation of mean
task-lengths \emph{across} many \emph{i.i.d.} copies of job $j_2$,
\ie job-wise variation,
while $\sigmaonesqrd$ captures the variation of task-lengths \emph{within} a
single run of job $j_2$,
\ie task-wise variation.
%
% \soccReviewEdit{A}{
\addnsdiSHP{We note that the parameters $\sigmanotsqrd$ and
$\sigmaonesqrd$ are \emph{not} used by the predictors below.}
%to understand the accuracy of both
%history-based and sampling-based predictors, whose goal is to estimate
%the mean task-length $x$ (and consequently the job runtime which equals
%$x$ times the number of tasks)
%of a new copy of job $j_2$. The predictors
%themselves may not use the knowledge of $\sigmanotsqrd$ and
%$\sigmaonesqrd$. 
%%  In practice,
%%  these parameters can be estimated offline from %historical data. 
%We will soon see that the performance of history-based %schemes
%mainly depends on $\sigmanotsqrd$, while the performance of sampling-based
%schemes mainly depends on $\sigmaonesqrd$.
%}

%In practice, all these parameters can be obtained from historical data.
%Specifically, $\sigmanotsqrd$ can be estimated from the data of mean
%task-lengths (each of which is an average among the tasks in a job) across jobs,
%while $\sigmaonesqrd$ can be estimated from the data of task-lengths among
%the tasks of the same job. Thus, we will refer to $\sigmanotsqrd$ and
%$\sigmaonesqrd$ as the {\em job-wise variations} and {\em task-wise variations},
%respectively. 
%We will see soon that the performance of history-based schemes
%will mainly depend on $\sigmanotsqrd$, while the performance of sampling-based
%schemes will mainly depend on $\sigmaonesqrd$.

%\todoaj{Discuss with prof. Lin and write following claims in a better way.\\}
%We will show that:\\
%(1) if $\sigmanotsqrd >> \sigmaonesqrd$, sampling will be effective.\\
%(2) if $\sigmanotsqrd << \sigmaonesqrd$, sampling will not be very effective.\\

\addnsdiSHP{Now, consider two options for estimating the mean task-length $x$:} (1) A history-based approach
(\S\ref{sec:accuracy:quantity:history}) and (2) a sampling-based approach where
we sample $m$ tasks from $j_2$ (\S\ref{sec:accuracy:quantity:sampling}).

\subsubsection{History-based Schemes}
\label{sec:accuracy:quantity:history}
Since no samples of job $j_2$ are used, the best predictor for its mean task
length is $\mu$.
In other words, the scheduling decision will be based on $\mu$ only. The difference between the
true mean task length, x, and $\mu$ is simply captured by the job-wise variance
$\sigmanotsqrd$.

%If $\mu>1$, then $j_1$ should
%go first. The total completion time in this case is $n(1+(1+\mu))$.  Now if
%$\mu<1$, then $j_2$ should go first. The total completion time in this case
%will be $n(\mu+(\mu+1))$.  So in both cases, the total completion time is
%\footnote{For sake of simplicity in analysis we have assumed that there is only one server.}:
%\begin{equation}
%\label{equation:historyExpectation}
%n(1+\mu+min(1,\mu))
%\end{equation}

\subsubsection{Sampling-based Schemes}
\label{sec:accuracy:quantity:sampling}
Suppose that we sample $m$ tasks from $j_2$. Collect the sampled task lengths into a vector:
\begin{center}
	$\yvec = \left( y_1, y_2, ..., y_m \right)$.
\end{center}
Then, based on our probabilistic model, we have
\begin{center}
	$P\left(y_i|x\right) = \frac{1}{\sqrt{2\pi}\sigmaone}e^{-\frac{\left(y_i - x\right)^2}{2\sigmaonesqrd}}$, \mbox{\hspace{0.1in}}
	$P\left(\yvec|x\right) = {\prod_{i=1}^{m}}\frac{1}{\sqrt{2\pi}\sigmaone}e^{-\frac{\left(y_i - x\right)^2}{2\sigmaonesqrd}}$
\end{center}
We are interested in an estimator of $x$ given $\yvec$. We have 
%%\commentaj{added a footnote here.}\footnote{
%\soccReviewEdit{A}{
%%We do not include the detailed derivation here due to space limit. However, it is 
%}
%$P\left(x|\yvec\right)$
\begin{center}
	$P\left(x|\yvec\right) = \frac{P\left(\yvec|x\right)\cdot P(x)}{P(\yvec)} = \frac{P(\yvec|x)\cdot P(x)}{\int_{x}P(\yvec|x)\cdot P(x)dx}$

	$= \frac{1}{\sqrt{2\pi}}\left[\frac{m}{\sigmaonesqrd} +
	\frac{1}{\sigmanotsqrd}\right]^\frac{1}{2} \cdot e^{-
	\left(\frac{m}{2\sigmaonesqrd} +
	\frac{1}{2\sigmanotsqrd}\right)\left(x -
	\frac{{\sum_{i=1}^{m}\frac{1}{\sigmaonesqrd}y_{i} +
	\frac{1}{\sigmanotsqrd}\mu}}{\frac{m}{\sigmaonesqrd} +
	\frac{1}{\sigmanotsqrd}}\right)}$,
\end{center}
where the last step follows from standard results on the posterior distribution
with Gaussian priors (see, \eg~\cite{jordanLecturePosterirorDistribution}).
%This confirms that $x|\yvec$ is also a normal distribution with:\\
In other words, conditioned on $\yvec$, $x$ also follows a normal distribution with
mean = $ \frac{{\sum_{i=1}^{m}\frac{1}{\sigmaonesqrd}y_{i} + \frac{1}{\sigmanotsqrd}\mu}}{\frac{m}{\sigmaonesqrd} + \frac{1}{\sigmanotsqrd}}$ 
and variance = $\frac{1}{\frac{m}{\sigmaonesqrd} + \frac{1}{\sigmanotsqrd}}$.

Note that this represents the estimator quality using the information of
both job-wise variations and task-wise variations. If the estimator
% ignores
is not informed of the
job-wise variations, we can take $\sigmanotsqrd  \rightarrow +\infty$, and
the conditional distribution of $x$ given $\yvec$ becomes normal with mean
$\frac{1}{m} \sum_{i=1}^{m} y_{i}$ and variance $\frac{\sigmaonesqrd}{m}$.

From here we can draw the following conclusions. First, whether history-based
schemes or sampling-based schemes have better prediction accuracy for an unknown
job depends on the relationship between job-wise variations $\sigmanotsqrd$ and
the task-wise variation $\sigmaonesqrd$. If the job-wise variation is large
but the task-wise variation is small, \ie $\sigmanotsqrd >>
\frac{\sigmaonesqrd}{m}$, then sampling-based schemes will have better
prediction accuracy. Conversely, if the job-wise variation is small but
the task-wise variation is large, \ie $\sigmanotsqrd <<
\frac{\sigmaonesqrd}{m}$, then  history-based schemes will have better
prediction accuracy. Second, while the accuracy of history-based schemes is
fixed at $\sigmanotsqrd$, the accuracy of sampling-based schemes improves as $m$
increases. Thus, when we can afford the overhead of more samples, the
sampling-based schemes become favorable. Our results from experimental
data below will further confirm these intuitions.

\if 0

So, now once sampling is done we can make decision based on:\\
\begin{center}
\xspace $\mathbb{E}[x|\yvec] \cdot (n-m)$. 
\end{center}
If $\mathbb{E}[x|\yvec](n-m) > n$, then $j_1$ should be scheduled first otherwise $j_2$.
The total completion time in this case will be:\\
$Z = \sum_{i=1}^{m} y_{i} + \mathbb{E}(x|\yvec)(n-m) + n + min[\mathbb{E}(x|\yvec)(n-m), n]$\\
and the expected total completion time will be:\\
\begin{equation}
\label{equation:samplingExpectation}
\mathbb{E}(Z) = n\cdot\mu + n + \mathbb{E}\left[ min\left(\frac{{\sum_{i=1}^{m}\frac{1}{\sigmaonesqrd}y_{i} + \frac{1}{\sigmanotsqrd}\mu}}{\frac{m}{\sigmaonesqrd} + \frac{1}{\sigmanotsqrd}}\cdot(n-m),n\right)\right]\\
\end{equation}
\fi

\if 0
\begin{figure}[tp] 
\centering
\subfigure[\vspace{-0.2in} sigma0 = 0.01 sigma1 = 1]
{
\includegraphics[width=0.9\linewidth]{figures/numerical_analysis/numerical_analysis_0pt5_1pt5_0pt01_sigma0-0pt01_sigma1-1_varying_mu.pdf}
\label{fig:numerical_analysis:variableMu}
}
\subfigure[\vspace{-0.2in} sigma0 = 1 sigma1 = 0.01]
{
\includegraphics[width=0.9\linewidth]{figures/numerical_analysis/numerical_analysis_0pt5_1pt5_0pt01_sigma0-1_sigma1-0pt01_varying_mu.pdf}
\label{fig:numerical_analysis:variableMu}
}
\caption{Numerical Analysis. Varying $\mu$ from 0.5 to 1.5 with step of 0.01. $m = 5$, $n = 100$.
	\todoaj{Make the y-label as "E[Total JCT]"}
	}
\vspace{-0.1in}
\label{fig:qunatitative_analysis}
\end{figure}
\fi

%\if 0
\begin{table}[tp]
\caption{Summary of trace properties.}
\label{table:traceSummary}
\centering
{\small
\vspace{-0.1in}
\begin{tabular}{|c|c|c|c|c|c|}
\hline
		\textbf{ Trace} & \textbf{Arrival} & \textbf{Resource} & \textbf{Resource}  & \textbf{Indiv. task} \\
			& \textbf{time} & \textbf{requested} & \textbf{usage} & \textbf{duration}\\
 
\hline
	  \textbf{2Sigma} & Yes & Yes & No & Yes\\
\hline
	  \textbf{Google 2011} & Yes & Yes & Yes & Yes\\
\hline
	  \textbf{Google 2019} & Yes & Yes & Yes & Yes\\
	  %\footnote{\addnsdiSHP{We have used jobs from cluster-G BEB tier.}{E1}}
\hline
%\hline
%		\textbf{ Trace} & \textbf{Arrival} & \textbf{Resource} & \textbf{Resource}  & \textbf{Individual} & \textbf{Task success}&{\bf Fraction of}\\
%			& \textbf{time} & \textbf{requested} & \textbf{usage} & \textbf{task duration} & \textbf{status} &{\bf thin jobs}\\
%	 
%\hline
%	  \textbf{2Sigma~\cite{2Sigma:trace}} & Yes & Yes & No & Yes & Yes & Relatively low\\
%\hline
%	  \textbf{Google~\cite{googleTraceGithub}} & Yes & Yes & Yes & Yes & Yes & Relatively High\\
%\hline
\end{tabular}
\vspace{-0.1in}
}
\end{table}
%\fi

\subsection{Trace-based Variability Analysis}
\label{sec:accuracy:trace}

Our theoretical analysis in \S\ref{sec:accuracy:quantity} provides 
insights on how the prediction accuracies of the two approaches depend
on the variation of job run times across time and space.
To understand how such variations fare against each other in practice,
we next measure the actual variations
%  in average task runtimes for jobs across history ($\sigmanotsqrd$) and in task
%  runtime across the tasks of the same job (${\sigmaonesqrd}$).
in three production cluster traces.
\addnsdiSHP{Table~\ref{table:traceSummary} summarizes the information available in the traces
that are used in our analysis.}{}

\paragraph{Traces.} 
Our first trace is provided by 2Sigma~\cite{2Sigma:website}.
% engineers collected from their cluster.
The cluster uses an internal proprietary job scheduler
running on top of a {Mesos cluster manager}~\cite{2Sigma:scheduler}. This trace
was collected over a period of 7 months, from January to July 2016, and from  441
machines and contains approximately 0.4 million jobs ~\cite{2Sigma:trace}.

We also include two publicly available traces from Google released in
May 2011 and May 2019~\cite{googleTraceGithub, googleClusterData2019},
collected from 1 and 8 Borg~\cite{borg} cells
over periods of 29 and 31 days, respectively.  The machines in the clusters are
highly heterogeneous, belonging to at least three different
platforms that use different micro-architectures and/or memory
technologies~\cite{workloadDiversity:atc18}. Further, according
to~\cite{googleClusterData2011-2Schema}, the machines in the same
platform can have substantially different clock rates, memory speed,
and core counts.
%   (The trace does not contain actual machine
%   properties. Other details can be found
%   in~\cite{googleClusterData2011-2Schema, googleClusterData2019}).
%  As mentioned in~\cite{borgTraceAnalysis2019}, Google 2019 trace is very
%  large and has 2.8 TiB in compressed form and hence Google made it
%  available only via \textit{bigquery}~\cite{web:googleBigquery}.
%  First, the Google
%    2019 trace had information about the job tier, \eg batch or
%    production. Since the SLO for batch tier jobs is to minimize the
%    average completion time, we chose batch tier jobs from the 2019
%    trace.  Second,
Since the original Google 2019 trace has data from 8 different
cells located in 8 different locations,
%  . From private
%    communication~\cite{personalCommunication:Nan} we learned that {at
%      Google, the job properties vary with the social behaviour of
%      different locations and it is hard to identify an average cell.
and given that we already have two other traces from the US, we chose 
the batch tier of {Cluster G} in the Google 2019 trace, which is located in
Singapore~\cite{googleClusterData2019Schema}, as our third trace to
diversify our trace collection.

\if 0
Google has publicly released two sets of traces, one collected
in May 2011 and another in May 2019~\cite{googleTraceGithub,
  googleClusterData2019}.  The traces have information collected over
a period of 29 (31) days from 1 (8) Borg~\cite{borg} cells for the
2011 (2019) trace and are released in csv (bigquery) format.  The
machines are highly heterogeneous; they belong to at least three
different platforms which use different micro-architectures and/or
memory technologies~\cite{workloadDiversity:atc18}, and according
to~\cite{googleClusterData2011-2Schema}, the machines in the same
platform can have substantially different clock rates, memory speed,
and core counts.  (The trace does not contain actual machine
properties. Other details can be found
in~\cite{googleClusterData2011-2Schema, googleClusterData2019}).  As
mentioned in~\cite{borgTraceAnalysis2019}, Google 2019 trace is very
large and has 2.8 TiB in compressed form and hence Google made it
available only via \textit{bigquery}~\cite{web:googleBigquery}.
\fi

We calculate the variations in task runtimes for each job across time and
across space as follows.

\paragraph{Variation across time.} To measure the variation in mean task
runtime for a job across the history, we follow the following prediction
mechanism defined in 3Sigma~\cite{3Sigma} to find similar jobs.

As discussed in \S\ref{sec:back:existing}, 3Sigma~\cite{3Sigma} uses
multiple features to identify a job and predicts its runtime using the feature
that gives the least prediction error in the past. We include all six features
used in 3Sigma:
\addnsdiSHP{{application name}, {job name}, {user name}
(the owner of the job), {job submission time (day and hour)},
and {resources requested (cpu and memory)} by the job.}{}
%These are the same set of features as listed in 3Sigma~\cite{3Sigma}.  

% hitory. However, they approximate it as storing entire history will take too much memory.}
For each feature, we define the set of similar jobs as
all the jobs executed in the history window
(defined below) that had the same feature value.
Next, we calculate the average task runtime of each job in the set. Then, we
calculate the {\em Coefficient of Variation} (CoV) of the average task runtimes
across all the jobs in the set. We repeat the above process for all the
features.  We then compare the CoV values thus calculated and pick the minimum
CoV.
% The above procedure is similar in
% principle to the prediction procedure of 3Sigma~\cite{3Sigma}.
Effectively, the above procedure selects the least possible variation
across history.

\paragraph{Varying the history length in prediction across time.}
%\questionaj{The figure \ref{fig:accuracy:trace} is on job count
%window. Will add a day wise figure for \ref{fig:accuracy:trace} too. But there
%we need to have only one fixed window size. We cannot have multiple windows
%there. What do you think we should keep the window size?}
3Sigma used the entire history for prediction.
Intuitively, the length of the history affects the trade-off between
the number of similar jobs and the staleness of the history information.
For this reason, we optimized 3Sigma
by finding and using the history length that gives the least variation.
Specifically, we define the length of history based
on a window size $w$, \ie the number of past consecutive days.
%  In particular, when we calculate the CoVs for a job on day $i+w$,
% the history window starts from day $i$ to day $i+w-1$.
In our analysis below, we vary $w$ among 3, 7, and 14 for the
three traces.

\begin{figure*}[tp]
%\vspace{-0.1in}
\centering
\subfigure[Task runtime -- 2Sigma]
{
\includegraphics[width=0.2\linewidth]{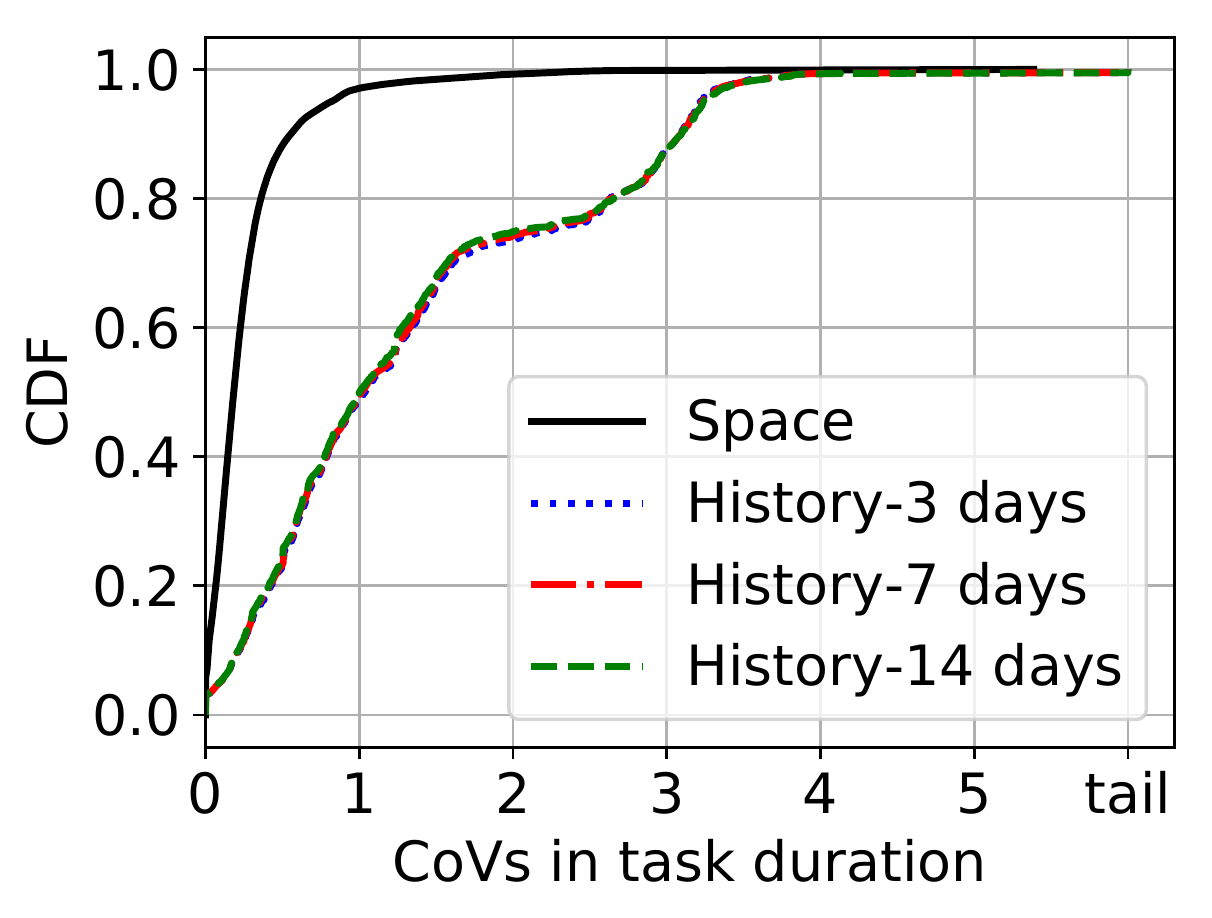}	% done
\label{fig:accuracy:trace_analysis_window:sp3:2Sigma:task_dur}
}
\hspace{-12pt}
\subfigure[Task runtime -- Google 11]
{
\includegraphics[width=0.2\linewidth]{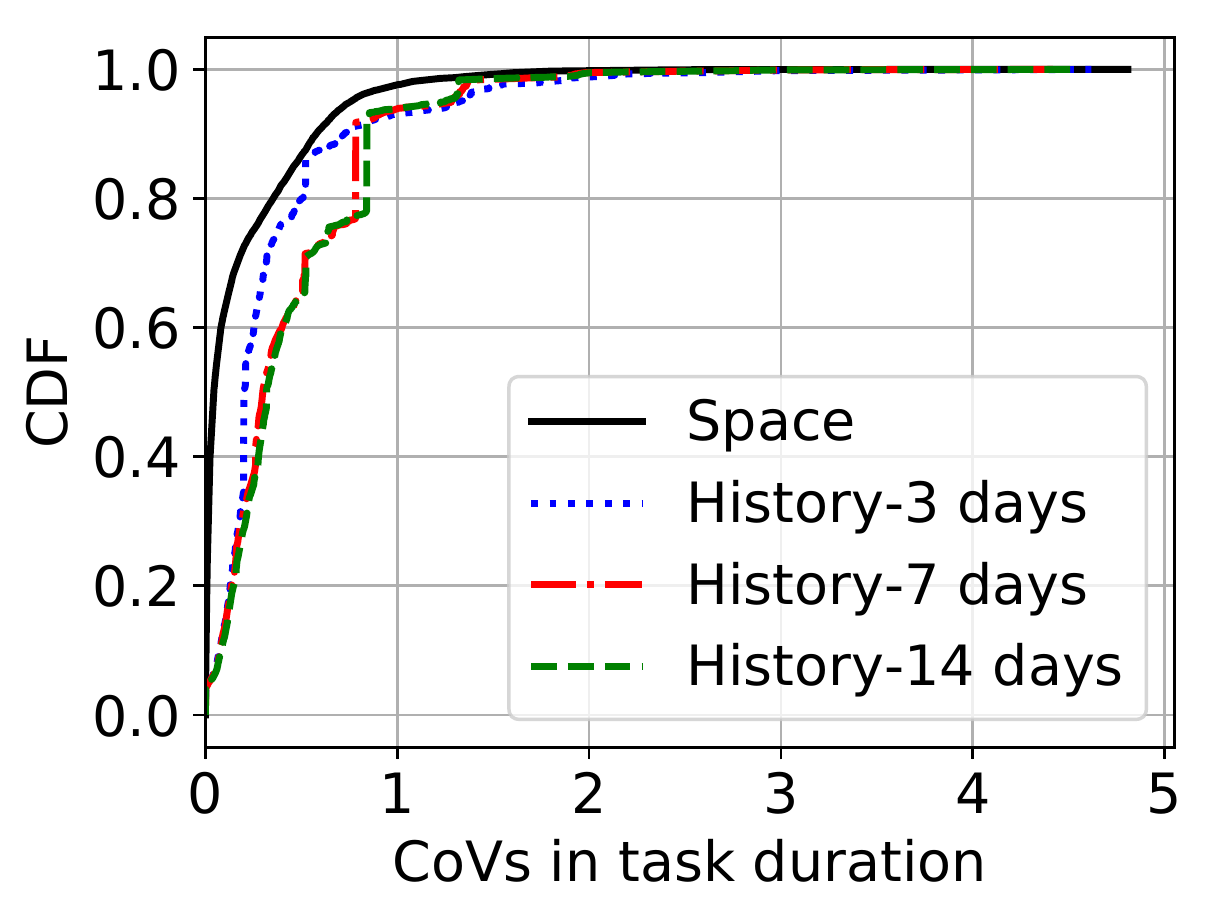}	% done
\label{fig:accuracy:trace_analysis_window:sp3:google11:task_dur}
}
\hspace{-12pt}
\subfigure[Task runtime -- Google 19]
{
\includegraphics[width=0.2\linewidth]{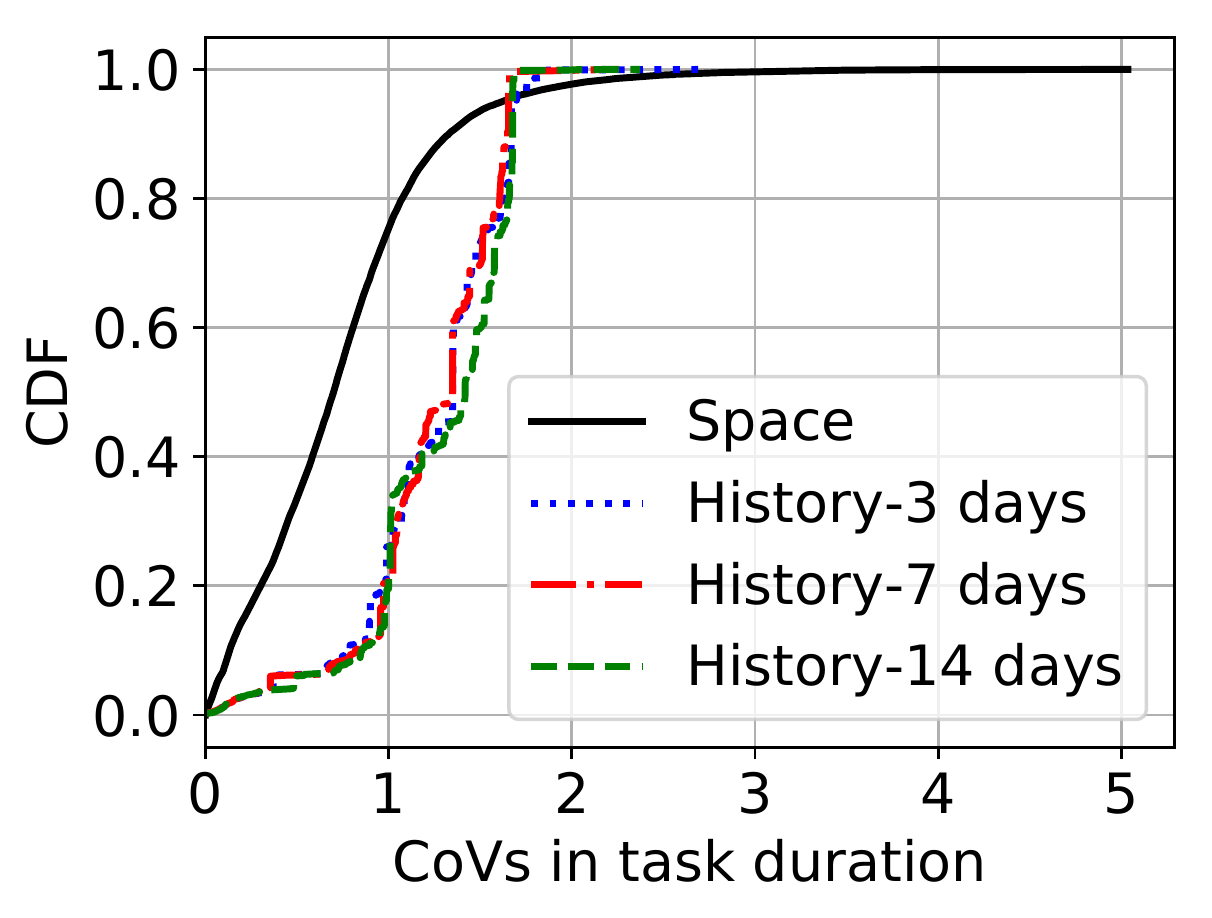}	% done
\label{fig:accuracy:trace_analysis_window:sp3:google19:task_dur}
}
\hspace{-12pt}
\vspace{-0.1in}
\subfigure[CPU usage -- Google 11]
{
\includegraphics[width=0.2\linewidth]{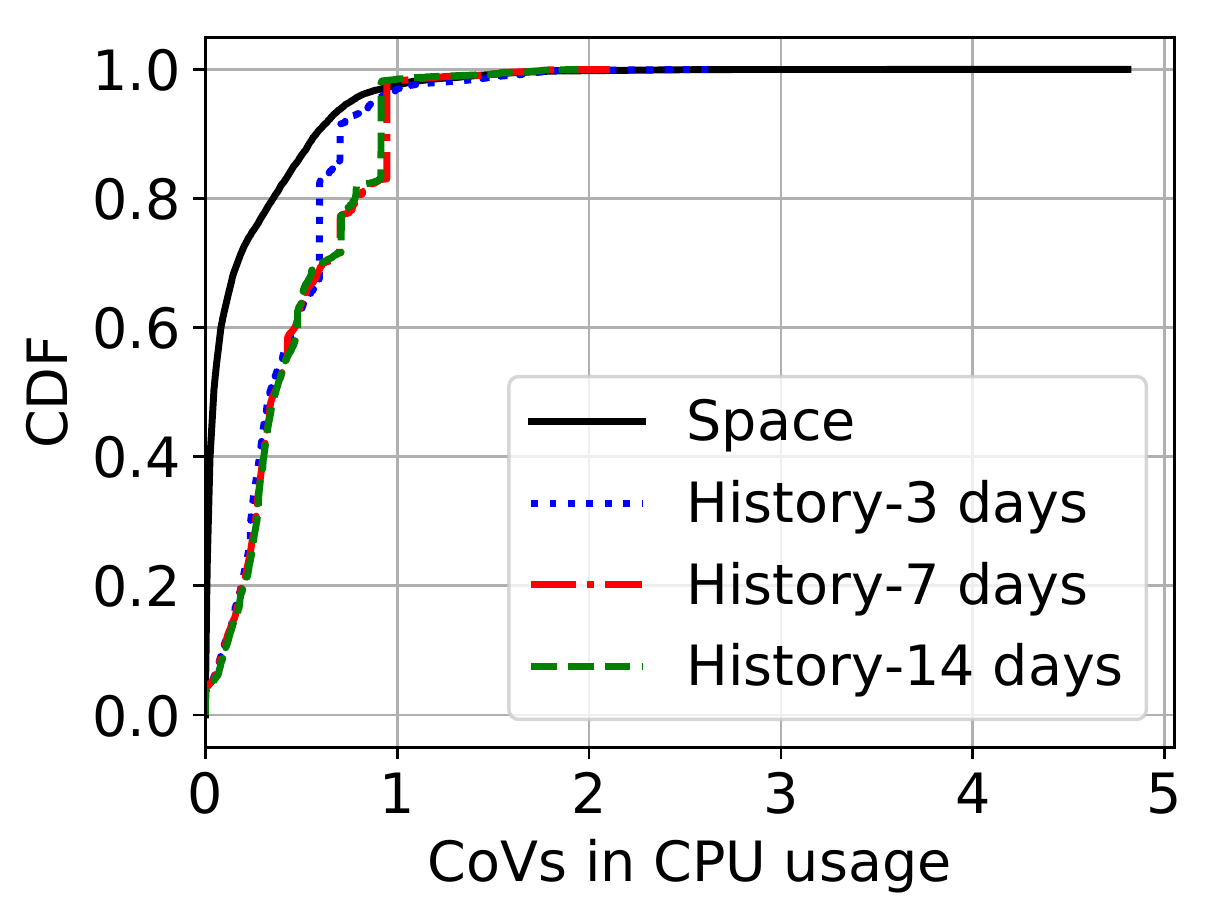}	% done
\label{fig:accuracy:trace_analysis_window:sp3:google11:cpuUsage}
}
\hspace{-12pt}
\subfigure[Disk IO time -- Google 11]
{
\includegraphics[width=0.2\linewidth]{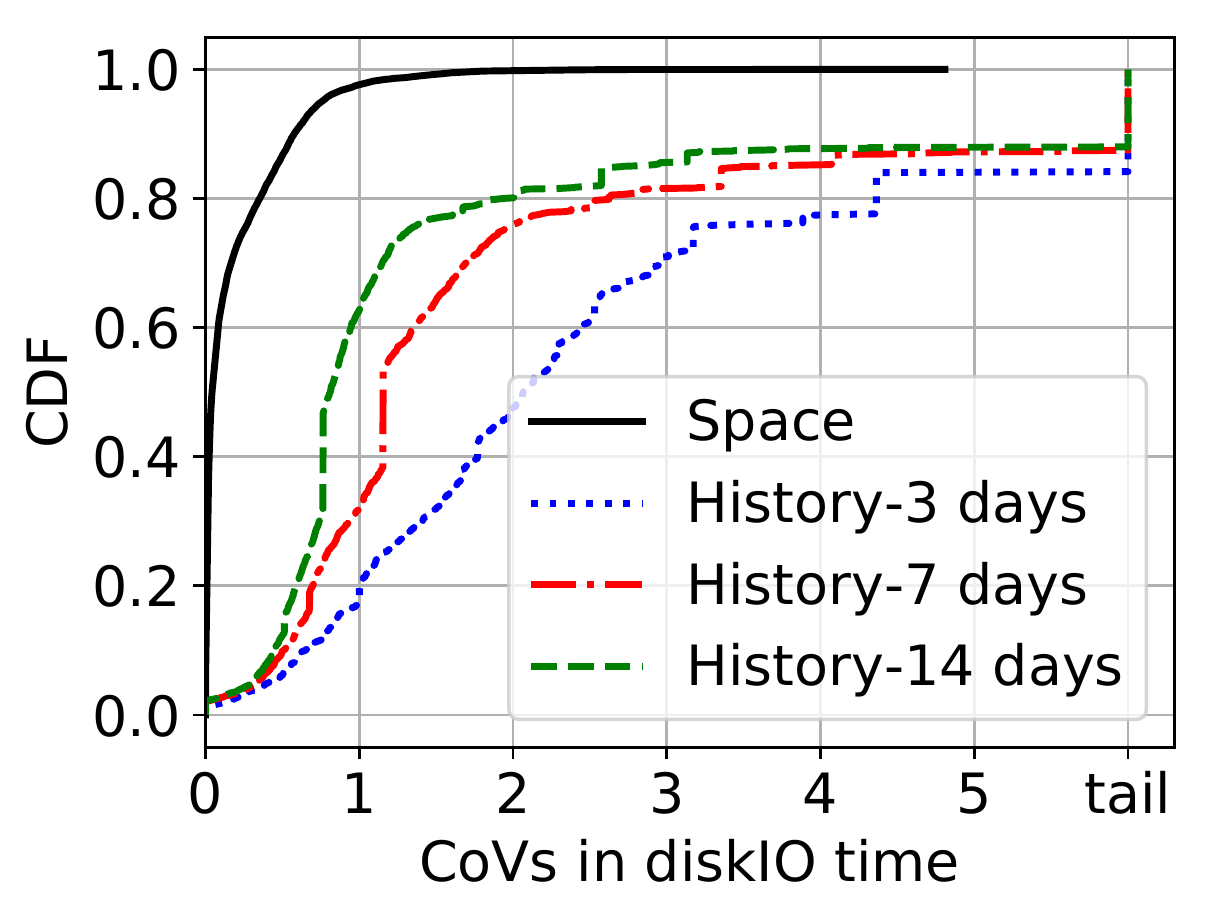}	% done
\label{fig:accuracy:trace_analysis_window:sp3:google11:diskIO}
}
\vspace{-0.1in}
\caption{CDF of CoV of runtime properties across space and across time with
	varying history windows, using the 2Sigma, Google 2011 and Google 2019
	traces.  Single-task jobs are excluded from the analysis across space.
        %	This is with 3\% sampling.
}
\vspace{-0.1in}
\label{fig:accuracy:trace_analysis_window:sp3}
\end{figure*}

\paragraph{Variation across space.} 
To measure the extent of variation across space, we look at the
CoV (CoV $= \frac{\sigma}{\mu}$) in the task
runtimes within a job.  As shown in \S\ref{sec:accuracy:quantity}, the variance in
the task runtime predicted from sampling is $\frac{\sigmaonesqrd}{m}$, where
$\sigmaonesqrd$ is the variance in the runtimes across all the tasks within the job
and $m$ is the number of tasks sampled.
Thus, we first estimate $\sigma_1^2$ from all tasks within the job. We then report the CoV of our task runtime prediction after sampling $m$ tasks as
%     Thus, the observed \textit{CoV} in the
%     task runtimes within a job after sampling $m$ tasks becomes
$\frac{\sigmaone/\sqrt{m}}{\mu}$.
%= \frac{cov}{\sqrt{m}}$,
%where $\sigmaone$ is the standard deviation in the runtimes of all the tasks of
%the job, $m$ is the number of tasks sampled (5\% of the total tasks) and $\mu$
%is the average task runtime.
Our complete scheduler design in \S\ref{sec:study:design}
uses an adaptive sampling algorithm which mostly uses 3\%
for the three traces.
% Empirically (\S\ref{sec:sim:numPilots}), we found that sampling 5\% of the
% total tasks gives considerably low variation in the predicted tasks runtimes.
Thus, for measuring the extent of variation across space here, we assume
a 3\% sampling ratio and plot
$\frac{\sigmaone}{(\sqrt{0.03\times numberOfTasksInJob}\ )\times\mu}$.

% {We start the analysis on the day after the largest window value
%   so that we can compare the CoVs for the same jobs when varying the history length.
%  }
\paragraph{Variability comparison.}
For consistency, all
analysis results here 
% CDF curves in Fig.~\ref{fig:accuracy:trace_analysis_window:sp3}
are for the same, shortest 
trace period that can be used for sliding-window-history based analysis, \eg the last 15
days under the 14-day window for the 29-day Google 2011 trace.
(The analysis then varies the length of the sliding window in history-based learning.)
%  Fig.~\ref{fig:accuracy:trace_analysis_window:sp3:2Sigma:task_dur}--Fig.~\ref{fig:accuracy:trace_analysis_window:sp3:google19:task_dur} show the CDFs of CoVs
%    the 14-day window for the Google 2011 trace.
  %  However, table
  %  \ref{table:accuracy:trace_analysis:covs:sp3} shows the value for the maximum
  %  possible period.}
\iftoggle{techreport}{ 
Figure~\ref{fig:accuracy:trace_analysis_window:2Sigma:task_dur:scatter}
visualizes the two CoVs for each of 70 randomly selected jobs from the 2Sigma
trace in the order of their arrival, also using the best window size of 14
days.}{}
\rm{For clarity, we first show the result for 70 jobs extracted at
random from the 2Sigma trace, plotted in the order of job arrival in
Figure~\ref{fig:accuracy:trace_analysis_window:2Sigma:task_dur:scatter}.
For each job, we plot the following two values on the y-axis:
(1) the CoV in average task runtime for the jobs
in the job's history window, for the feature that gives the least CoV,
using 30-day history window which was found to give the least
variation across history as shown in
Figure\ref{fig:accuracy:trace_analysis_window:2Sigma:task_dur}; and
(2) the CoV in task runtimes of the job.  We see that for both traces, the
variation across history is higher than the variation across tasks for more
than 85\% of the jobs.
}

\iftoggle{techreport}{ 
\begin{figure}[tp]
\vspace{-0.1in}
%\includegraphics[width=1.0\textwidth]{figures/trace_analysis/slidingWindow_analysis_the_scatter_plot_avg_task_dur_in_google11_initial_history_window_3_days.pdf} %done
%\label{fig:accuracy:trace_analysis_window:google11:task_dur:scatter}
%\includegraphics[width=6.5in]{figures/trace_analysis/slidingWindow_analysis_the_scatter_plot_avg_task_runtime_in_2Sigma_initial_history_window_30_days.png} %done
\mbox{\hspace{-0.3in}}\includegraphics[width=1.2\columnwidth]{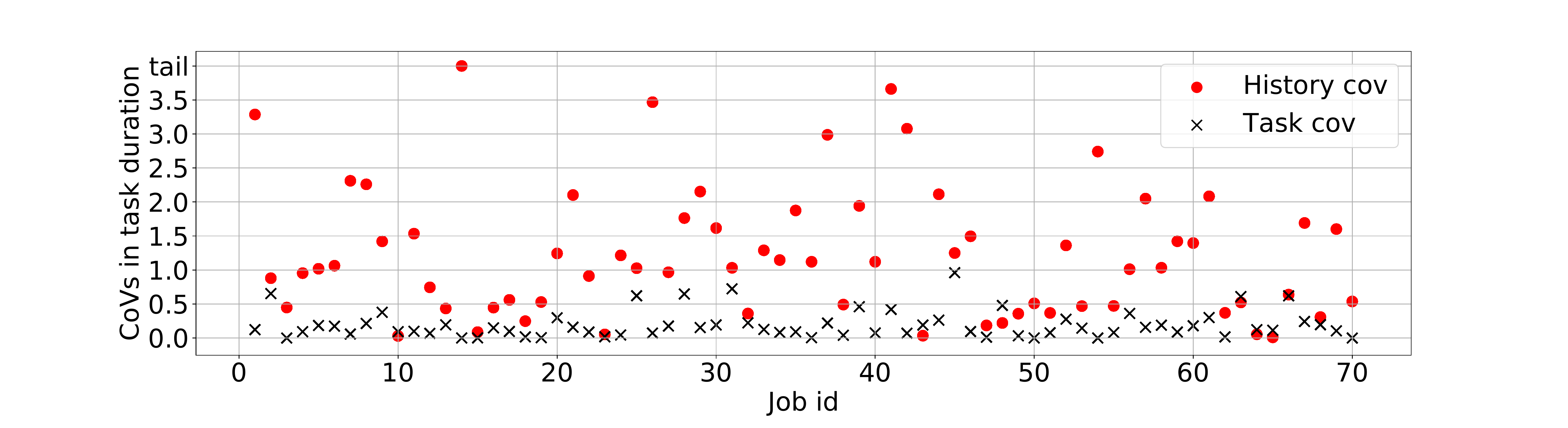} %done
%\label{fig:accuracy:trace_analysis_window:2Sigma:task_dur:scatter}
\vspace{-0.25in}
	\caption{CoVs across time and space
	for 70 jobs selected randomly from the 2Sigma trace.
% 	As evident from Figure
%	\ref{fig:accuracy:trace_analysis_window:2Sigma:task_dur} 30-day history
%	window gives lowest CoVs across time so we choose 30 days as history
        %	period to calculate CoVs across time.
        The x-axis represents job ids in the order of their arrival.}
\vspace{-0.1in}
\label{fig:accuracy:trace_analysis_window:2Sigma:task_dur:scatter}
\end{figure}
}{}

%      The high variability across history can be
%      explained as follows. For the Google trace, more than 67\% of the jobs studied
%      were non-recurring, \ie no exact same job was found for them in history, and
%      only 9\% jobs had more than 20\% of all historical jobs as recurring instances.
%      For this analysis, we
%      follow~\cite{corral, morpheus}, and identify two jobs as recurring instances
%      when they have the same jobname and are being executed by the same application.
%      We did
%      not find recurring jobs in the 2Sigma trace. That is what this statement is
%      saying that we were not able to any such information from the 2 Sigma trace.}
%      The lack of a large number of recurring jobs in the histoy window in turn leads
%      to high error in prediction.

Fig.~\ref{fig:accuracy:trace_analysis_window:sp3:2Sigma:task_dur}--Fig.~\ref{fig:accuracy:trace_analysis_window:sp3:google19:task_dur}
show the CDFs of CoVs in task duration measured across space and across history
for multiple history window sizes for the three traces.
\if 
{For the 2Sigma (Google 2011) trace, the CoV across history is higher than the
CoV across tasks for 81\% (77\%) of the jobs.  In particular, for more than
30\% of the jobs, the CoV across history is at least 7.13 (26.81) times higher than
the CoV across tasks.  }
\fi
  We see that in general using a shorter sliding window reduces the prediction
  error of 3Sigma, and the CoVs across tasks are moderately lower than the CoVs
  across history for the Google 2011 trace but significantly lower for 2Sigma
  and Google 2019 traces.  For example, for the 2Sigma trace, the CoV across
  history is higher than the CoV across tasks for 85.40\% of the jobs
  (not seen in Fig.~\ref{fig:accuracy:trace_analysis_window:sp3:2Sigma:task_dur}
  as jobs are ordered differently in different CDFs)
  and for
  more than 30\% of the jobs, the CoV across history is at least 12.10$\times$
  higher than the CoV across tasks.
\if 0
{ aj: 2STrace values for 14 days window. 85.40 for gt1 and 12.12 f30 percentile.
Space COVs P50 (P90) 1.00 (3.10)
}
\fi

\if
{For the 2Sigma (Google 2011) [Google 2019] trace, the CoV across history is higher than the
CoV across tasks for 85.68 (71.94) [80.64]\% of the jobs.  In particular, for more than
30\% of the jobs, the CoV across history is at least 12.54 (18.22) [2.89] $\times$ higher than
the CoV across tasks.}
\fi

Table~\ref{table:accuracy:trace_analysis:covs:sp3} summarizes the results,
where the CoVs across time correspond to the best history window size, \ie 3
days for both Google traces and 14 days for the 2Sigma trace.  As shown in the table,
% the 50th (90th) percentile CoV, denoted as
the P50 (P90) CoV across
history are 1.00 (3.10) for the 2Sigma trace, 
0.20 (0.73) for the Google 2011
trace, and 1.35 (1.67) for the Google 2019 trace.
In contrast, the P50 (P90) CoV value across the task duration of the same set
of jobs is much lower, 0.18 (0.55) for the 2Sigma trace,
0.04 (0.58) for the
Google 2011 trace,
and 0.70 (1.33) for the Google 2019 trace.
%   The P50 (P90) task duration CoV value for the Google 2019
%  trace, 0.70 (1.33), is higher compared to those of the other two traces, but is
%  still much lower when compared to the CoV across history for the same trace,
%  1.35 (1.67).

%\comment{any insights for why?}

%   \commentaj{Yes, I think it could be due to clusterG's high Machine utilization.
%   This means tasks have easy access to resources and hence they do not get
%   swapped in and out and finish within their time. I did some study on this
%   earlier, I am looking into it more today. Additionally, for 2019 trace this is
%   not the case with all the clusters. I looked up for cluster B and cluster H
%   task CoVs. P50 (P90) values for them are 0.09 (0.51) and 0.13 (0.64)
%   respectively. History CoVs for cluster B are 0.39 (2.12)}
%
% {For the 2Sigma (Google) trace, for more than 30\% of the
%   jobs, the CoV across history is at least 10 (35) times higher than the CoV across tasks.
  % for next 20\% it is somewhere between 10-4 (35-15). CoV across history is higher than CoV
% across tasks only for 13\% (16\%) of the jobs.}
%

\rm{The CoV values across tasks is much lower because the tasks of a job run the
same code with the same flags, settings and priority, as mentioned in the trace
schema released by
Google~\cite{googleClusterData2011-2Schema,googleClusterData2019Schema} and
confirmed by 2Sigma engineers~\cite{personalCommunication:MarkAstley}.
% Hence the predictions made by sampling tasks should have a low error.
}

{Fig.~\ref{fig:accuracy:trace_analysis_window:sp3:google11:cpuUsage} and
Fig.~\ref{fig:accuracy:trace_analysis_window:sp3:google11:diskIO} further show the
CDF of CoVs for CPU usage and Disk IO time for the Google 2011 trace (such resource
usage is not available in the 2Sigma trace).
%\addaj{The CoVs here were also measured in the same way as for the task durations.}
%  We use the same
%  methodology for plotting these figures as we did for the task durations.
% We plot in Fig.~\ref{fig:accuracy:trace_analysis_window}
% the variation across space and time for all the jobs in the two traces.
%  For the Google trace, we measure the variation in
%  three runtime properties: Task duration, CPU usage and Disk IO
%  time. In the 2Sigma trace, resource usage is not available and hence
%  we only measured the variation in task duration.
%
The figures show that the variation in the values of these properties when
sampled across space is also considerably lower compared to the variation
observed over time.
}

\if 0
\begin{table}[tp]
%\vspace{-0.05in}
  \caption{CoV in task runtime across time and across space for the
	the 2Sigma, Google 2011, and Google 2019 traces. \commentaj{The 2019 trace values are not correct.}}
\label{table:accuracy:trace_analysis:covs}
\centering
{\small
\vspace{-0.1in}
\begin{tabular}{|c|c|c|c|c|c|}
\hline
Trace       & \multicolumn{2}{|c|}{CoV over Time} &
\multicolumn{2}{|c|}{CoV over Space}  \\
\cline{2-5}
			& P50  & P90& P50 & P90 \\
%\hline
\hline
	2Sigma & 0.63 & 2.99 & 0.15 & 0.84 \\
	 %Trace &  & &      &\\
\hline
	Google 2011 & 0.50 & 2.87 & 0.04 & 0.58 \\
	%Trace &    &&&\\
\hline
	Google 2019 & 1.35 & 1.67 & 0.70 & 1.33 \\
	%Trace &    &&&\\
\hline
%\vspace{-0.2in}
\end{tabular}
}
\vspace{-0.1in}
\end{table}
\fi

\begin{table}[tp]
%\vspace{-0.05in}
  \caption{CoV in task runtime across time and across space for the
    the 2Sigma, Google 2011, and Google 2019 traces.
    % with 3\% sampling.
    %    \commentaj{the 2019 trace values are not correct.}
  }
\label{table:accuracy:trace_analysis:covs:sp3}
\centering
{\small
\vspace{-0.1in}
\begin{tabular}{|c|c|c|c|c|c|}
\hline
Trace       & \multicolumn{2}{|c|}{CoV over Time} &\multicolumn{2}{|c|}{CoV over Space}  \\
            \cline{2-5}
			& P50  & P90& P50 & P90 \\
%\hline
\hline
	%2Sigma & 0.63 & 2.99 & 0.12 & 0.65 \\
	2Sigma & 1.00 & 3.10 & 0.18 & 0.55 \\
	 %Trace &  & &      &\\
\hline
	%Google 2011 & 0.50 & 2.87 & 0.03 & 0.45 \\
	Google 2011 & 0.20 & 0.73 & 0.04 & 0.58 \\
	%Trace &    &&&\\
\hline
	Google 2019 & 1.35 & 1.67 & 0.70 & 1.33 \\
	%Trace &    &&&\\
\hline
%\vspace{-0.2in}
\end{tabular}
}
\vspace{-0.2in}
\end{table}

\subsection{Experimental Prediction Error Analysis}
\label{sec:accuracy:experiment}

\if 0
To validate the analysis from \S\ref{sec:accuracy:quantity} 
that lower task-wise variation than job-wise
variation (\S\ref{sec:accuracy:trace}) will translate into better
prediction accuracy of sampling-based schemes over history-based
schemes, we next implement a sampling-based predictor \slearn and
experimentally compare it against a state-of-the-art history-based
predictor \primarybasepredict~\cite{3Sigma} in estimating the job
runtimes.
\fi

Recall from our analysis in \S\ref{sec:accuracy:quantity}
that lower task-wise variation than
job-wise variation (\S\ref{sec:accuracy:trace}) will translate into better prediction
accuracy of sampling-based schemes over history-based schemes. While
our analysis in \S\ref{sec:accuracy:quantity}
assumes normal distribution, we believe that a
similar conclusion will hold in more general settings. To validate
this, we next implement a sampling-based predictor \slearn,
and experimentally compare it against a
state-of-the-art history-based predictor 3Sigma~\cite{3Sigma} in estimating the
job runtimes directly on production job traces.

%The experimental setup used for this comparison is described below.
\paragraph{Workload characteristics.}
Since the three production traces described in
\S\ref{sec:accuracy:trace} are too large,
as in 3Sigma~\cite{3Sigma},
we extracted smaller traces for experiments
% We extracted roughly 1250 jobs each from the three production traces
using the procedure described below.
%  We denote the traces extracted
% from 2Sigma~\cite{2Sigma:trace}, Google 2011 version~\cite{googleTraceGithub}
% and Google 2019 version~\cite{googleClusterData2019}
% These extracted jobs form the traces that we will experiment with,
\if 0
We denote the extracted traces which consist of 
% roughly 
1250 jobs each
as 2STrace, GTrace11 and GTrace19, respectively.
\fi

Since the history-based predictor 3Sigma needs a history trace,
we followed the same process as in~\cite{3Sigma} to extract the training
trace for 3Sigma and the execution trace for all predictors, in three steps.
(1) We divided each original trace in chronological
order in two halves.
(2) 
We compressed 2Sigma jobs to 150 tasks or fewer,
% Each job in the 2Sigma trace goes under this process for the simulation trace. 
by applying a compression ratio of original cluster size/150.
Since the Google traces do not have many wide jobs yet the
original clusters are very wide, with ~12.5K machines, we dropped jobs
with more than 150 tasks \footnote{\addnsdiSHP{This is to avoid potential bias
towards \name. A job with more than 150 tasks will 
have to be scheduled in more than one phase, which will be in 
favor of \name by diminishing the sampling overhead.}{D6}}. 
(3) We next selected the execution trace following the process below
from the second half; these became 2STrace, GTrace11 and GTrace19,
respectively.
(4) We then selected jobs from the first half of each original trace
that are feature-clustered with those jobs in the execution trace to form
the "history" trace for 3Sigma.

We extracted the execution trace from each of the above-mentioned second halves 
by randomly selecting 1250 jobs with equal probability.
Then, for each extracted trace, we adjust the arrival time of the
jobs so that the average cluster load matches that in the original
trace~\cite{2Sigma:trace,googleTraceGithub,googleClusterData2019}.
Table~\ref{table:accuracy:load:stats} summarizes the workload
per window of the extracted traces, where a window is defined
as a 1000-second interval sliding by 100 seconds at a time,
and the load per window is the total runtime of all the jobs arrived in that window,
normalized by the total number of CPUs in the cluster times the window length, \ie 1000s.
We see that for all three traces, the average system load is close to 1, though 
the load fluctuates over time, which is preserved by the random uniform job extraction.

\begin{table}[tp]
  \caption{Statistics for system load per 1000s sliding window.}
  \label{table:accuracy:load:stats}
  \vspace{-0.1in}
  \centering
      {
\small
\begin{tabular}{|c|c|c|c|c|c|}
\hline
		 Trace       & Average &  P50 & P90\\
%\hline
\hline
	2STrace & 1.05  & 0.13 & 2.47 \\
\hline
	GTrace11 & 1.01  & 0.29 & 1.49 \\
\hline
	GTrace19 & 1.04  & 0.09 & 0.91 \\
\hline
\end{tabular}
}
  \vspace{-0.1in}
%  \vspace{-0.2in}	% AJ_spacecut
\end{table}

\paragraph{Prediction mechanisms and experimental setups.}
%\label{subsec:setup}
We implement the \primarybasepredict predictor following its description in~\cite{3Sigma}.
After learning the job runtime distribution (\S\ref{sec:accuracy:trace}),
it uses a utility function of the estimated job runtime
associated with every job to derive its estimated runtime from the
distribution, by integrating the utility function
over the entire runtime distribution.
Since our goal is to minimize the average JCT, we used a utility function that is inversely
proportional to the square of runtime. 
% To ensure correct execution of \primarybasepredict,
We kept all the default settings
% in a private communication~\cite{personalCommunication:JunWoo} with
we learned from the authors of 3Sigma~\cite{3Sigma}.
% and kept all the settings and parameters in our implementation the same 
% as in 3Sigma~\cite{3Sigma}.

As in \S\ref{sec:accuracy:trace}, \slearn
samples
% the number of sampled tasks assigned to a job in \lTechnique is
$max(1, 0.03 \cdot S)$ tasks per job, where $S$ is the number of tasks in the job.
%  We used 5\% sampling fraction to keep the experimental analysis
% the same as the trace analysis
% (\S\ref{sec:accuracy:trace}).
{We only show the results for wide jobs (with \thinLimit or more
  tasks) as in the complete \slearn design (\S\ref{sec:design:gs}),
  only wide jobs go through the sampling phase.  }

%\begin{figure*}[tp]
%\centering
%\subfigure[2STrace]
%{
%\includegraphics[width=0.33\linewidth]{figures/simulation/prediction_error_new_2STrace.pdf}
%%\caption{Job runtime learning accuracy for the 2Sigma trace.}
%\label{fig:sim:estimationAccuracy:2STrace}
%}
%\hspace{-0.25in}
%\subfigure[GTrace11]
%{
%\includegraphics[width=0.33\linewidth]{figures/simulation/prediction_error_new_GTrace11.pdf}
%\label{fig:sim:estimationAccuracy:GTrace11}
%}
%\hspace{-0.25in}
%\subfigure[GTrace19]
%{
%\includegraphics[width=0.33\linewidth]{figures/simulation/prediction_error_new_GTrace19.pdf}
%\label{fig:sim:estimationAccuracy:GTrace19}
%}
%\vspace{-0.15in}
%	\caption{Job runtime prediction accuracy \updated{all - 16th Sep 2020}}
%\vspace{-0.15in}
%\label{fig:sim:estimationAccuracy}
%\end{figure*}

\begin{figure*}[tp]
\centering
\subfigure[2STrace]
{
\includegraphics[width=0.30\linewidth]{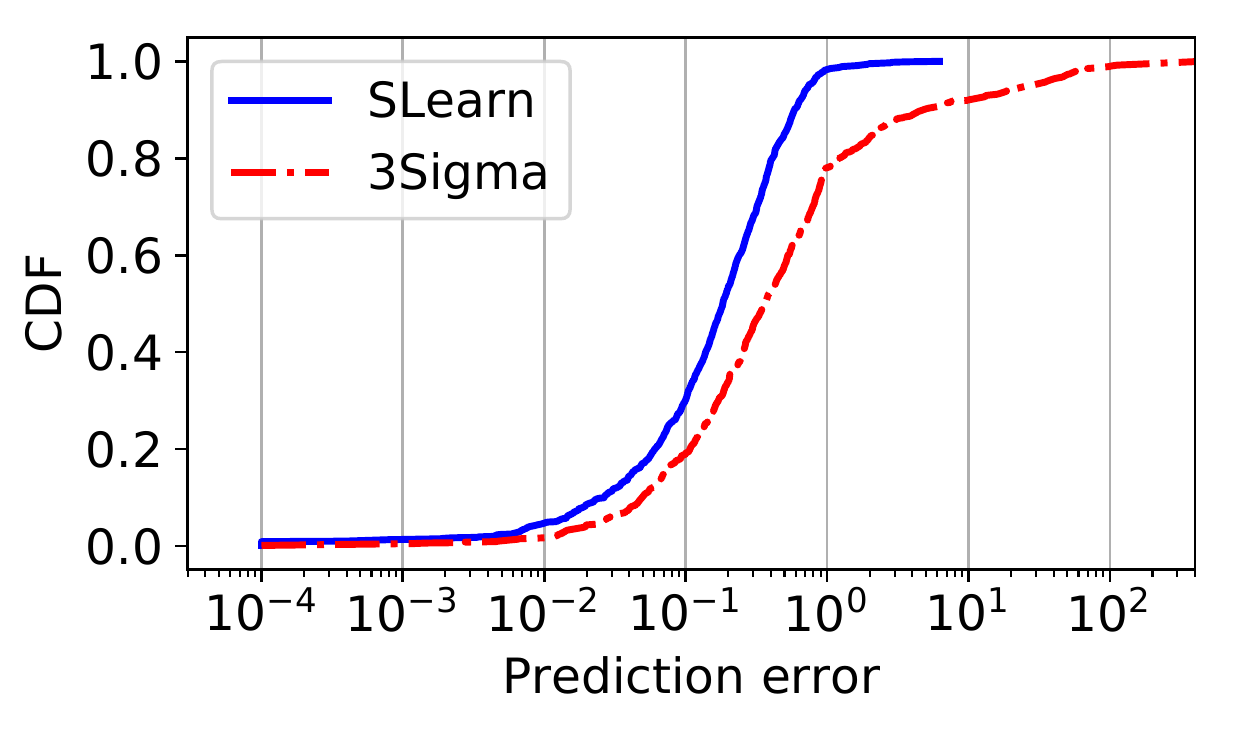}
\label{fig:sim:estimationAccuracy:2STrace}
}
%\hspace{-0.25in}
\subfigure[GTrace11]
{
\includegraphics[width=0.30\linewidth]{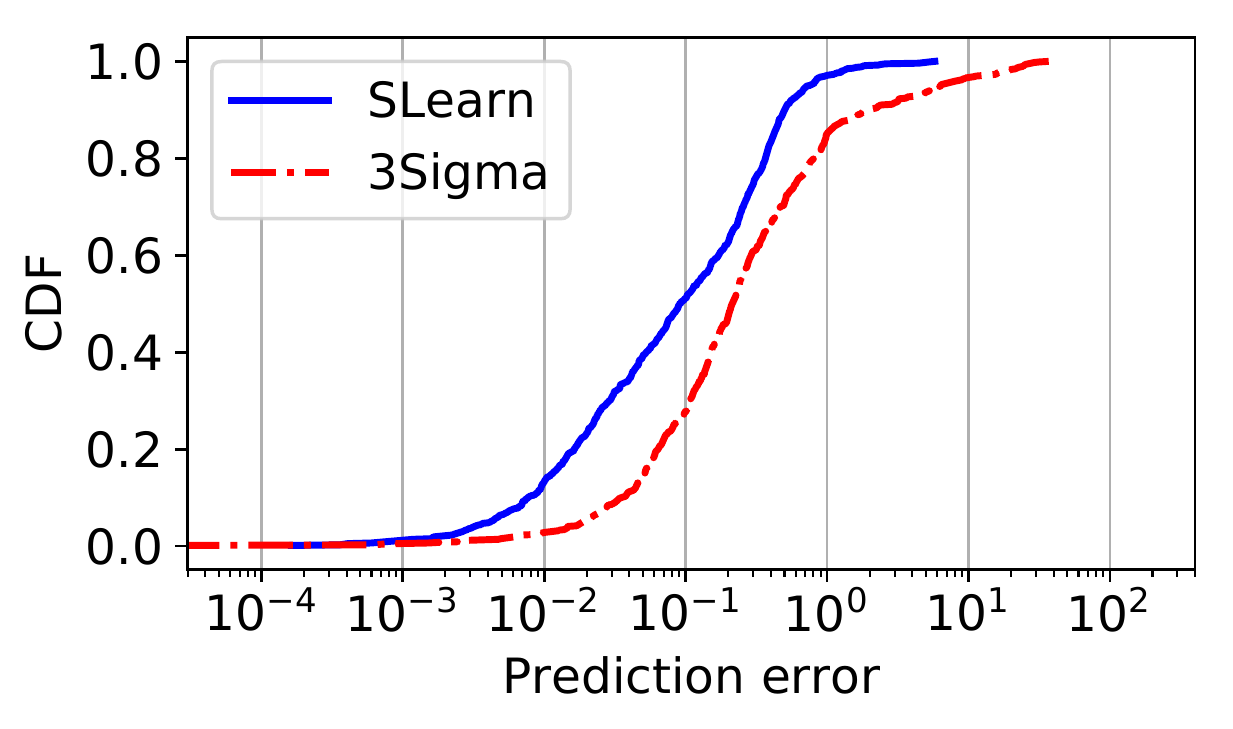}
\label{fig:sim:estimationAccuracy:GTrace11}
}
%\hspace{-0.25in}
\subfigure[GTrace19]
{
\includegraphics[width=0.30\linewidth]{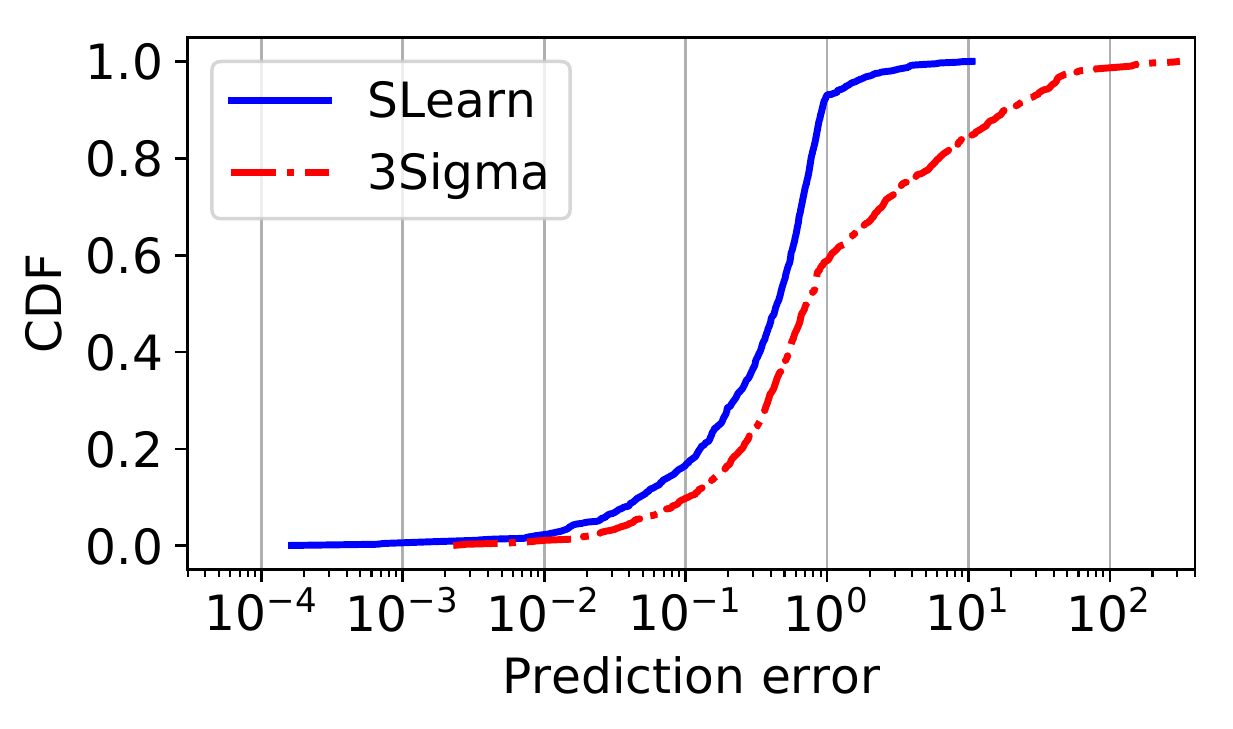}
\label{fig:sim:estimationAccuracy:GTrace19}
}
\vspace{-0.15in}
\caption{Job runtime prediction accuracy.
  % For \slearn we assumed the sampling ratio to be 3\%. \updated{all - 25th Sep 2020}
}
\vspace{-0.15in}
\label{fig:sim:estimationAccuracy}
\end{figure*}

\paragraph{Results.}
Fig.~\ref{fig:sim:estimationAccuracy} shows the CDF of percentage error in the predicted
job runtimes for the three traces.
% where error = $\frac{abs(predicted\_dur\ -\ actual\_dur)}{actual\_dur}\times100$.
{We see that \slearn has much better prediction accuracy than
  \primarybasepredict.} For 2STrace,
GTrace11, and GTrace19,
the P50 prediction error
are 18.30\%, 9.15\%, 21.39\% for \namepredict
but 36.57\%, 21.39\%, 71.56\% for \primarybasepredict, respectively,
and the P90 prediction error
% for 2STrace, GTrace11, and GTrace19
are 58.66\%, 49.95\%, 92.25\% for \namepredict
but 475.78\%, 294.52\%, 1927.51\% for \primarybasepredict, respectively.

%% file: case-study.tex
\section{Integrating Sampling-based Learning with Job Scheduling: A Case Study}
\label{sec:study}

In this section, we answer the second key question about the sampling-based
learning: {Can delaying scheduling the
  remaining tasks till completing the sampled tasks be
  compensated by the improved prediction accuracy?
%  the job performance, \eg completion time?
% \deadlineCS{or meeting deadline (SLO)}?
}
We answer it through extensive simulation and testbed experiments.

Our approach is to design a generic scheduler, denoted as \gs, that schedules
jobs based on job runtime estimates to optimize a given performance metric, 
average job completion time (JCT).  We then plug into \gs different
prediction schemes to compare their end-to-end performance.
\rm{In particular, we compare four predictors:
(1) the sampling-based predictor \slearn,
(2) the distribution based predictor proposed in 3Sigma~\cite{3Sigma},
(3) a point estimate predictor,
and
(4) a LAS estimator.
(5) an Oracle estimator, which always predicts with 100\% accuracy.
We further compare with a FIFO-based scheme, where the scheduler
simply prioritizes jobs in the order of their arrival.
}

%\addaj{We do not evaluate against Kairos~\cite{kairos:socc2018} as the policy
%is inherently preemptive and doesn't provide same liveness guarantee.}

\input design

%\subsection{Simulation}
\input simulation

\if 0
\subsubsection{Evaluation}
\label{sec:study:testbed}
\input testbed
\fi

\if 0
\section{Case Study - II}
\label{sec:study2}

\subsection{Scheduling for Meeting Deadlines}
\label{sec:study2:design}
\fi
%\input deadline-design

%% file: design.tex
\subsection{Scheduler and Predictor Design}
\label{sec:study:design}

%  We first discuss the design of \gs
%  and the baseline predictors and policies we are comparing \slearn with.

\subsubsection{Generic Scheduler \gs}
\label{sec:design:gs}

\gs replaces the scheduling component of a cluster manager like YARN~\cite{yarn:web}. 
% \comment{It does not deal with job and resource life-cycle management.???}
The key scheduling objective of \gs is to minimize the average JCT.
%without any prior knowledge of job runtimes.
Additionally, \gs aims to avoid starvation.
\rm{so that all jobs can continually make progress.}

The scheduling task in \gs is divided into two phases, (1) job runtime estimation,
and (2) efficient and starvation-free scheduling of jobs whose runtimes have been
estimated. 
We focus here on the scheduling mechanism 
and discuss the different job runtime estimators in the following sections.

%  \paragraph{Runtime estimators: }
%  We can plug different schemes into \gs to estimate runtimes. We describe
%  designs of the baseline predictors used in our experiments and the \namepredict
%  in \S\ref{sec:design:baselines} and \S\ref{sec:design:namepredict}. All these
%  predictors are trained or designed to predict average task runtime for a job.

\paragraph{Inter-job scheduling. }
Shortest job first (SJF) is known to be optimal in minimizing the average JCT
when job execution depends on a single resource.
%Since a distributed job has many tasks, in \gs we use a simple heuristic of
%ordering jobs based on the total runtime, \ie $average$ $task$ $runtime$
%$\times$ $number$ $of$ $tasks$ (\textit{impact of the job}) which has been
%shown to be effective~\cite{aalo:sigcomm15}.
Previous work has shown
% {Aalo~\cite{aalo:sigcomm15} (Varys
that scheduling distributed jobs
% (\eg coflows)
even with prior knowledge is NP-hard (\eg~\cite{varys:sigcomm14}),
and an effective online heuristic is to order the distributed jobs
based on each job's total size~\cite{jajooPhilae, philaeTechReport, aalo:sigcomm15, jajooSaath} \iftoggle{nsdi22}{}{\cite{jajoo2020exploiting}}.
In \gs we use a similar heuristic;
the jobs are ordered based on their total estimated runtime, \ie $mean$ $task$ $runtime$
$\times$ $number$ $of$ $tasks$.
% (\textit{impact of the job}).

\paragraph{Starvation avoidance.}
SJF is known to cause starvation to long jobs. Hence, {in \gs we adopt
a well-known multi-level priority queue structure to avoid job 
starvation~\cite{feedback:jacm1968, raiLAS:sigmetrics2003, nuyens:survey2008, aalo:sigcomm15, graviton:hotcloud16}}.  
% Unlike classin non-clairvoyant schedulers --
% least-attained service (LAS) in single links \cite{raiLAS:sigmetrics2003,
% nuyens:survey2008} and multi-level feedback queues (MLFQ) in operating systems
% \cite{tss:afips1962, feedback:jacm1968, OSConcepts}
%    -- that perform fair sharing in presence of similar flows/tasks to provide interactivity, our
%   solution improves the average CCT even in presence of identical flows.
%
Once \gs receives the runtime estimates of a job, it assigns the job
to a priority queue based on its runtime.  Within a queue, we use
FIFO to schedule jobs. Across the queues, we use weighted sharing of
resources, where a priority queue receives a resource share according
to its priority.
% In \gs, the weights for resource share and priority distribution of queues are
% configurable and can be determined by the policy in use. 

In particular, \gs uses $N$ queues, $Q_0$ to $Q_{N-1}$, with each queue having
a lower queue threshold Q$^{lo}_{q}$ and a higher threshold Q$^{hi}_{q}$ for
job runtimes. We set Q$^{lo}_0$ = 0, Q$^{hi}_{N-1}$ = $\infty$, Q$^{lo}_{q+1}$ =
Q$^{hi}_{q}$. A queue with a lower index has a higher priority.
\gs uses exponentially growing queue thresholds, \ie
Q$^{hi}_{q+1}$ = E $\cdot$ Q$^{hi}_{q}$.
%  By default, \gs sets E = 10, N = 10, shown to be effective in our
%  evaluation (\S\ref{sec:study:sim}).
To avoid any bias, we use the multiple priority queue structure with the same
configuration when comparing different job runtime estimators.
%Since \name needs to separate thin jobs and jobs which are in sampling phase
%we add two additional queues, thin queue and sampling queue, with \name.

%\paragraph{Failure tolerance and recovery}
{\paragraph{Basic scheduling operation.} \gs keeps track of
resources being used by each priority queue. It offers the next available resource
to a queue such that the weighted sharing of resources among the queues 
for starvation avoidance is maintained. Resources offered to a queue are
always offered to the job at the head of the queue.}

% \vspace{-0.1in}

\subsubsection{\slearn}
\label{sec:design:namepredict}

\rm{Since \slearn learns job runtimes online by sampling pilot tasks, it
needs to interact with the scheduler.}
To seamlessly integrate \slearn
with \gs, we need to use one
of the priority queues for scheduling
sampled tasks. We denote it as the sampling queue.

%   \paragraph{Fast sampling.}
%   As pointed above that non-sampling tasks of a job keep waiting until all
%   sampling tasks are over. Hence it becomes very important to make the process of
%   sampling fast. To speed up the process of sampling, \name treats the sampling
%   queue as the second highest priority queue.

\paragraph{Fast sampling.}
One design challenge is how to determine the priority for the sampling queue
w.r.t. the other priority queues.
On one hand, sampled tasks should be given high priority so that the job
runtime estimation can finish quickly. On the other hand, the jobs whose
runtimes have already been estimated should not be further delayed by learning
new jobs. To balance the two factors, we use the second
highest priority in \gs as the sampling queue.

\paragraph{Handling thin jobs.}
%\label{sec:design:thin}
Recall that in \name, when a new job arrives, \name only schedules its pilot
tasks, and delays other tasks until the pilot tasks finish
and the job runtime is estimated. Such a design choice can inadvertently lead
to higher JCTs for thin jobs, \eg a two-task job would experience serialization
of its two tasks. To avoid JCT degradations for thin jobs, we place a job
directly in the highest priority queue if its width is under a threshold thinLimit.
% (set to \thinLimit in \name; \S\ref{sec:sim:thin} evaluates the sensitivity to this design provision).
%\soccReviewEdit{C}{In principle the thinLimit should be dynamically adjusted
%based on observed workload in a cluster. We leave this as future work and in
%this paper we} set it to \thinLimit;

\paragraph{Basic operations.}
% Fig.~\ref{fig:design:arch} shows the \slearn architecture.
Upon the arrival of a new job, the cluster manager asynchronously communicates
the job's information to \gs, which relays the information to \slearn.  If the
number of tasks in the job is under thinLimit, \slearn assigns it
to the highest priority queue; otherwise, the job is assigned to the
sampling queue, where a subset of its tasks (\textit{pilot tasks}) will be scheduled to run.
% We do not schedule the tasks of a job other than the pilot tasks until the
% completion of the pilot tasks to avoid unnecessary resource blocking and
% slowing down other potentially shorter jobs.
Once a job's runtime is estimated from sampling, it is placed in the
priority queue corresponding to its runtime estimate where the rest of its tasks
will be scheduled.

\if 0
\begin{figure}[tp]
  \centering
  \includegraphics[page=2, width=0.95\linewidth]{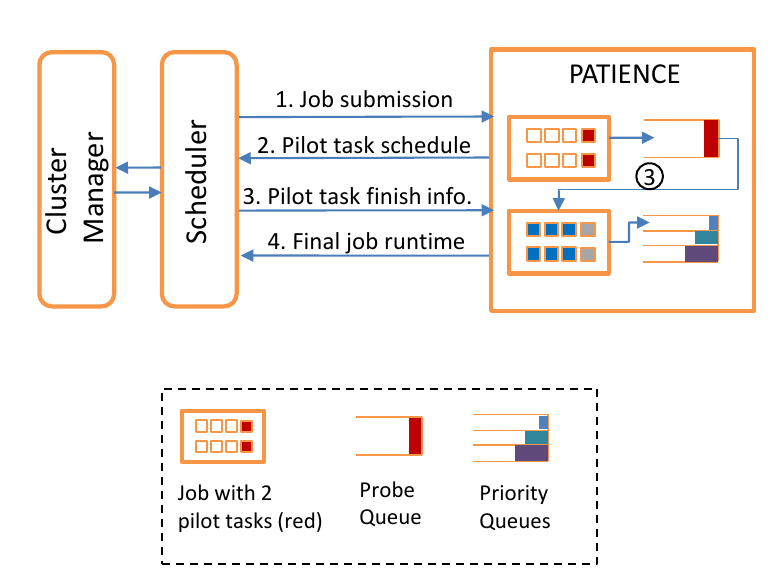}
  \caption{\name architecture.}
  \label{fig:design:arch}
	\vspace{-0.1in}
\end{figure}
\fi

\paragraph{How many and which pilot tasks to schedule?}
When a new job arrives, \name first needs to determine the number of
pilot tasks.
%  The number of pilot tasks affects the trade-off between
%  the job runtime estimation accuracy and scheduling overhead.
Sampling more tasks can give higher estimation accuracy, 
%   especially in the
%   presence of task skew, which leads to better scheduling order. However,
%   sampling more tasks 
but also consumes more resources early on, which can potentially delay other
jobs, if the job turns out to be a long job and should have been scheduled to
run later under SJF.
Further, we found the best sampling ratio appears to vary across difference traces.
%   \editaj{We experiment with varying the number of pilot
%   tasks to find the one that strikes a good balance between estimation accuracy
%   and overhead. Once the number of pilot tasks is fixed, the next step is to pick
%   tasks for piloting.}
%
To balance the trade-off, we use an adaptive algorithm to dynamically
determine the sampling ratio, as shown in
Figure~\ref{fig:design:AdaptiveSamplingAlgo}. The basic idea of the
algorithm is to suggest a sampling ratio that has resulted in the lowest
job completion time normalized by the job runtime based on the recent past.
To achieve this, for every value in a defined range of possible sampling ratios (between
1\% and 5\%), it maintains a running score ($srScoreMap$), which is
the average normalized JCT of $T$ recently finished jobs that used the corresponding
sampling ratio. In practice we found a $T$ value of 100 works reasonably well.
During system start-up, it tries sampling ratios of 2\%, 3\%, and 4\%
for the first $3T$ jobs (Line 2--7).
It further tries sampling ratios of 1\% and 5\% if
going down from 3\% to 2\%
or going up from 3\% to 4\%
reduces the normalized JCT.
Afterwards, for each new job, 
it uses the sampling ratio that has the lowest running score.
Finally, upon each job completion, the score map is updated (Line 16--24).

Once the sampling ratio is chosen,
\slearn selects pilot tasks for a job randomly.

\paragraph{How to estimate from sampled tasks?}  Several methods such
as bootstrapping, statistical mean or median can be used to predict job
properties from sampled tasks.  In \gs, we use empirical mean
to predict the mean task runtime.
% $t_{avg}$ of a job and then the job runtime as $N_{tasks}\cdot t_{avg}$.

\paragraph{Work conservation.}
% By default, \slearn schedules non-sampling tasks
%of a job only after all its sampling tasks are over. This can lead to
When the system load is low,
some machines may be idle while the non-sampling tasks are waiting for
the sampling tasks to finish. In such cases, \slearn schedules
non-sampling tasks of jobs to run on otherwise idle machines.
%  that are still in the sampling phase when a
%  machine becomes available and there are no jobs in non-sampling priority queues.
In work conservation, the jobs are scheduled in the FIFO order of their arrival.
\input algo
%\vspace{-0.1in}
\subsubsection{Baseline Predictors and Policies}
\label{sec:design:baselines}
We compare \slearn's effectiveness against four different baseline predictors
and two policies:
%  (1) \primarybasepredict,  (2) \pointestimator , (3) \oracle , and two policies (4) \las
%and (5) \fifo. All baseline  predictors predict average task runtime.
{\bf (1) \primarybasepredict:} as discussed in \S\ref{sec:accuracy:experiment}.
{\bf (2) \primarybasepredictTL:} same as \primarybasepredict but
handles thin jobs in the same way as \slearn; they are directly placed in the highest priority queue. This is to isolate the effect of thin job handling.
\if 0
As discussed in \S\ref{sec:accuracy:experiment},
\primarybasepredict~\cite{3Sigma} 
predicts the historical distribution of runtimes 
% instead of just a point estimate 
and estimates a job's runtime by integrating some utility function over the distribution.
To minimize the average JCT, we used a utility function that is inversely
proportional to the square of runtime.  We kept all the other settings
and parameters in our implementation of \primarybasepredict the same 
as in~\cite{3Sigma}, which were confirmed with the authors of
3Sigma~\cite{personalCommunication:JunWoo}.
\fi
{\bf (3) \pointestimator: }
% \pointestimator is another history-based predictor. We keep its design the
same as \primarybasepredict, with the only difference being that, instead of integrating a
utility function over the entire runtime history, it predicts a
point estimate (median in our case) from the history.
%  We keep other settings and mechanism to pick the best feature
%  for \pointestimator similar to default 3Sigma-predict~\cite{3Sigma}.
{
{\bf (4) \las: }
% We also compare \slearn against a
The Least Attained Service~\cite{raiLAS:sigmetrics2003} policy
% which is an effective way to
approximates SJF online without explicitly learning job sizes,
and is most recently implemented in the Kairos~\cite{kairos:socc2018} scheduler.
% The policy is inherently preemptive and does not provide the same
% liveliness guarantee as \slearn with \gs.
%  Further, \slearn is a learning based technique and Kairos is a
%  non-learning approach.
\if 0
Since we were not able to access the simulation source code of Kairos at the
provided url~\cite{kairosScheduler} or from the authors,
we reimplemented \las.
%  in order to evaluate \slearn thoroughly
%
In our implementation, \las starts all the jobs in the highest priority queue
and schedules a task only for a duration up to the threshold of the current queue
of the job. When a task finishes executing for the assigned duration, \las
updates its parent job's priority.
\fi
\las uses multiple priority queues
and the priority is inversely proportional to
the service attained so far, \ie the total execution time so far. We use the
sum of all the task execution time to be consistent with all the other schemes.
%  \slearn, because other baselines and \slearn predict average task duration and
%  priority is calculated on basis of product of average duration and width \ie
%  total duration.%sum
%of all task runtimes.
}
{\bf (5) \fifo: } 
{The FIFO policy in YARN simply prioritizes jobs in the order of their arrival.}
%  Upon arrival, it appends each job to the tail of the highest priority
%  queue.
Since FIFO is a starvation free policy, there is no need for
multiple priority queues.
{\bf (6) \oracle: }
\oracle is an ideal predictor that always predicts with 100\% accuracy.

%% file: algo.tex
%Its \textbf{objective:} is to calculate sampling ratio that is expected to
%yeild best in runtime and prediction accuracy tradeoff for the system condition
%at that instance. It \textbf{outputs} the best sampling percentage
%({\textit{sp}}) to be used.\\
%Following are \textbf{defination of variables and abbreviations} used in the
%pseudocode in fig.~\ref{fig:design:AdaptiveSamplingAlgo}: \textit{$T$} -
%Maximum number of past jobs to keep in record corresponding to any sampling
%percentage.  \textit{$SP$ or $sp$} - denotes sampling percentage. \textit{$spScoreMap$}\\
\begin{center}
\begin{figure}[tp]
%\vspace{-0.2in}
\begin{small}
  \centering
\begin{minipage}{1\textwidth}
\captionsetup[algorithm]{font=footnotesize}
\noindent
\fbox{%
\begin{varwidth}{\dimexpr\linewidth-2\fboxsep-2\fboxrule\relax}
%%\begin{algorithm} %issue is with use of algorithm package. It is conflicting with floatrow also
%%\caption{Algo}
{\small
\begin{algorithmic}[1]
\Procedure{GetCurrentSamplingPercentage}{Job j}
  %\For{q in Q}
  %\EndFor
  \If {j in First $T$ jobs}
  \State \Return 3
  \ElsIf {j in Second $T$ jobs}
  \State \Return 2
  \ElsIf{j in Third $T$ jobs}
  \State \Return 4
  \EndIf
  %\Else
      \State minScore = getMinValue(srScoreMap)
      \If {minScore.SR == 2} % \&\& srScoreMap[1].value == NULL}
           \If {1.1*minScore.value < srScoreMap[3].value}
     	\State \Return 1
         \EndIf
       \EndIf
       \If {minScore.SR == 4} % \&\& srScoeMap[5].value == NULL}
          \If {srScoreMap[3].value > 1.1*minScore.value}
     	\State \Return 5
         \EndIf
       \EndIf
    \State \Return minScore.SR
    \EndProcedure
   \Procedure {UpdateScoreOnJobCompletion}{Job j}
  \State sr = j.sr  \Comment{Get j's sampling ratio.}
  \State normalizedJCT = j.jct  \Comment{Get j's normalized JCT.}
  \State UpdateScoresMap(sr, normalizedJCT)
\EndProcedure
\Procedure {UpdateScoreMaps}{sr, normalizedJCT}
    \If {Len(jobWiseSrScoresMap[sr])>$T$} 
      \State Drop first element of jobWiseSrScoresMap[sr]
    \EndIf
  \State jobWiseSrScoresMap[sr].append(normalizedJCT)
  \State srScoreMap[sr].value = mean(jobWiseSrScoresMap[sr])
  \EndProcedure
\end{algorithmic}
}
\label{algo:AdaptiveSampling}
%%\end{algorithm}
\end{varwidth}%
}
\end{minipage}
\caption{Adaptive sampling algorithm in \name.}
\vspace{-0.1in}	%	AJ_spacecut
\label{fig:design:AdaptiveSamplingAlgo}
\end{small}
\end{figure}
\end{center}
\vspace{-0.1in}	%	AJ_spacecut

%% file: simulation.tex
\subsection{Experimental Results}
\label{sec:study:sim}

We evaluated \slearn's performance against the six baseline schemes discussed
above by plugging them in \gs and execute the \numTraces traces (2STrace, GTrace11,
and GTrace19) 
\addnsdiSHP{using large scale simulations 
and on a 150-node testbed cluster in Azure (\S\ref{subsec:testbed}).}{E6}

\begin{figure*}[tp]
\centering
	\subfigure[2STrace]{
	\includegraphics[width=0.31\textwidth]{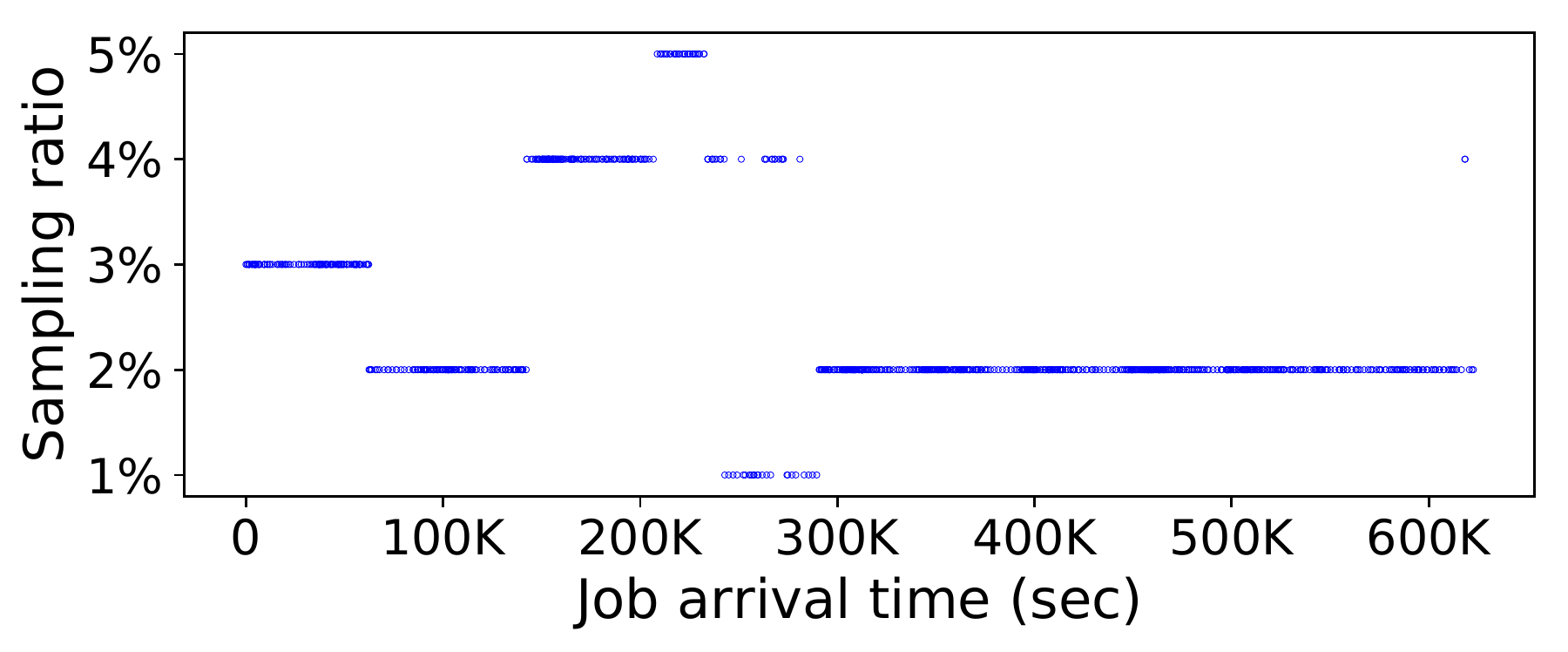}
	\label{fig:sim:numPilots:2STrace}
	\vspace{-0.1in}
}
%\hfill
	\subfigure[GTrace11]{
\vspace{-0.2in}
	\includegraphics[width=0.31\textwidth]{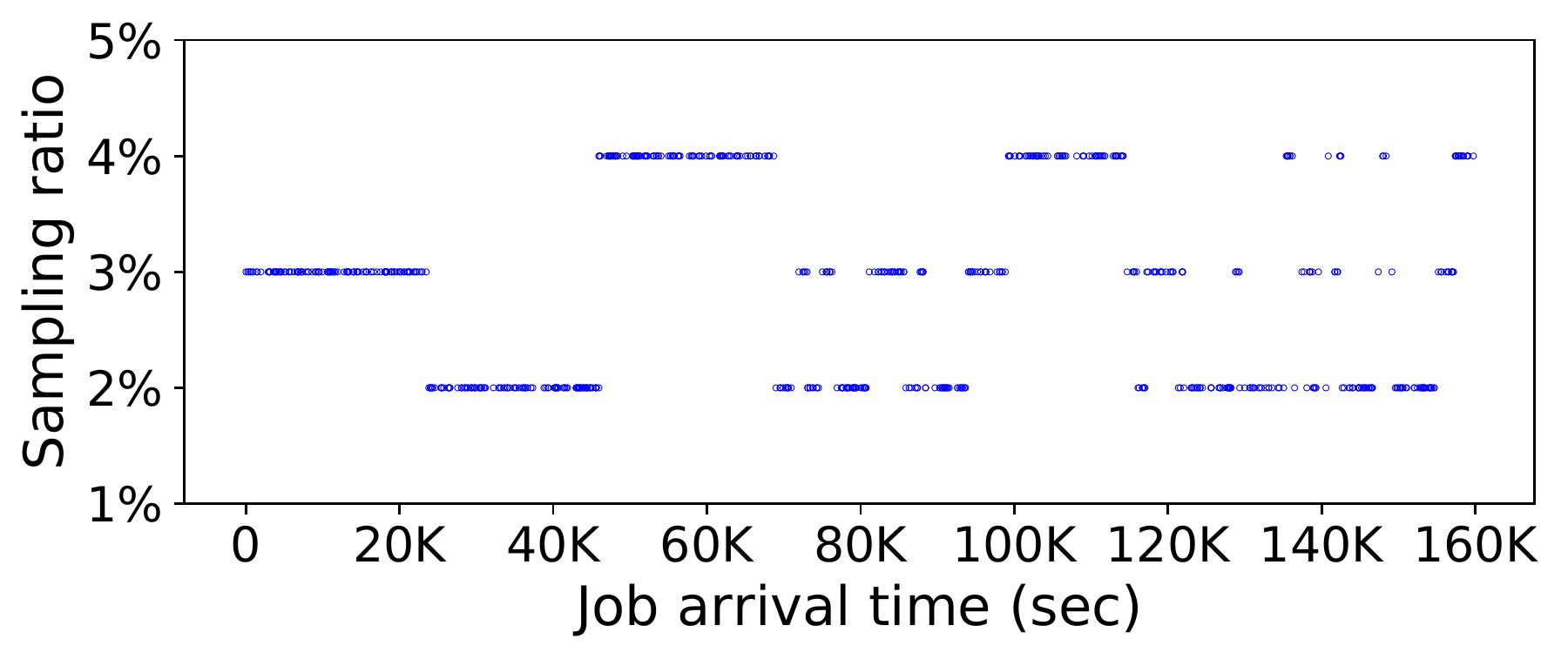}
	\label{fig:sim:numPilots:GTrace11}
	\vspace{-0.1in}
}
%\hfill
	\subfigure[GTrace19]{
	\includegraphics[width=0.31\textwidth]{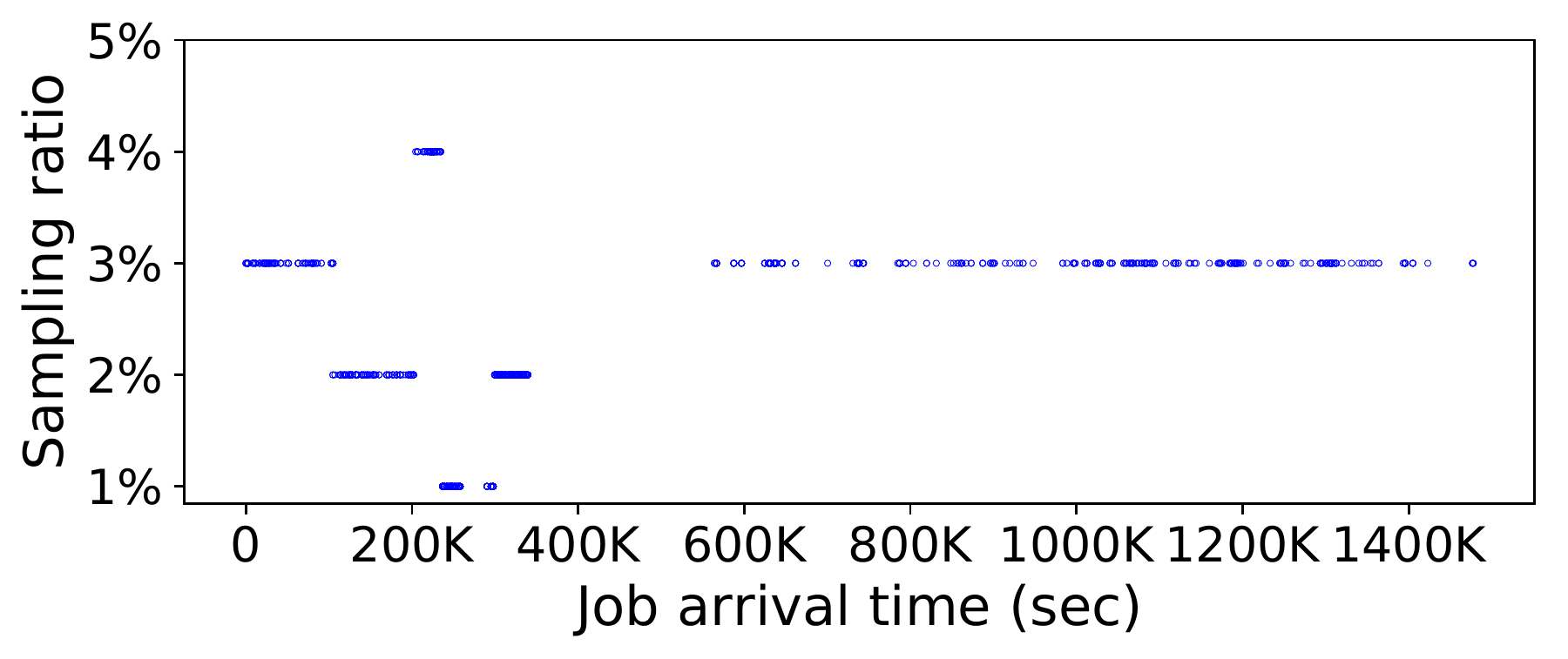}
	\label{fig:sim:numPilots:Trace19}
	\vspace{-0.1in}
}
\vspace{-0.2in}
\caption{Sampling ratios selected by the adaptive sampling algorithm.
  The duration of initial $3T$ jobs appear varying due to uneven arrival times.
  % \updated{all - 25th Sep 2020}
}
\label{fig:sim:numPilots}
\vspace{-0.1in}
\end{figure*}

\subsubsection{Experimental Setup}
\label{sec:sim:setup}

\paragraph{Cluster setup.}
% Our simulated cluster uses the same number of nodes (150) as in our real cluster.
We implemented \gs, \slearn and baseline estimators with 11 KLOC of Java
and python2. We used an open source java patch for Gridmix
\cite{gridmixpatch:junwoo} and open source java implementation of
NumericHistogram~\cite{numericHistogramJavaPatch} for Hadoop. We used some
parts from DSS, an open source job scheduling simulator \cite{dssScheduler}, in
simulation experiments.

% {We emulated a Yarn cluster with \gs replacing the scheduling module }
We implemented a proxy scheduler wrapper that plugs into the resource
manager of YARN~\cite{yarn:web} and conducted real cluster experiments
on a 150-node cluster in \removesl{Microsoft}{MS} Azure~\cite{azure:web}.

Following the methodology in recent work on cluster job
scheduling~\cite{3Sigma,jamiasvu,stratus:socc2018}, 
% we model jobs as
% mapper-only jobs. 
we implemented a synthetic generator based on the Gridmix
implementation to replay 
% mapper-only 
jobs that follow the arrival time and task
runtime from the input trace. The Yarn master runs on a standard DS15 v2 server with
20-core 2.4 GHz Intel Xeon E5-2673 v3 (Haswell) processor and 140GB memory, andthe
slaves run on D2v2 with the same processor with 2-core and 7GB memory.
%a Standard F4s server with 4-core 2.4 GHz Intel Xeon E5-2673 v3 (Haswell)
%processor and 8GB memory. The local agents run on D2v2 with the same processor

\paragraph{Parameters.}
The {default parameters} for priority queues in \gs in the experiments
are: starting queue threshold ($Q^{hi}_0$) is $10^6$ ms, exponential threshold
growth factor ($E$) is 10, number of queues ($N$) is set to 10, and the weights for
time sharing assigned to individual priority queues decrease exponentially by a
factor of 10.
Previous work (\eg~\cite{aalo:sigcomm15}) and our own evaluation have shown that the
scheduling results are fairly insensitive to these configuration parameters. We
omit their sensitivity study here due to page limit.
% The default inter-job scheduling policy is SJF.
%In \slearn, by default the number of pilot tasks assigned to wide jobs is
%$max(1, ceil(0.05 \cdot S))$, where $S$ is the number of tasks,
\slearn chooses the number of pilot tasks for wide jobs using
the adaptive algorithm described in \S\ref{sec:design:namepredict}
and the threshold for thin jobs is set to \thinLimit.
We evaluate the effectiveness of adaptive sampling
in \S\ref{sec:sim:numPilots} and the sensitivity to thinLimit in \S\ref{sec:sim:thin}.

\if 0
In order to achieve accurate replication of \primarybasepredict we, discussed
the default settings in a private
communication~\cite{personalCommunication:JunWoo} with the authors of the
3Sigma~\cite{3Sigma}.  We kept all the other settings and the parameters in
our implementation of \primarybasepredict same as used in the
3Sigma~\cite{3Sigma}. 
\fi

%\questionaj{How to write about utility function of 3Sigma?\\}
%\questionaj{Where should we write about things that we are predicting average
%task duration?}

%To make this
%article self-contained we briefly describe it here. It uses four estimation
%techniques namely, average, median, rolling average and average of X(=20) recent job
%runtimes and it uses following features, application name, user name, job name,
%resource requested and submission time (hour, day etc.) of the job.  The
%\primarybasepredict tracks accuracy of each, feature value and estimation pair
%technique \cite{3Sigma} by the help of a running metric. It always uses the
%pair which is expected to give the most accurate prediction.

\paragraph{Performance metrics.}
%   {The primary goal of this paper is to show that
%   sampling-based prediction can be a viable alternative to history-based
%   prediction. Thus, the specific choice of the runtime property to optimize is
%   somewhat orthogonal to our main focus. We pick  the job completion time as our
%   specific objective as it is an important metric that has also been studied in
%   prior work \cite{DontCryOverSpilledRecords}, \cite{kairos:socc2018} and
%   \cite{cora:infocom2015}. Consistent with this objective, we mainly focus on
%   estimating task running times.} 
% \commentaj{Note: In the past we have had got a negative feedback about the choice of
% JCT as the metric.\\}
We measure three performance metrics in the
evaluation: JCT speedup, defined as the ratio of a JCT under a baseline
scheme over under \slearn, the job runtime
estimation accuracy, and job waiting time.

\if 0
\soccReviewEdit{A, B}{\paragraph{Job model.}
\label{sec:sim:jobmodel}
For all our experiments in this paper, each job consists of a single phase of
parallel tasks. We assume one-phase model because (1) the same model is
assumed in previous work \cite{jamiasvu}, \cite{3Sigma} and
\cite{stratus:socc2018}; (2) it is practical to implement - in each phase the
application manager submits to YARN scheduler a request for the tasks belonging
to this phase, and the YARN scheduler then decides how to schedule them. (3) as
stated in \cite{stratus:socc2018} as well, the traces used in our experiment do not contain
multi-phase information.  Our sampling-based learning method can be applied to
multi-phase jobs (DAGs), e.g., within each phase. Note that scheduling (of
multi-phase or single-phase jobs) is orthogonal to and can be decoupled from
learning job/task size.}
\fi

\if 0
Since we also need training data for history-based
baseline scheduler, we divided the source traces into two halves in
chronological order and extracted the training data from the first half and the
test trace from the second half. Since our cluster size is 150 nodes, we
filtered out jobs with more than 150 tasks. Jobs were selected at random from
the base traces. We assume each task in the trace requires one node and zero
memory.  Training data is 3 times as large as the test trace.
\fi

\paragraph{Workload.}
\label{sec:sim:workload}
We used the same training data for history-based estimators and the test
traces (2STrace, GTrace11 and GTrace19)
% are extracted from two production traces.
as described in
\S\ref{sec:accuracy:experiment}.
\if 0
We could not use the Mustang and Trinity traces released
in~\cite{workloadDiversity:atc18} as they do not contain information about
individual task runtimes.
\fi
%\commentaj{Leaving a place holder here for why not google 2019 trace if not used finally.}

\subsubsection{Effectiveness of Adaptive Sampling}
\label{sec:sim:numPilots}

\begin{figure*}[tp]
\vspace{-0.15in}
\centering
\subfigure[\vspace{-0.2in} 2STrace]
{
\includegraphics[width=0.30\linewidth]{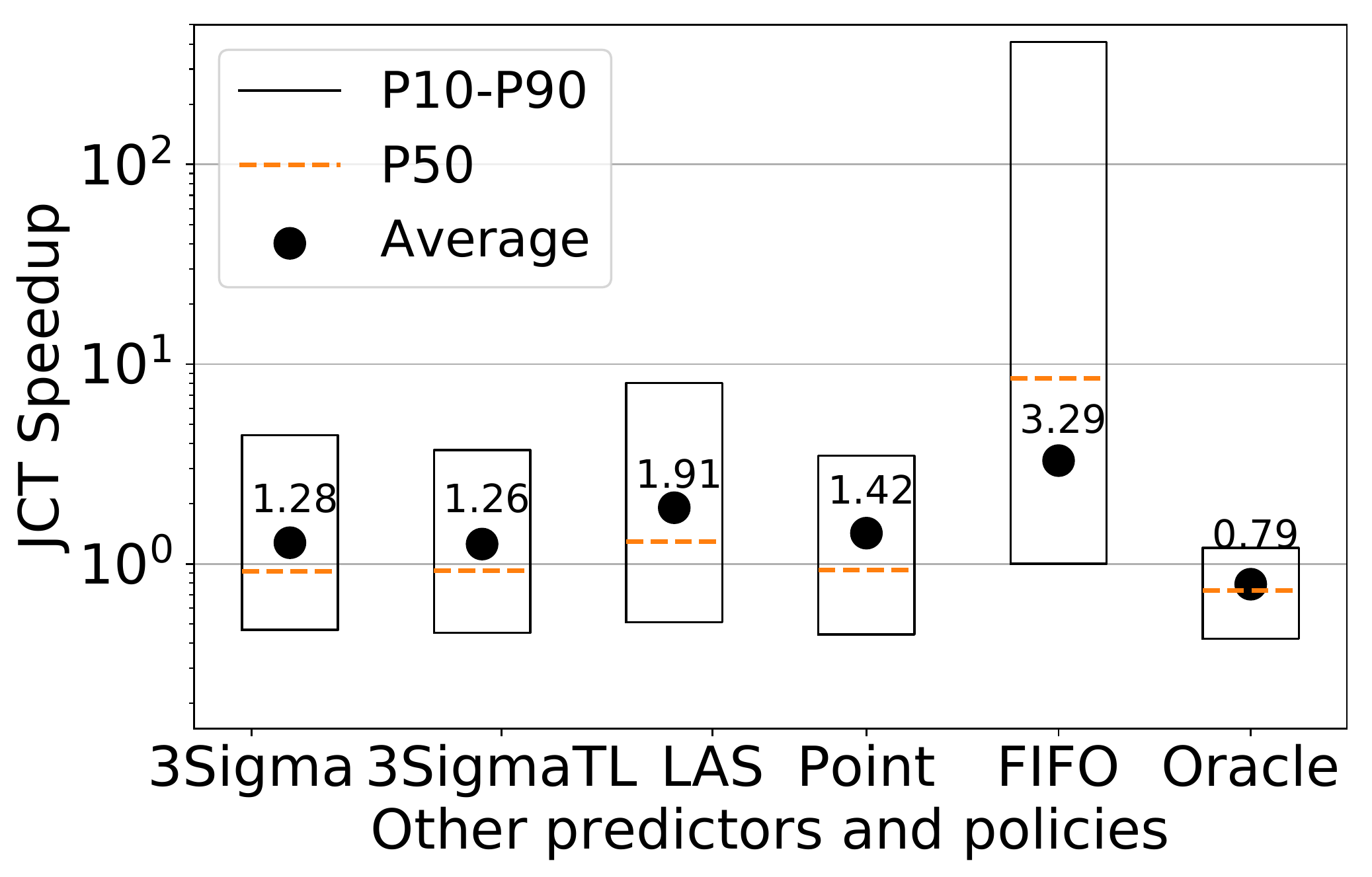}
\label{fig:sim:jctOthers:2STrace}
}
\hfill
\subfigure[\vspace{-0.2in} GTrace11]
{
\includegraphics[width=0.30\linewidth]{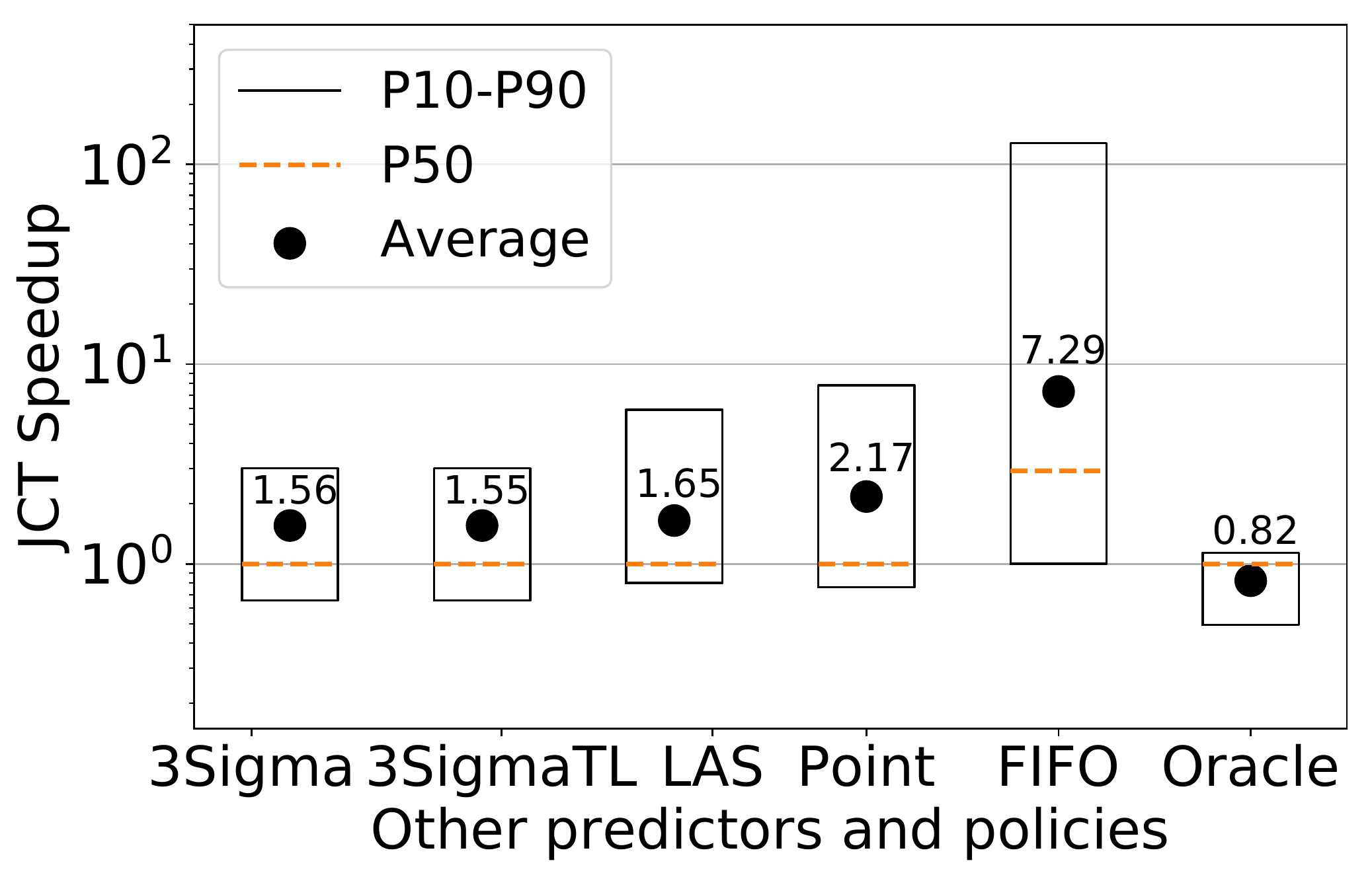}
\label{fig:sim:jctOthers:GTrace11}
}
\hfill
\subfigure[\vspace{-0.2in} GTrace19]
{
\includegraphics[width=0.30\linewidth]{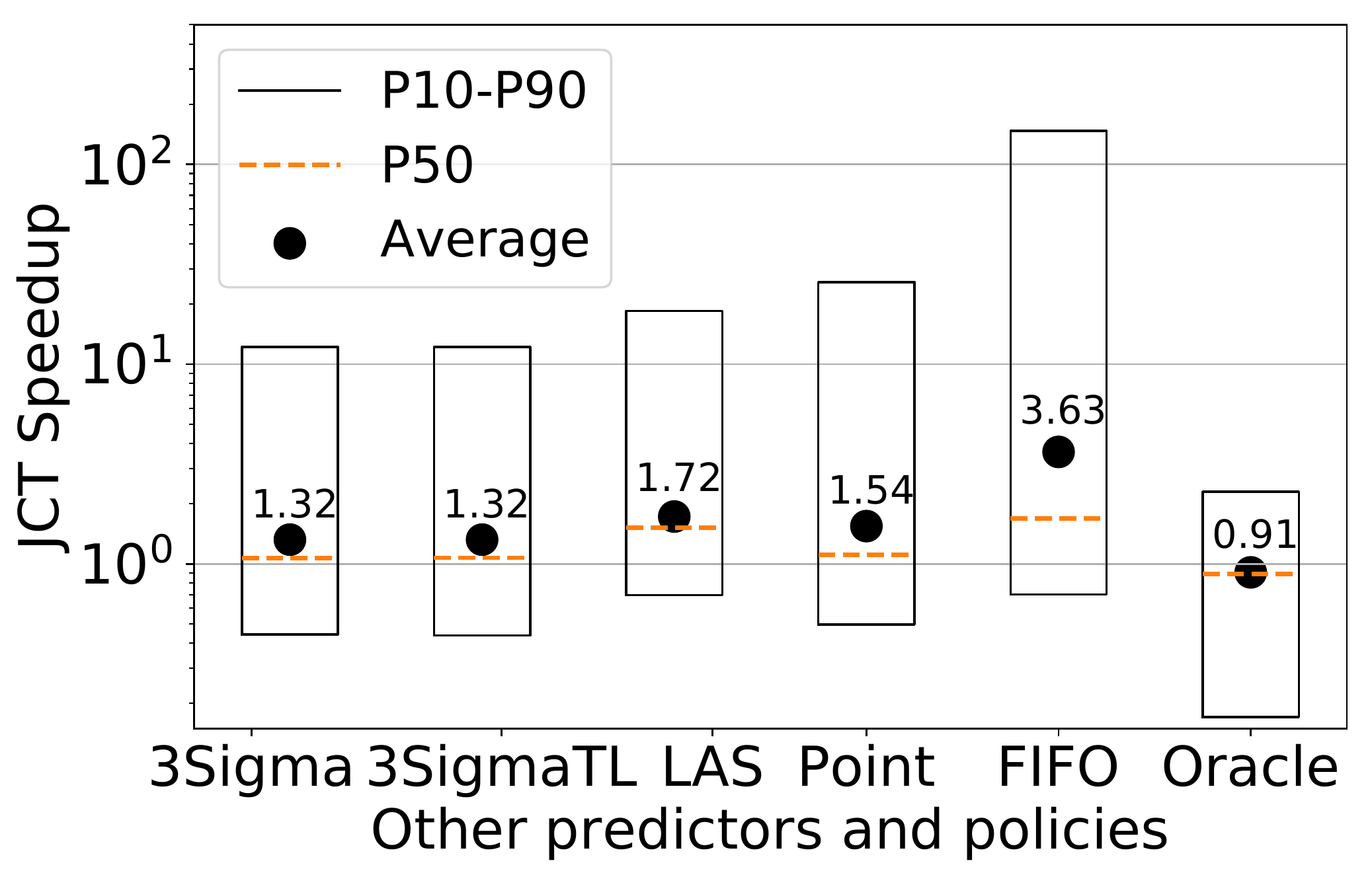}
\label{fig:sim:jctOthers:GTrace19}
}
\vspace{-0.2in}
\caption{JCT speedup using \slearn as compared to other baseline schemes for
	the three traces. \updated{all - 16th Sep 2020}
}
\vspace{-0.1in}
\label{fig:sim:jctOthers}
\end{figure*}

\begin{table}[tp]
\centering
\caption{Performance improvement of \slearn over \primarybase
  under adaptive sampling and fixed-ratio sampling.
  %\updated{all - 16th Sep 2020}
  % {Percentile average task waiting times are calculated after removing thin jobs.}
}
\label{table:sim:numPilots:2STrace}
\vspace{-0.25in}
{\small	
		\begin{tabular}{cp{0.14in}p{0.14in}p{0.14in}p{0.14in}p{0.14in}p{0.14in}p{0.14in}}\\
\hline
			 & \multicolumn{7}{c}{Fraction of tasks chosen as pilot tasks }\\
			&1\%&2\%&3\%&4\%&5\%&10\%&Adap.\\
\hline
		& \multicolumn{7}{c}{2STrace}\\
\hline
			P50 pred. error (\%) \hspace{-0.1in}&19.4&19.0&19.0&18.7&18.4&16.9&19.0\\
			Avg. JCT speedup ($\times$)\hspace{-0.1in}&1.24&1.23&1.27&1.26&1.27&1.28&1.28\\
			P50 speedup ($\times$)\hspace{-0.1in}&0.93&0.92&0.93&0.92&0.93&0.91&0.92\\
\hline
		& \multicolumn{7}{c}{GTrace11}\\
\hline
			P50 pred. error (\%)\hspace{-0.1in}&14.4&14.0&13.6&13.1&12.7&9.09&13.7\\
			Avg. JCT speedup ($\times$)\hspace{-0.1in}&1.52&1.55&1.54&1.56&1.58&1.51&1.56\\
			P50 speedup ($\times$)\hspace{-0.1in}&1.00&1.00&1.00&1.00&1.00&1.00&1.00\\
\hline
		& \multicolumn{7}{c}{GTrace19}\\
\hline
			P50 pred. error (\%)\hspace{-0.1in}&55.7&53.8&47.1&46.5&42.1&36.1&51.8\\
			Avg. JCT speedup ($\times$)\hspace{-0.1in}&1.31&1.31&1.31&1.32&1.28&1.24&1.32\\
			P50 speedup ($\times$)\hspace{-0.1in}&1.07&1.07&1.05&1.05&1.01&1.00&1.07\\
\hline
	\end{tabular}
}
\vspace{-0.2in}
\end{table}

\if 0
\soccReviewEdit{A, C}{The results for GTrace show the same trend as for 2STrace.
The speedup is 1.40$\times$,
1.28$\times$, and 1.19$\times$,
when sampling 5\%, 10\% and 20\% tasks, respectively.
The speedup increase from sampling 5\% to sampling 10\% 
tasks is slightly faster for GTrace compared to
2STrace. ????? This is because GTrace has a relatively higher fraction of thin tasks
and hence, by increasing the sampled tasks, the negative impact of sampling on
thin tasks increases more.  However, the trends are similar for both traces.
Also, prediction error shows an improving trend. The error decreases as the
number for sampling tasks is increased. For 5\% tasks, P50 prediction error
(50PE) is 7.85\%; for 10\% it is 6.09\%; and for 20\% it is 4.26\%}
\fi
In this experiment, we evaluate the effectiveness of our adaptive
algorithm for task sampling. Fig.~\ref{fig:sim:numPilots}
shows how the sampling ratio selected by the adaptive algorithm for each
job varies between 1\% and 5\% over the duration of the three traces. 
We further compare average JCT speedup and P50
speedup under the adaptive algorithm with those under a fixed sampling
ratio, ranging between 1\% and 10\%.
Table~\ref{table:sim:numPilots:2STrace} shows that 
the adaptive sampling algorithm leads to the best speedups for 2STrace
and GTrace19 and is about only 1\% worse than the best for GTrace11.
Interestingly, we observe that no single sampling ratio works the best for
all traces. Nonetheless, the adaptive algorithm always chooses one that
is the best or closest to the best in terms of JCT speedup.
More importantly, we see that the adaptive algorithm does not always
use the sampling ratio with the best prediction accuracy, which shows
that it effectively balances the tradeoff between
prediction accuracy and sampling overhead.
\subsubsection{Prediction Accuracy}
\label{sec:sim:accuracy}

\slearn achieves more accurate estimation of job runtime over 
\primarybasepredict\ --
the details were already discussed in \S\ref{sec:accuracy:experiment}.

%\vspace{-0.2in} % AJ_spacecut
\subsubsection{Average JCT Improvement}
\label{sec:sim:averageJCT}

We now compare the JCT speedups achieved using \slearn over using the
five baseline schemes defined in \S\ref{sec:design:baselines}.
% (1) \primarybase, (2) \pointestimator, (3) \oracle, (4) \las, and (5) \fifo.
%  All experiments use the default parameters discussed in the setup, including
%  $K, E, S$, unless otherwise stated.

Fig.~\ref{fig:sim:jctOthers:2STrace}
shows the results for 2STrace. We make the following observations.
(1)
% \oracle predicts with 100\% accuracy and thus all the jobs are placed in
% their correct queues.  Therefore,
%  The JCT of \oracle serves as a lower-bound for all other schemes.
%
Compared to \oracle, \slearn achieves an average and P50 speedups of
0.79$\times$ and 0.73$\times$, respectively.
%  In other words, the JCT of \slearn is not increased significantly even
%  though
This is because \slearn has some estimation error; it places 10.91\% of wide
jobs in the wrong queues, 3.54\% in lower queues and 7.37\% in higher queues.
(2) \slearn improves the average JCT over \primarybase by 1.28$\times$.
% and the P50 speedup is 0.92$\times$.
This significant improvement of \slearn comes
from much higher prediction accuracy compared to \primarybasepredict
(Fig.~\ref{fig:sim:estimationAccuracy}).
(3) The improvement of \slearn over \primarybasepredictTL, 1.26$\times$,
is similar to that over \primarybasepredict,
confirming thin job handling only played a small role in
the performance difference of the two schemes.
To illustrate \slearn's high prediction accuracy,
we show in Table~\ref{table:sim:correctQueue} the fraction of wide jobs that were
placed in correct queues by \slearn and \primarybasepredict.
We observe that \slearn consistently
assigns more wide jobs to correct queues
than \primarybasepredict for all three traces.
(4) Compared to \pointestimator, \rm{which uses a point estimate generated from
  historical data,} \slearn improves the average JCT by 1.42$\times$.
 %and the P50 speedup is 0.93$\times$.
%
Again, this is because \slearn estimates runtimes with higher accuracy. 
(5) Compared to \las, \slearn achieves an average JCT speedup of
1.91$\times$ and P50 speedup of 1.29$\times$. This is because \las
pays a heavy penalty in identifying the correct queues of jobs
by moving them across the queues incrementally.
% In that process, many large jobs are scheduled before small jobs which results in
% needless delay for small jobs.
(6) Lastly, compared with \fifo,
%  where all the jobs are scheduled in the order of arrival,
\slearn achieves an average JCT speedup of 3.29$\times$ and P50
speedup of 8.45$\times$.

\begin{figure*}
	%\hspace{-0.2in}
\begin{minipage}[t]{0.24\textwidth}
	\includegraphics[width=\linewidth]{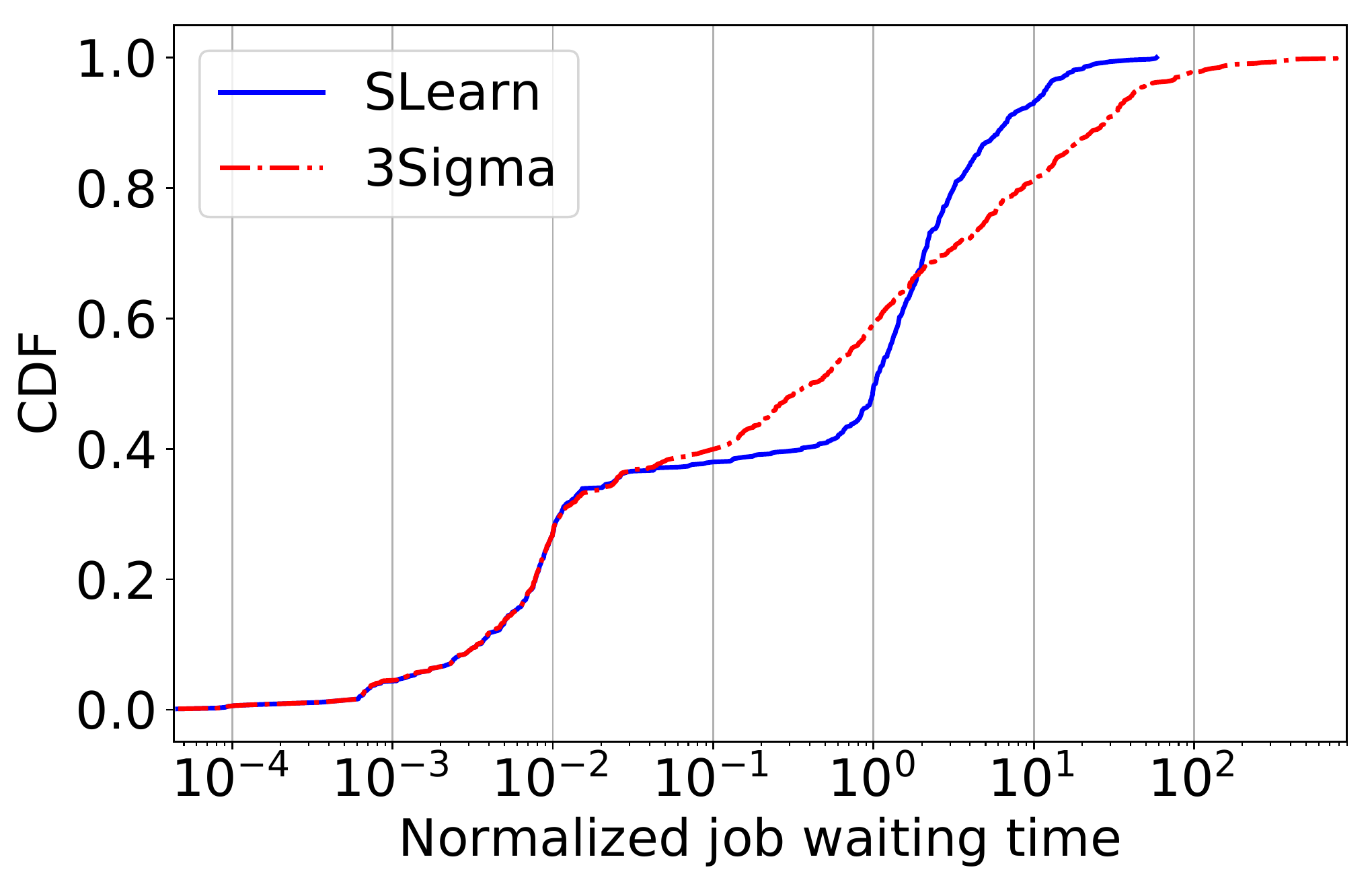}
	\vspace{-0.30in}
	\captionof{figure}{CDF of waiting times for wide jobs in GTrace11. \updated{all - 23rd Sep 2020}}
	\label{fig:sim:waitingTimes:GTrace11}
\end{minipage}
\hfill
\begin{minipage}[t]{0.24\textwidth}
	\includegraphics[width=\linewidth]{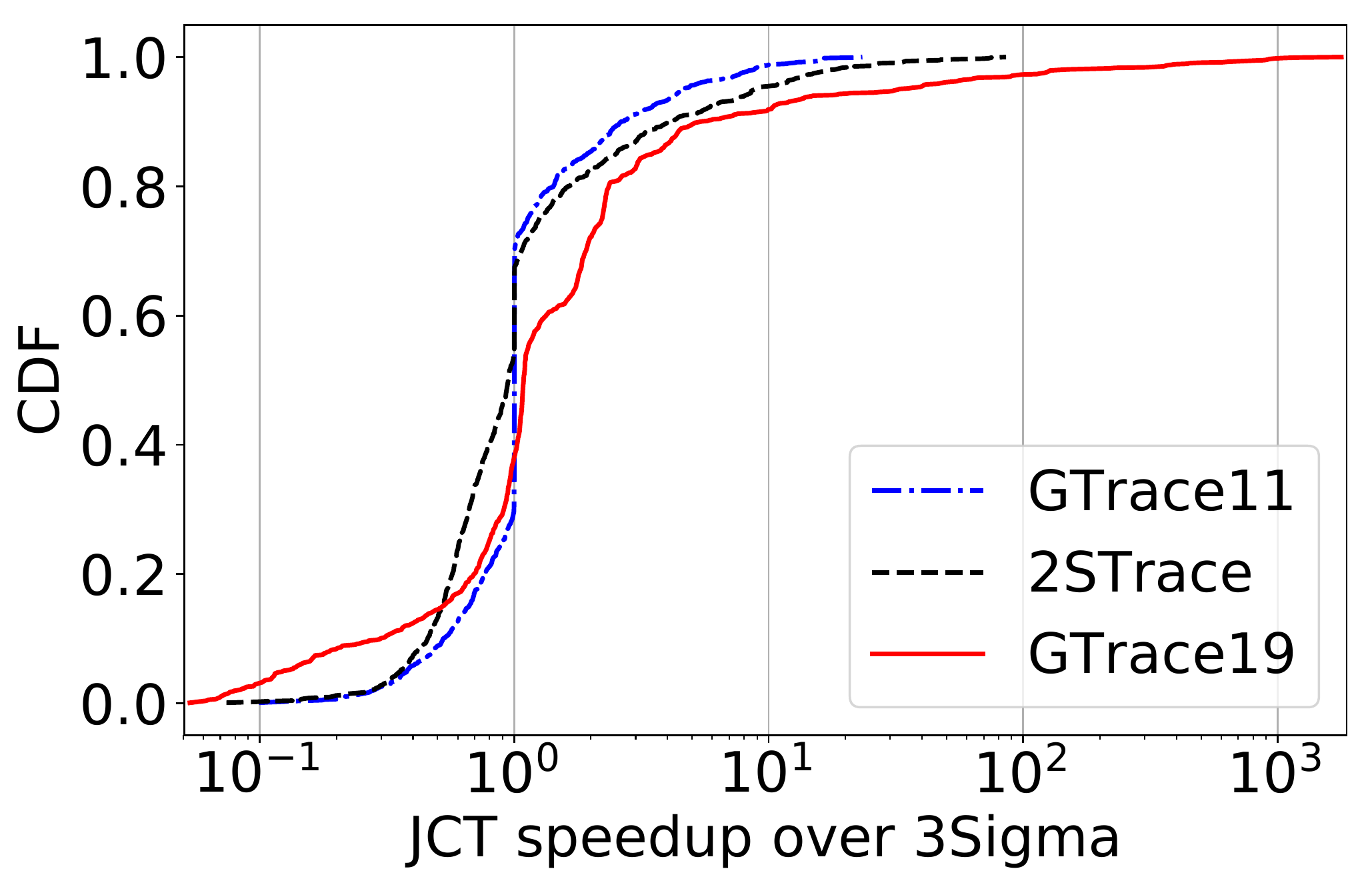}
	\vspace{-0.30in}
	\captionof{figure}{[Testbed] CDF of speedup: \slearn vs \primarybasepredict.}% for 2STrace, GTrace11, and GTrace19. \updated{23rd Sep 2020}}
	\label{fig:testbed:speedup:cdf}
\end{minipage}
\hfill
\begin{minipage}[t]{0.24\textwidth}
	\includegraphics[width=\linewidth]{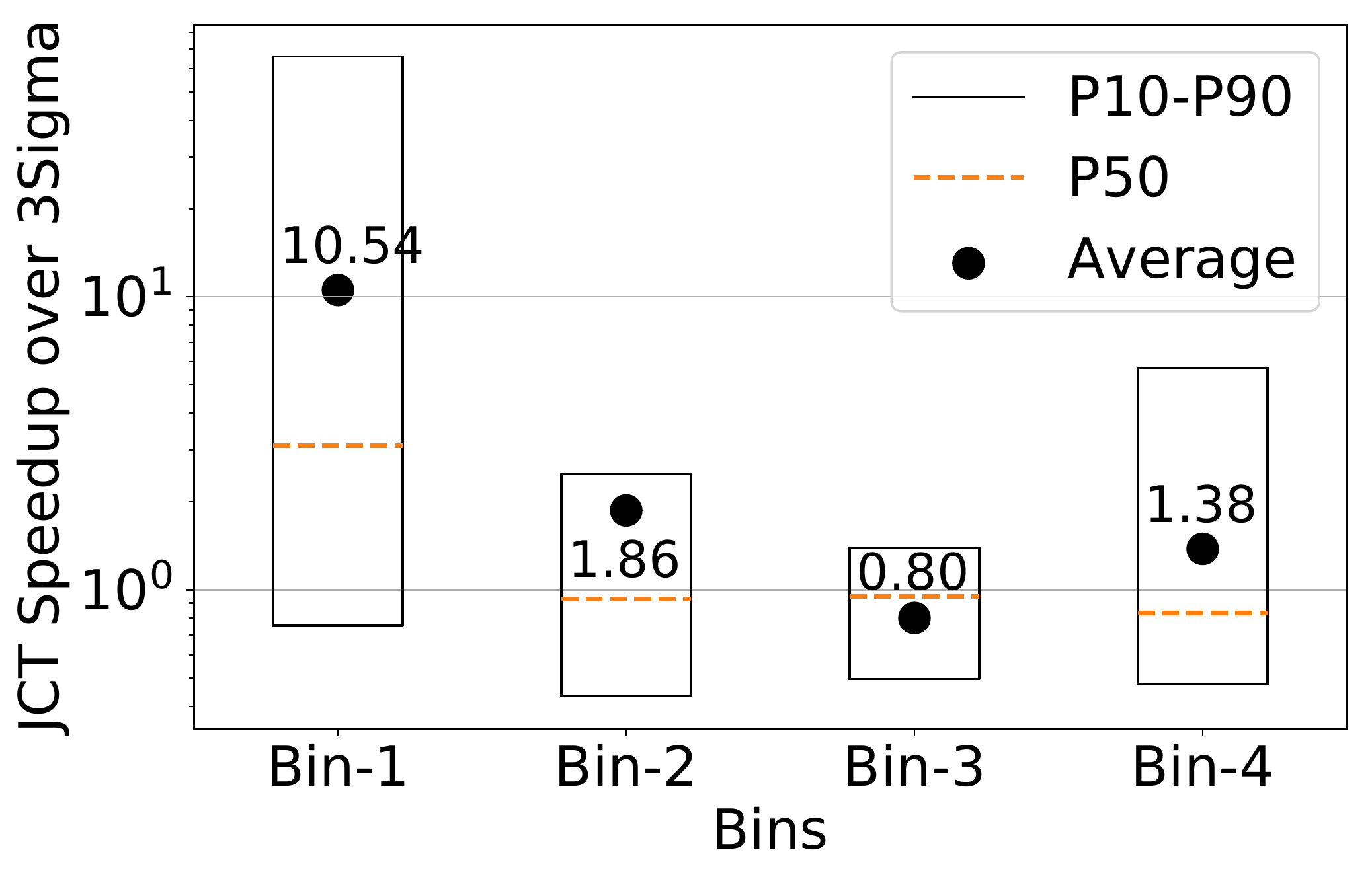}
	\vspace{-0.30in}
	\captionof{figure}{Performance breakdown into the bins in Table~\ref{table:sim:bin:2STrace}.}
	\label{figs:sim:bin}
\end{minipage}
\hfill
\begin{minipage}[t]{0.24\textwidth}
    \includegraphics[width=\linewidth]{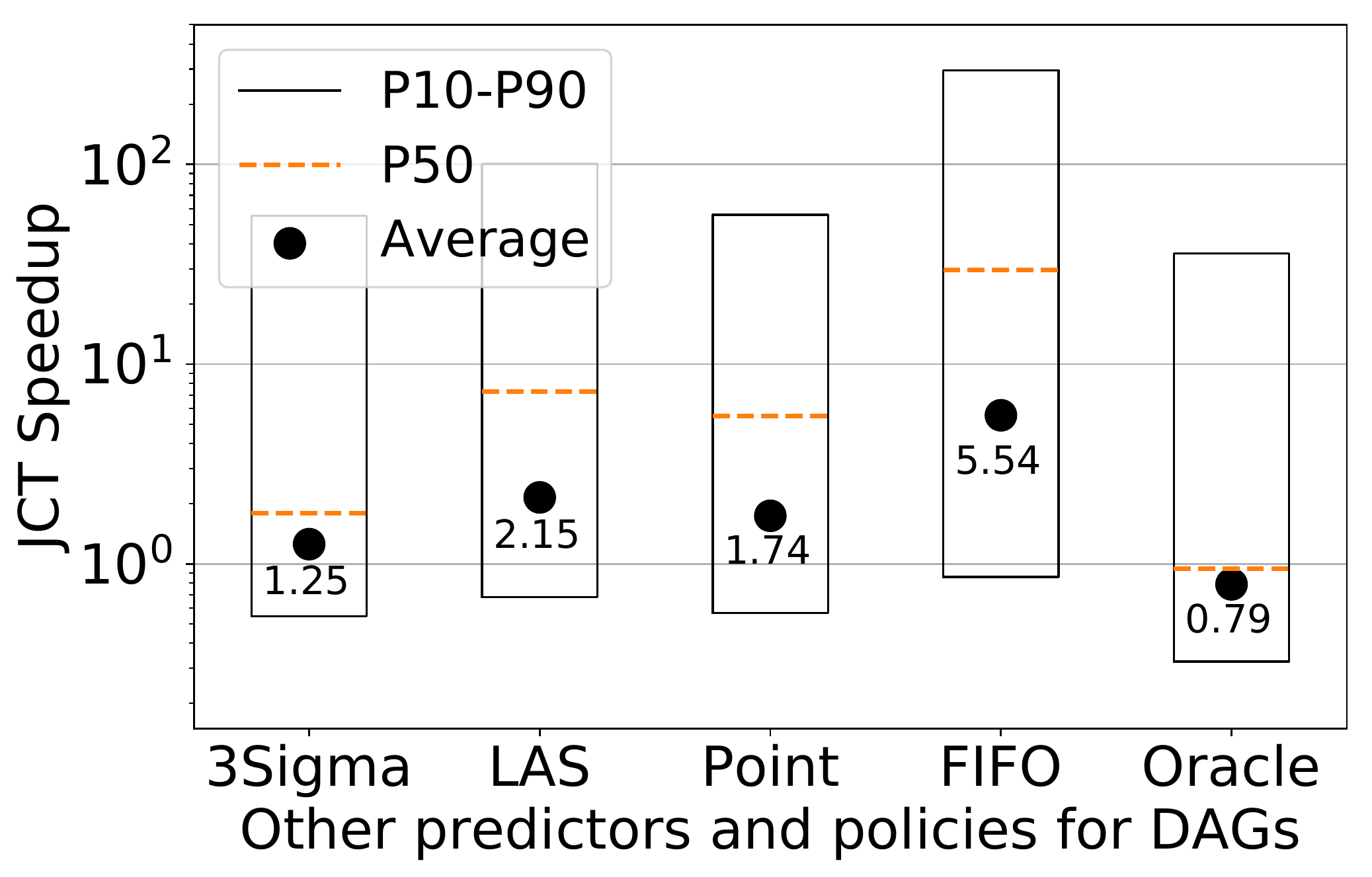}
	\vspace{-0.30in}
	\captionof{figure}{JCT speedup using \slearndag over baselines for \dagtrace.}
    \label{fig:sim:jctOthers:GTrace19-DAG}
\end{minipage}
\vspace{-0.2in}
\end{figure*}

\begin{table}[tp]
  \caption{Percentage of the wide jobs that had correct queue assignment.
    \updated{all - 16th Sep 2020 TableStructure - 10 Jul 2020}
    }
\vspace{-0.1in}
\label{table:sim:correctQueue}
  \centering
      {\small
	\begin{tabular}{|c|c|c|c|c|c|} 
	  \hline
		Prediction&	\slearn &\primarybasepredict \\%&\oracle\\  
		%Prediction&	\slearn &\primarybasepredict &\pointestimator\\  
		Technique&&\\%&\\
	  \hline
		2STrace &89.09\%&73.84\%\\%&100\%\\
		GTrace11 &86.45\%&76.20\%\\%&100\%\\
		GTrace19 &73.96\%&58.07\%\\%&100\%\\
		%Queue Assign-&89.38\%&68.39 (73.85)\%&61.02 (62.8)\%\\
		%ment Accuracy&&&\\
	  \hline
	\end{tabular}
      }
\vspace{-0.1in}
\end{table}

\if 0
A recent trace analysis paper~\cite{workloadDiversity:atc18} showed
the distribution of job runtimes can vary from one cluster to another.
To evaluate \slearn's robustness to trace distribution, we compared
\slearn against all four baseline schemes using GTrace.  
\fi

Fig.~\ref{fig:sim:jctOthers:GTrace11} shows the results for GTrace11.  
% We see that 
Scheduling under \slearn again outperforms all other schemes. In particular,
using \slearn improves the average JCT by
1.56$\times$ compared to using \primarybasepredict,
1.55$\times$ compared to using \primarybasepredictTL,
2.17$\times$ compared to using Point-Est,
and 1.65$\times$ compared to using the LAS policy.
Fig.~\ref{fig:sim:jctOthers:GTrace19} shows that
scheduling under \slearn outperforms all other schemes for GTrace19 too.
In particular, using \slearn improves
the average JCT by 1.32$\times$, 1.32$\times$, 1.54$\times$, and 1.72$\times$
compared to using \primarybasepredict, \primarybasepredictTL,
% the Point-Estimate predictor,
\pointestimator and 
the LAS policy, respectively.

In summary, our results above show that \slearn's higher estimation accuracy
outweighs its runtime overhead from sampling, and as a result achieves much
lower average job completion time than history-based predictors and the LAS policy for
the three production workloads.

\if 0
\commentaj{Cluster G from 2019 has very high error. However, that is not the
case with all the cluster of 2019. I tried with other cluster H with all tier
jobs. That also has a speed up 1.25$\times$ and P50 (P90) error for \slearn is
1.23 (9.39)\% and for \primarybasepredict it is 8.63 (53.60)\%. With cluster B
beb-tier jobs the speedup is 1.9$\times$ and P50 (P90) error for \slearn is
7.65 (58.68)\% and for \primarybasepredict it is 28.72 (144.01)\%. I think we
can use this result somewhere as cluster G 2019 has very high error for both
\slearn and \primarybasepredict.}
\fi

\vspace{-0.05in}
\subsubsection{Impact of Sampling on Job Waiting Time}
\label{sec:sim:waitingTimes}

% A key question about sampling-based learning is whether sampling pilot tasks
% first will cause the remaining tasks of a job to wait too long, thus affecting
% its JCT.  To answer this question,
To gain insight into why sampling pilot tasks first under \slearn does not hurt
the overall average JCT, 
we next compare the {\em normalized waiting time} of jobs,
calculated as the average waiting time of its tasks under
  the respective scheme, divided by the mean task length of the job.
%  We compare the job waiting times under \slearn with that under
%  \primarybasepredict.

Fig.~\ref{fig:sim:waitingTimes:GTrace11} shows the CDF of the
normalized job waiting time under \slearn and
\primarybasepredict. We see that the CDF curves can be divided into
three segments.
%%% Total size of the segment in following is on the basis of \slearn 289, 241, and 260
(1) The first segment, where both SLearn and 3Sigma have normalized waiting
time (NWT) less than 0.04, covers 36.58\% of the jobs, and 35.57\% of the jobs are
common. The jobs have almost identical NWT, much lower than 1 under both
schemes. This happens because during low system load periods, \eg lower than 1, 
the scheduler will schedule all the tasks to run under both scheme; under
\slearn it schedules non-sampled tasks of jobs to run before their sampled
tasks complete due to work conservation.
(2) The second segment, where both schemes have NWT between 0.04 and
1.90, covers 30.51\% of the jobs, and 20.38\% of the jobs are common.  Out of
these 20.38\%, 29.81\% have lower NWT under \slearn and 70.19\% have lower NWT
under \primarybasepredict. This happens because when the system load is
moderate, the jobs experience longer waiting time under \slearn
than under \primarybasepredict because of sampling delay.
(3) The third segment, where both schemes have NWT above 1.90, cover
32.91\% of the jobs, and 24.68\% of jobs are common. Out of these 24.68\%,
83.08\% have lower waiting time under \slearn and 16.92\% under
\primarybasepredict. This happens because when the system load is relatively high,
although jobs incur the sampling delay under \slearn, they also experience
queuing delay under \primarybasepredict, and the more accurate prediction of
\slearn allows them to be scheduled following Shortest Job First
more closely than under \primarybasepredict.

A detailed 
% timeline 
analysis of how the system load of the trace
affects the relative job performance under the two predictors can be
found in the Appendix\iftoggle{techreport}{}{ in \cite{slearnTechReport}}.
%~\cite{slearnTechReport}.
% of \iftoggle{techreport}{.}{ of the full report~\cite{slearnTechReport}.}

\subsubsection{Testbed Experiments}
\label{subsec:testbed}

We next perform end-to-end evaluation of \slearn and 3Sigma on our 150-node
Azure cluster.  Fig.~\ref{fig:testbed:speedup:cdf} shows the CDF of JCT
speedups using \slearn over \primarybasepredict using 2STrace, GTrace11 and
GTrace19.  \slearn's performance on the testbed is similar to that observed
in the simulation.  In particular, \slearn achieves {average} JCT speedups of
1.33$\times$, 1.46$\times$, and 1.25$\times$ over \primarybase for the 2STrace,
GTrace11, and GTrace19 traces, respectively.
%  \addaj{The experimental traces have job arrivals spread over a very long
%  period. So, to make the testbed experiments practically feasible, we compressed
%  all the inputs and queue parameters along time. In the comppressed traces, the
%  difference between the arrival of the first and last job is an hour.}

%\vspace{-0.1in}
\subsubsection{Binning Analysis}
\label{sec:sim:binning}

To gain insight into how different jobs are affected by \slearn over
\primarybase, we divide the jobs into four bins  in
Table~\ref{table:sim:bin:2STrace} for 2STrace
% , \ref{table:sim:bin:GTrace11}, \ref{table:sim:bin:GTrace19},
and show the JCT speedups for each bin in Fig.~\ref{figs:sim:bin}.
The results for the other two traces are similar and are omitted due to page limit.

\begin{table}
  \caption{Breakdown of jobs based on total duration and width (number
    of tasks) for 2STrace. 
\editnsdiSHP{Shown in brackets are a bin’s
share in term of job count and total job runtime.}{Shown in brackets are a bin's fraction of all the jobs in the trace
    in terms of job count and total job runtime.}{A2}
}
% The 	numbers in brackets denote the fraction of jobs in that bin for the 2STrace. \updated{10th May 2019 for 2STrace and on 13th for GTrace}}
  \label{table:sim:bin:2STrace}
\vspace{-0.1in}	
  \centering
      {\small
	\begin{tabular}{|c|c|c|c|c|} 
	  \hline

		& width $<$ \thinLimit (thin) & width $\geq$ \thinLimit (wide) \\
	  \hline
		size $< 10^3 s $ (sm) & bin-1 (4.55\%, 0.01\%) & bin-2 (28.73\%, 0.06\%) \\
	  \hline
	  	size $\geq 10^3 s $ (lg) & bin-3 (14.29\%, 5.41\%) & bin-4 (52.43\%, 94.52\%) \\
	  \hline
	\end{tabular}
      }
\vspace{-0.1in}	
%\vspace{-0.2in}	%	AJ_spacecut	
\end{table}

We make the following observations.
(1) \slearn improves the JCT
for 82.46\% of the jobs in Bin-1 and the average JCT speedup for the
bin is 10.54$\times$. This happens because the jobs in this bin are
thin and hence \slearn assigns them high priorities, which is also the
right thing to do since these jobs are also small.  
(2) For bin-2, \slearn achieves an average JCT speedup of 1.86$\times$
from better prediction accuracy of \slearn.
The speedups are lower than for Bin-1 as the jobs have to undergo sampling.
%  bin are wide, so they undergo sampling and have to pay the cost of sampling
%  hence the improvements here are not as sound as bin-1.
However, Bin-1 and Bin-2
make up only 0.01\% and 0.06\% of the total job runtime and thus have little
impact on the overall JCT.
% Bin-2 also makes up a very small fraction of the total runtime,
% 0.06\%, and has little impact on the total JCT.
(3) 
Bin-3, which has 14.29\% of the jobs and accounts for 5.41\% of the total
job size, has a slowdown of 20.00\%.
% The primary reason for the slowdown of \slearn in Bin-3
The main reason is that \slearn treats thin jobs in the FIFO order,
whereas \primarybase schedules them based on predicted
sizes. 
\if 0
Additionally, \primarybase misplaces 80.35\% of the largest
62.57\% jobs of Bin-3 in higher priority queues, which results in
speeding up these jobs under \primarybase.
\fi
(4)
Bin-4, which accounts for a majority of the job and total job size,
%  which has 52.43\% of the jobs and accounts for 94.52\% of the total job size,
has an average speedup of 1.38$\times$, which contributes to the
overall speedup of 1.28$\times$.
The job speedups come from more accurate job runtime
estimation of \slearn over \primarybase.
\addnsdiSHP{Finally, we note that while for the 2Sigma trace, the majority of thin
  jobs are large, for the Google 2011 (Google 2019) trace, only
  1.90\% (1.60\%) of the total number of jobs are thin and large and they make up only 0.5\% (0.5\%) of the total job runtime.}{A2}.

\subsubsection{Sensitivity to Thin Job Bypass}
\label{sec:sim:thin}

Finally,
%   we evaluate the effect of \slearn's policy to bypass thin jobs from
%   sampling and schedule them with the highest priority. We conducted two
%   experiments: (1) evaluating \slearn against all the baselines on
%   a ``Wide-Only'' 2STrace, (2) evaluating \slearn's sensitivity to thinLimit.
we evaluate \slearn's sensitivity to thinLimt. Table~\ref{table:sim:sa:tl} shows that for GTrace11 and GTrace19, the average
JCT speedup barely varies with thinLimit, but
for 2STrace,
% also fluction is not high when the limit is decrease by 1. However,
there is a big dip when increasing thinLimit
to 4 or 5.  This is because a significant number of jobs in 2STrace
have width 4, which causes the number of thin jobs
to increase from 
\addnsdiSHP{18.84\% to 58.50\% when increasing thinLimit from 4 to 5.}{E7}

\begin{table}
	\caption{Sensitivity analysis for thinLimit. Table shows average JCT speedup over \primarybase. \updated{all - 16th Sep 2020}}
\vspace{-0.1in}	
	\label{table:sim:sa:tl}
  \centering
      {\small
	\begin{tabular}{|c|c|c|c|c|c|} 
	  \hline
		thinLimit&	2 & 3 & 4 & 5 & 6\\
	  \hline
	  2STrace&	1.23x&1.28x&1.14x&0.97x&0.84x\\%1.45
	  \hline
	  %	2STrace P(50)&	1.40x&1.41x&1.44x&1.40x&1.35x\\
          %GTrace&	1.53x&1.49x&1.40x&1.40x&1.41x\\
          GTrace11&	1.54x&1.56x&1.55x&1.54x&1.53x\\
	  \hline
          GTrace19&	1.33x&1.32x&1.32x&1.30x&1.29x\\
	  \hline
	\end{tabular}
      }
\vspace{-0.1in}
\end{table}

\if 0
\begin{table}
  \caption{Sensitivity analysis for thinLimit on GTrace. Table shows average JCT speedup over \primarybase. \updated{2S - 12th May 2020}.
    }
\label{table:sim:sa:tl:GTrace}
  \centering
      {\small
	\begin{tabular}{|c|c|c|c|c|c|} 
	  \hline
          thinLimit &		2 & 3 & 4 & 5 & 6 \\
	  \hline
          Avg. JCT speedup	&	1.31x&1.43x&1.42x&1.42x&1.40x\\
%	  \hline
	  %	P(50)&	0.44x&0.93x&0.94x&0.94x&0.74x\\
	  \hline
	\end{tabular}
      }
\end{table}
\fi
\if 0
\begin{table}
  \caption{Sensitivity analysis for thin node percentage on the GTrace.
    }
  \label{table:sim:sa:tnp}
  \centering
      {\small
	\begin{tabular}{|c|c|c|c|c|} 
	  \hline
thinLimit &		10 & 20 & 30 & 40 & 50\\
	  \hline
Avg. JCT speedup	 	0.72x&1.20x&1.49x&1.66x&1.65x\\
	  \hline
	\end{tabular}
      }
\end{table}
\fi

\vspace{-0.05in}
\if 0
\paragraph{Sampling node percentage} We next vary the threshold for the
sampling node percentage from 10 to 50. Table~\ref{table:sim:sa:snp}
show the results.
\fi

\if 0
\paragraph{Sampling percentage} In this experiment, we vary the sampling percentage 
of queues decreases. As shown in Table~\ref{table:sim:sa:sp}. 
\fi

\if 0
\paragraph{Thin job bypassing limit ($T$)}
In this experiment, we vary the thinLimit (T) in \slearn for bypassing jobs from
the probing phase. The results in Fig.~\ref{table:sim:sa:tl} shows that the
average JCT remains almost the same as T increases.  However, the P50 speedup
increases till $T = \thinLimit$, and tapers off after $T = \thinLimit$.
The key reason for the JCT improvement until $T = \thinLimit$ is that
all tasks of the thin job (with width $< \thinLimit$) are scheduled immediately
upon arrival which improves their JCT (\S\ref{sec:design:thin}).
\fi

\if 0
\paragraph{Thin node percentage} We next vary the threshold for
the thin node percentage from 10 to 50. Table.~\ref{table:sim:sa:tnp}
\fi

%% file: dag.tex
\if 0 
\section{Discussions}

Despite these encouraging findings about learning in space,
we envision several motivations for exploring combining history-
and sampling-based learning. 
%   (1) For workloads with mixed recurring and first-time jobs,
%   sampling-based learning can be applied to
%   % to estimate the job runtime
%   first-time jobs
%   while history-based learning can be applied to
%   % to estimate the job runtime
%   recurring jobs.
(1) History-based learning can be used to establish a prior
distribution, and sampling-based approach can be used to refine the
posterior distribution. Such a combination may potentially be more accurate
than using either history or sampling alone.
% For example, knowing the distribution of task lengths can help develop better
% max task length predictors.
(2) Though not seen in the production traces used in our study,
in case task-wise variation and job-wise variation fluctuate, adaptively switching
between the two prediction schemes may also help.
(3) Sampling-based learning can be integrated with history-based learning
to schedule multi-phase DAG jobs.
We present our preliminry work on this below.

\fi
\section{Scheduling for DAG Jobs}
\label{sec:dag}

In earlier sections, we have focused on the benefits of sampling-based prediction.
On the other hand, we envision that there are situations where it would be 
beneficial to combine sampling-based and history-based predictions. Below, we 
present our preliminary work applying such a hybrid strategy for scheduling DAG jobs.
\addnsdiSHP{We will discuss several other use cases of a hybrid strategy  in \S\ref{sec:discuss}.}{}
Note that for multi-phase DAG jobs, simply applying sampling-based prediction to 
each phase in turn cannot estimate the whole DAG runtime ahead of time.
%   applied in each phase to optimize the performance
%   % (\eg completion time)
%   of each phase.  An interesting question is how to apply sampling
%   if we wish to 
%Below we present a design that combines history-based
%prediction with sampling-based prediction to 
Instead, our hybrid design below aims to learn the runtime
properties and optimize the performance of a multi-phase DAG job \emph{as a
  whole} (\eg~\cite{jockey:eurosys2012, AltruisticScheduling}).

% To evaluate the performance of sampling based prediction for DAGs we desgined
% \primarybasedag, a derivative of the \primarybase and \slearndag, which uses a
% hybrid approach of sampling and history for prediction.

\paragraph{Hybrid learning for DAGs (\slearndag).}
The key idea of \slearndag is to adjust history-based prediction of the runtime of 
DAG jobs using sampling-based learning of its first stage.
Upon arrival of a new DAG job, we 
estimate the runtime of its first stage using 
sampling-based prediction 
as described in \S\ref{sec:design:namepredict}, 
denoted as $d_{s}$. We also estimate the duration of
this stage using history-base \primarybase, denoted as $d_{h}$, and compute 
the adjustment ratio of $\frac{d_{s}}{d_{h}}$. For each of the
remaining stages of the DAG, we predict their runtime using \primarybase and then
multiply it with the adjustment ratio.
In a nutshell, this hybrid design reduces the error of history-based prediction due
to staleness of the learning data, while avoiding the delay of
sampling across all other stages. 

\paragraph{History-based learning for DAGs (\primarybasedag).}
This is a straight-forward extension of \primarybase. Upon arrival of a DAG job,
it predicts independently the runtime for each stage using the \primarybase and sums up the
estimated runtime of all stages as the estimated runtime of the entire DAG.

We similarly extended other baselines described in \S\ref{sec:design:baselines} for DAG job.

\paragraph{Experimental setup.}
We evaluated \slearndag against \primarybasedag by replaying cluster
trace in simulation experiments based on \gs
(\S\ref{sec:design:gs}). We kept the simulation setup and parameters
the same as used in the other experiments.  In particular, a DAG is placed
in the corresponding priority queue based on its estimated total runtime.

\paragraph{DAG Traces.}
The only publically available DAG trace we
could find is a trace from Alibaba\cite{alibabaDAGTrace}, which could not
be used as it does not contain features required for history-based
prediction using \primarybase. 
% We equested trace authors for more information however couldn't get any.
Instead, we followed the ideas in previous work, \eg Branch
Scheduling~\cite{branchScheduling:IWQoS19}, to generate a synthetic DAG trace
of about 900 jobs using the Google 2019 trace~\cite{googleClusterData2019},
denoted as \dagtrace. The number of stages in DAGs in the \dagtrace was randomly
choosen to be between 2-5 and each stage is a complete job from the Google 2019
trace. The jobs that are part of the same DAG have the same \textit{jobname} and
the same \textit{username}.

%\subsection{Results and Explanation}
\if 0
\begin{table}
	\caption{Performance comparison of \slearndag over \primarybasedag and other baselines.}
\vspace{-0.1in}	
	\label{table:dag:performance}
  \centering
      {\small
	\begin{tabular}{|c|c|c|c|c|c|} 
	  \hline
		Baselines&	P50 & P90 & Average \\
	  \hline
	  \primarybasedag&	1.77x&51.43x&1.26x\\
	  \hline
	  \lasdag&	6.98x&99.86x&2.15x\\
	  \hline
          \pointestimatordag&	4.47x&52.96x&1.74x\\
	  \hline
          \fifodag&	29.45x&287.02x&5.54x\\
	  \hline
          \oracledag&	0.93x&1.12x&0.79x\\
	  \hline
	\end{tabular}
      }
\vspace{-0.1in}
\end{table}
\fi

%\begin{figure}
%\centering
%\includegraphics[width=\linewidth]{figures/simulation/Perfo%rmanceCompareGTrace19-DAG.pdf}
%	\caption{JCT speedup using \slearndag as compared to %other baseline schemes for \dagtrace. All the baselines are %DAG versions.}
%\label{fig:sim:jctOthers:GTrace19-DAG}
%\end{figure}
\paragraph{Results.}
The results in Fig.~\ref{fig:sim:jctOthers:GTrace19-DAG} show that \slearndag achieves significant speedup over other designs. 
The speedup is 1.26$\times$ over \primarybasedag, 2.15$\times$ over \lasdag, and 1.74$\times$ over \pointestimatordag.
Looking deeper, we find that our sampling-based prediction still yields higher prediction accuracy: the P50 prediction error is 33.90\% for \slearndag, compared to 47.21\% for \primarybasedag.
On the other hand, for DAG jobs the relative overhead of sampling (e.g, the delay) is lower since only the first stage is sampled. 
Together, they produce speedup comparable to earlier sections.

% \addnsdiSHP{We note that while these results show that sampling and history can be combined to effectively schedule DAG jobs, the results are preliminary. We plan to explore this idea in depth in future work.}{E}
%(1) Better prediction accuracy of sampling; 31.56\% higher in the median case.
%(2) Delay incurred due to sampling is now shared among multiple stages.

%% file: discussion.tex
\addnsdiSHP{
\section{Discussions and Future Work}
\vspace{-0.1in}
\label{sec:discuss}

\vspace{-0.05in}
\paragraph{Combining history and sampling.} 
%  We have shown in \S\ref{sec:dag}
% that combining sampling- and history-based learning can 
In addition to improving the scheduling of DAG jobs (\S\ref{sec:dag}),
we discuss several additional motivations for combining history-
and sampling-based learning. 
(1) For workloads with both recurring and first-time jobs,
sampling-based learning can be used
to estimate properties 
%the job runtime 
for first-time jobs, while history-based learning can be used 
%to estimate the job runtime 
%to estimate properties 
for recurring jobs.
%(2) For mixed workloads, sampling-based learning can be used
%to estimate the mean task length to minimize the total completion
%time for jobs without deadlines, while history-based learning can be
%used to estimate the max task length for deadline jobs, if it
%can achieve comparable accuracy as 
%sampling-based learning but without incurring any runtime overhead.
%   deadlines needs knowledge of both the average task
% length and the maximum task length. Our initial study has shown that,
% without any knowledge about the task length distribtuion, the
% prediction accuracy for max task length is about the same for both
%history and sampling-based schemes. However, since 
(2) When the workload has both thin and wide jobs,
history-based learning can be used for estimating the runtime for
thin jobs, while sampling-based learning is used for wide jobs.
(3) History-based learning can be used to establish a prior
distribution, and sampling-based approach can be used to refine the
posterior distribution. Such a combination is potentially  more accurate
than using either 
% history or sampling
approach alone. For example,
knowing the prior distribution of task lengths can help to develop better
max task-length predictors, which can be useful for jobs with deadlines.
(4) Though not seen in the production traces used in our study,
in cases when task-wise variation and job-wise variation fluctuate, 
adaptively switching between the two prediction schemes may also help.
(5) When the cluster is heterogeneous, an error adjustment using history, similar to what we did in \S\ref{sec:dag}, can be applied.

\paragraph{Dynamic adjustment of ThinLimit.}
ThinLimit is a subjective threshold. It helps in segregating jobs for which waiting 
time due to sampling overshadows the improvement in prediction accuracy. The optimal choice of this limit
will depend on the cluster load at the moment and hence can be adaptively chosen like the 
sampling percentage (Fig.~\ref{fig:design:AdaptiveSamplingAlgo} on page~\pageref{fig:design:AdaptiveSamplingAlgo}).

\paragraph{Heterogeneous clusters.}
% Our 150-node test cluster was homogeneous in terms of
% machines. 
Extending sampling-based learning to heterogeneous clusters requires
adjusting the task sampling process.
One idea is to schedule pilot tasks on homogeneous servers and then scale their
runtime to different types of servers using the ratio of machine speeds.}{A1, A3, C2, C5, D1/2}

%% file: conc.tex
\vspace{-0.2in}
\section{Conclusions}
\label{sec:conc}
\vspace{-0.1in}
% The ability to accurately estimate job runtime properties allows a cluster job
% scheduler to effectively schedule jobs.
In this paper, we performed a
comparative study of task-sampling-based prediction and history-based
prediction commonly used in the current cluster job schedulers. Our study
answers two key questions: (1) Via quantitative, trace and experimental
analysis, we showed that the task-sampling-based approach can predict job
runtime properties with much higher accuracy than history-based schemes.  (2)
Via extensive simulations and testbed experiments
% on a 150-node cluster in Microsoft Azure
of a generic cluster job scheduler,
we showed that although sampling-based learning delays
non-sampled tasks till completion of sampled tasks,
such delay can be more than compensated by the improved accuracy over the prior-art
history-based predictor, and as a result reduces the average JCT by
1.28$\times$,
1.56$\times$, and 1.32$\times$ 
for three production cluster traces.
These results suggest task-sampling-based prediction offers a 
promising alternative to the history-based prediction in facilitating cluster
job scheduling. 

%  \deadlineCS{We also found that sampling and history based
%  approaches have similar accuracy when it comes to predicting the maximum task length.}

%% file: paper_acknowledgement.tex
\paragraph{Acknowledgement}
We thank our shepherd Sangeetha Abdu Jyothi
and the anonymous reviewers for their
helpful comments. This work was supported in part by
NSF grant 2113893.

%% file: appendix.tex
\section*{Appendix: How does the system load affect the speedups of \slearn over \primarybase?}
\label{sec:sim:intuitionSpeedup}

\begin{figure*}[t]
	\includegraphics[width=1.0\linewidth]{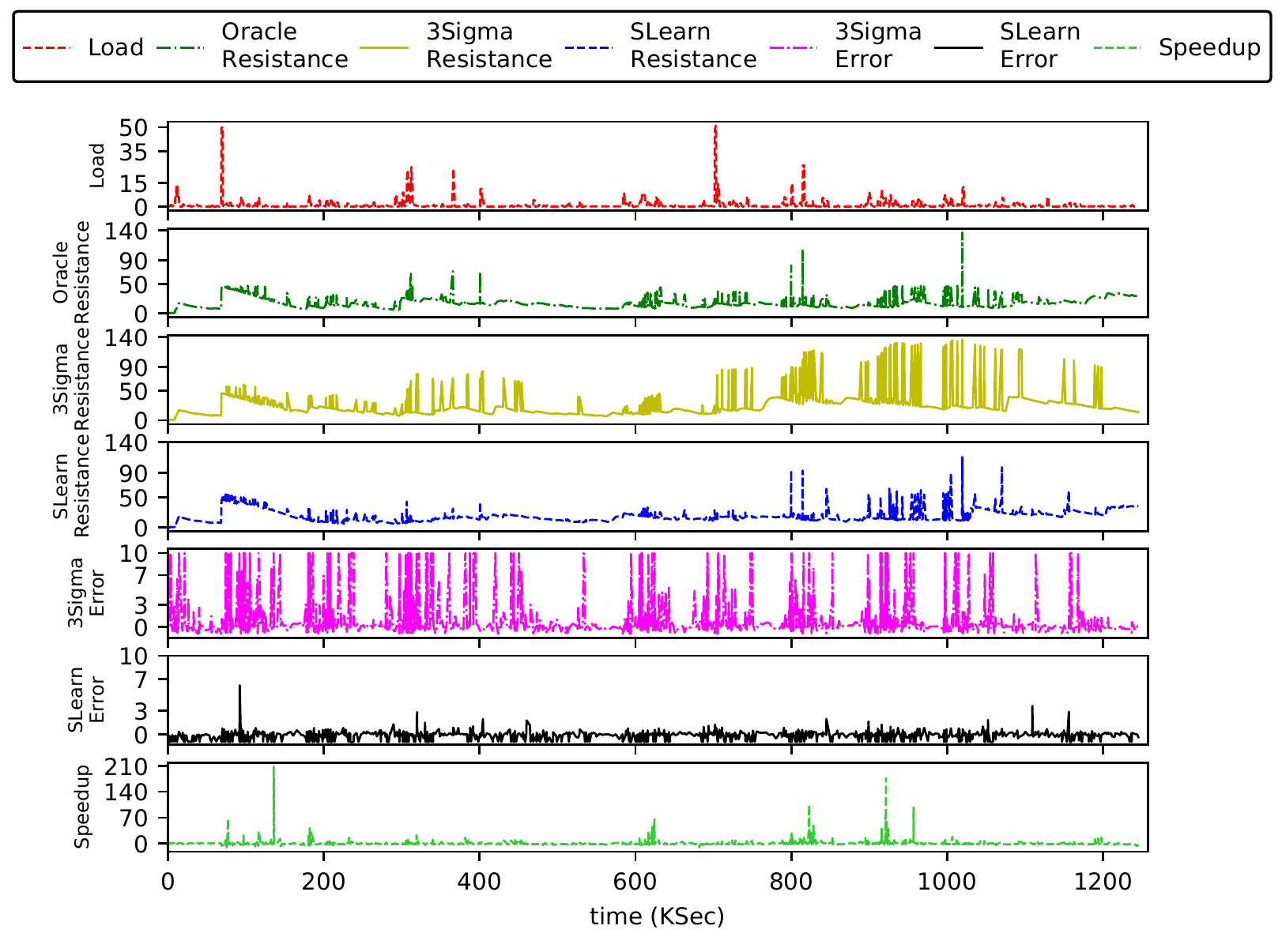}
	\caption{Correlation between load, \textit{resistance}, estimation error and speedup for 2STrace. \updated{16th Sep 2020}}
	\label{figs:sim:intuitionSpeedup:allInOne}
\end{figure*}

% \S\ref{sec:sim:numPilots} - \S\ref{sec:sim:averageJCT} explain better
% performance under \slearn over \primarybase by experimental results.
In this section, we provide an intuitive explanation for \slearn's
JCT speedup over \primarybasepredict in Section 5.2.4 for 2STrace in Figure 5(a).
Figure~\ref{figs:sim:intuitionSpeedup:allInOne} shows seven timeline
values comparing \slearn and \primarybase for the 2STrace as follows:

\begin{itemize}
\item The top curve shows the total workload arrived in the past 1000
  seconds, in terms of execution duration. The values
  are plotted in steps of 1000 seconds along the x-axis.  A unit along
  the y-axis corresponds to the workload that needs 1000 seconds
  of the entire cluster's compute capacity.
  Thus a workload of 1 in steady state implies no queue build-up under 100\% utilization of the whole cluster.
\item The next three curves show the \resistance faced by newly arrived jobs under
	\oracle, \primarybase and \slearn, respectively, where
	\resistance for a job is defined as the amount of higher
	priority workload existing at the time of its arrival,
	{including} the remaining duration of the
	already scheduled tasks. A unit along the y-axis for these
	curves also corresponds to the workload that needs 1000
	seconds of the entire cluster's compute capacity. For wide
	jobs  (\ie with 3 or more tasks),
        under \slearn we show the \resistance value corresponding to the moment
	when the job's size estimation is over and it has been placed
	in its estimated priority queue. {The \resistance values are plotted along
	the x-axis corresponding to each job's arrival time.}
\item The next two curves correspond to the percentage prediction error in
	\primarybasepredict and \slearn, respectively.  They show signed
	error which are capped at 1000, \eg a value of -20 on error
	curves means the job was estimated to be 20\% smaller and a
	value of 1000 means job was estimated at least 1000\%
	larger. The values are plotted along the x-axis corresponding
	to each job's arrival time.
      \item The bottom curve shows the job speedup (positive values)
        or slowdown (negative values) of \slearn compared to
        \primarybase, plotted along the x-axis corresponding to each
        job's arrival time.  Thus all values are either above 1, showing the speedups of
        jobs under \slearn over under \primarybase, or
        below -1, showing the speedups of jobs under \primarybase over under
        \slearn.
%  	\commentaj{Almost all are above or below 1. 3.6\% \ie approximately 45
%  		jobs have speedup = 1. So for those 45 points the value will be
%  		1. The least speedup is 0.1 \ie a slow down of 10 $\times$ and
%  		that will be shown as -10.}
 
%	\questionaj{Should we also add speedup of \slearn vs. \oracle and
		%	\primarybase vs. \oracle. If yes, should we then move
		%	the \primarybase vs. \slearn speedup curve.}
\end{itemize}

With the above definitions of the curves, we next discuss how these
curves in Fig.~\ref{figs:sim:intuitionSpeedup:allInOne} demonstrate
provides insights to when and why \slearn outperforms \primarybase.

\begin{table*}[h!]
\caption{Fraction of overestimated jobs and incorrect queue placement for 2STrace. Job performance in the third and seventh column is relative to the \oracle. \updated{2S - 16th Sep 2020}}
  \label{table:sim:misplacedJobs:all}
\vspace{-0.1in}	
  \centering
      {\small
	\begin{tabular}{|c||c|c|c|c||c|c|c|c|} 
	  \hline
	& Overestim- & Misplaced  & Slowed  & Average (P50) & Underesti- & Misplaced  & Speedup & Average (P50)\\
	  & ated jobs & overestimated & misplaced & Positive error & mated jobs & underestimated & misplaced & Negative error\\
          & & jobs & jobs &&  jobs &  jobs && \\
	  \hline
		%\primarybase & 59.37\% & 21.09\% & 17.72\% & 1253.60 (66.38)\% & 40.63\% & 10.22\% & 8.46\% & -39.33 (-31.59)\%\\
		\primarybase & 59.78\% & 17.50\% & 12.19\% & 898.5\% (48.00)\% & 40.22\% & 8.65\% & 6.88\% & -37.0\% (-28.57)\%\\
	  \hline
	  	\slearn & 43.75\% & 3.54 \% & 2.85\% & 30.65\% (18.19)\% & 55.45\% & 7.37\% & 3.64\% & -26.79\% (-20.69)\% \\
	  \hline
	\end{tabular}
      }
\vspace{-0.1in}	
\end{table*}

\begin{itemize}
\item The speedup curve (bottom) shows the speedup under \slearn over under \primarybase
  happens when the workload is high,
  \eg between 600s and 620s, and 800s to 840s.
  Conversely, when the workload is below 1, \eg between 400 and 600s,
  the two scheme perform similarly and there is no speedup
  of either scheme.
  In such cases, task sampling  \slearn did not hurt jobs because
  non-sampled tasks did not have to wait for completion of sampled tasks
  due to work conservation (\S\ref{sec:design:namepredict}).
\item Intuitively, under any scheme, a job's completion time is
  roughly proportional to its own total runtime (which is independent of the scheduling)
  plus the \resistance it sees upon arrival,
  because the \resistance value indicates the amount of workload that
  needs to be scheduled before the arriving job gets to run.
\item The \resistance value, in turn, depends on the recently arrived
  workload and the prediction error and hence the scheduling decision for them.
\item  First, if more workload has
arrived in the recent past, it is likely that
a newly arrived job will face higher \resistance.
This is shown by the strong correlation between the load curve
and the \oracle resistance curve.
\item Second, high runtime prediction error can lead to high \resistance.
When the job runtime is estimated by the predictor to be larger than its
 actual size, it may be misplaced in a lower priority queue. If the
  error is more than 1000\% then the job will definitely be placed in
  a lower priority queue.
  In such cases, the job will likely face
  higher \resistance than it would have with accurate estimation.
  Conversely, when the job runtime is underestimated, it may be placed in a higher
  priority queue. Though such a job will finish faster than otherwise,
  it will create more \resistance for other jobs that are
  actually smaller than it and thus slow them down.
\item The above impact of prediction error on \resistance can be seen in
Fig.~\ref{figs:sim:intuitionSpeedup:allInOne}.
Since the prediction accuracy of \slearn is high,
it has less impact on the \resistance and
as a result its \resistance (fourth curve) is very similar to that of \oracle (second curve)
\footnote{\addnsdiSHP{We note that there can be some exceptions where
    jobs face lower \resistance under \slearn compared to under
    \oracle, \eg between time 200 - 400 KSec in
    Fig.~\ref{figs:sim:intuitionSpeedup:allInOne}. This happens
 because of mis-prediction in \slearn, \eg when it underestimates 
the runtime of some jobs and places them in lower priority queues than otherwise,
the subsequently arriving job will experience lower \resistance.
}{Shep}}.
In contrast, the \resistance curve for \primarybase (third curve) has
many spikes, \eg between 800s and 1050s, which happen when the workload (top curve) is high
and it has high positive prediction error (fifth curve).
\item
  Finally, we can see that where ever there is higher \resistance
  under \primarybase (third curve) compared to under \slearn (fourth curve), \eg between 800s and
  1000s, jobs experience speedups under \slearn over under
  \primarybase.
\end{itemize}

%The data shown in table \ref{table:sim:overestimatedJobs} and
%\ref{table:sim:underestimatedJobs} also establishes the intuitive observations
%made from the curves.  \primarybase overestimates 59.37\% of the total jobs.
%Among the overestimated jobs, 36.02\% are misplaced at a lower priority queue,
%and among the misplaced jobs 82.84\% slowdown when compared to \oracle.
%Whereas \slearn overestimates 43.75\% of the total jobs. Among those only
%8.09\% are misplaced at a lower priority queue and are slowed down as compared
%to \oracle.
%%On the other side \primarybase underestimates 40.63\% of the jobs.

% As visible from Fig.~\ref{figs:sim:intuitionSpeedup:allInOne}, most of the jobs
% have low prediction error under \slearn whereas a significant number of jobs
% have large prediction error under \primarybase. This explains why a larger
% number of jobs, as shown in

While the above explanation using
Fig.~\ref{figs:sim:intuitionSpeedup:allInOne} is based on the performance
of \slearn and \primarybase relative to that of \oracle,
Table~\ref{table:sim:misplacedJobs:all} gives a direct
comparison of the scheduling behavior of the jobs under the two schemes in terms of runtime
overestimation/underestimation,
prediction error, and the resulting misplacement to the
priority queues.  We see that a larger number jobs are misplaced under
\primarybase compared to \slearn which led to the overall lower performance
under \primarybase.
%\questionaj{The current values in table \ref{table:sim:overestimatedJobs} and
%\ref{table:sim:underestimatedJobs} for \slearn are only for wide jobs and for
%\primarybase they are for all the jobs. What should we be doing here?}

%\iftoggle{techreport}{\vspace{-3in}}{}
In summary, whether a job finishes faster under \slearn compared to
\primarybase depends on two factors: the recent workload and the runtime
prediction error. Due to higher prediction error of \primarybase compared to \slearn, during
high workload, jobs are more likely to be misplaced to the priorty queues and hence face
higher \resistance, which results in longer average completion time under
\primarybase.

%% file: paper.bbl
\begin{thebibliography}{10}

\bibitem{2Sigma:website}
2sigma hedge fund.
\newblock www.twosigma.com.

\bibitem{2Sigma:scheduler}
2sigma's proprietary job scheduler.
\newblock
  https://www.twosigma.com/insights/article/cook-a-fair-preemptive-resource-scheduler-for-compute-clusters/.

\bibitem{alibabaDAGTrace}
Alibaba cluster trace.
\newblock https://github.com/alibaba/clusterdata.

\bibitem{hadoop:web}
Apache hadoop.
\newblock http://hadoop.apache.org.

\bibitem{yarn:web}
Apache hadoop yarn.
\newblock
  https://hadoop.apache.org/docs/current/hadoop-yarn/hadoop-yarn-site/YARN.html.

\bibitem{hive:web}
Apache hive.
\newblock http://hive.apache.org.

\bibitem{spark:web}
Apache spark.
\newblock http://spark.apache.org.

\bibitem{googleTraceGithub}
Cluster trace from google - 2011.
\newblock
  https://github.com/google/cluster-data/blob/master/ClusterData2011\_2.md.

\bibitem{googleClusterData2011-2Schema}
A document released by google containing schema and details of the cluster
  trace released by google.
\newblock https://drive.google.com/open?id=0B5g07T \_gRDg9Z0lsSTEtTWtpOW8.

\bibitem{dssScheduler}
Dss scheduler.
\newblock https://github.com/epfl-labos/DSS.

\bibitem{googleClusterData2019}
Google cluster-usage traces, retrieved 21st july 2020.
\newblock
  https://research.google/tools/datasets/google-cluster-workload-traces-2019/.

\bibitem{googleClusterData2019Schema}
Google cluster-usage traces, retrieved 21st july 2020.
\newblock https://drive.google.com/file/d/
  10r6cnJ5cJ89fPWCgj7j4LtLBqYN9RiI9/view.

\bibitem{numericHistogramJavaPatch}
Hadoop patch for numeric histogram.
\newblock https://issues.apache.org/jira/browse/YARN-2672.

\bibitem{azure:web}
Microsoft azure.
\newblock http://azure.microsoft.com.

\bibitem{gridmixpatch:junwoo}
A patch for gridmix.
\newblock https://issues.apache.org/jira/browse/YARN-2672.

\bibitem{personalCommunication:MarkAstley}
Personal communication with a 2sigma engineer regarding properties of the
  2sigma trace used.

\bibitem{2Sigma:trace}
A private trace collected by 2sigma engineers from their clusters.
\newblock www.twosigma.com.

\bibitem{jordanLecturePosterirorDistribution}
Resutls on the posteriro distribution with gaussian priors.
\newblock
  https://people.eecs.berkeley.edu/~jordan/courses/260-spring10/lectures/lecture5.pdf.

\bibitem{shufflewatcher}
Faraz Ahmad, Srimat~T. Chakradhar, Anand Raghunathan, and T.~N. Vijaykumar.
\newblock Shufflewatcher: Shuffle-aware scheduling in multi-tenant mapreduce
  clusters.
\newblock In {\em 2014 {USENIX} Annual Technical Conference ({USENIX} {ATC}
  14)}, pages 1--13, Philadelphia, PA, 2014. {USENIX} Association.

\bibitem{workloadDiversity:atc18}
George Amvrosiadis, Jun~Woo Park, Gregory~R. Ganger, Garth~A. Gibson, Elisabeth
  Baseman, and Nathan DeBardeleben.
\newblock On the diversity of cluster workloads and its impact on research
  results.
\newblock In {\em 2018 {USENIX} Annual Technical Conference ({USENIX} {ATC}
  18)}, pages 533--546, Boston, MA, 2018. {USENIX} Association.

\bibitem{Apollo:osdi2014}
Eric Boutin, Jaliya Ekanayake, Wei Lin, Bing Shi, Jingren Zhou, Zhengping Qian,
  Ming Wu, and Lidong Zhou.
\newblock Apollo: Scalable and coordinated scheduling for cloud-scale
  computing.
\newblock In {\em 11th {USENIX} Symposium on Operating Systems Design and
  Implementation ({OSDI} 14)}, pages 285--300, Broomfield, CO, 2014. {USENIX}
  Association.

\bibitem{scope:2008}
Ronnie Chaiken, Bob Jenkins, Per-AAke Larson, Bill Ramsey, Darren Shakib, Simon
  Weaver, and Jingren Zhou.
\newblock Scope: Easy and efficient parallel processing of massive data sets.
\newblock {\em Proc. VLDB Endow.}, 1(2):1265--1276, August 2008.
\newblock http://dx.doi.org/10.14778/1454159.1454166.

\bibitem{aalo:sigcomm15}
Mosharaf Chowdhury and Ion Stoica.
\newblock Efficient coflow scheduling without prior knowledge.
\newblock In {\em Proceedings of the 2015 ACM Conference on Special Interest
  Group on Data Communication}, SIGCOMM '15, pages 393--406, New York, NY, USA,
  2015. ACM.

\bibitem{varys:sigcomm14}
Mosharaf Chowdhury, Yuan Zhong, and Ion Stoica.
\newblock Efficient coflow scheduling with varys.
\newblock In {\em Proceedings of the 2014 ACM Conference on SIGCOMM}, SIGCOMM
  '14, pages 443--454, New York, NY, USA, 2014. ACM.

\bibitem{stratus:socc2018}
Andrew Chung, Jun~Woo Park, and Gregory~R. Ganger.
\newblock Stratus: Cost-aware container scheduling in the public cloud.
\newblock In {\em Proceedings of the ACM Symposium on Cloud Computing}, SoCC
  '18, pages 121--134, New York, NY, USA, 2018. ACM.

\bibitem{feedback:jacm1968}
Edward~G Coffman and Leonard Kleinrock.
\newblock Feedback queueing models for time-shared systems.
\newblock {\em Journal of the ACM (JACM)}, 15(4):549--576, 1968.

\bibitem{IfYouAreLateDontBlameUs:socc14}
Carlo Curino, Djellel~E. Difallah, Chris Douglas, Subru Krishnan, Raghu
  Ramakrishnan, and Sriram Rao.
\newblock Reservation-based scheduling: If you're late don't blame us!
\newblock In {\em Proceedings of the ACM Symposium on Cloud Computing}, SOCC
  '14, pages 2:1--2:14, New York, NY, USA, 2014. ACM.

\bibitem{mapreduce:osdi04}
Jeffrey Dean and Sanjay Ghemawat.
\newblock Mapreduce: Simplified data processing on large clusters.
\newblock In {\em OSDI'04: Sixth Symposium on Operating System Design and
  Implementation}, pages 137--150, San Francisco, CA, 2004.

\bibitem{kairos:socc2018}
Pamela Delgado, Diego Didona, Florin Dinu, and Willy Zwaenepoel.
\newblock Kairos: Preemptive data center scheduling without runtime estimates.
\newblock In {\em Proceedings of the ACM Symposium on Cloud Computing}, SoCC
  '18, pages 135--148, New York, NY, USA, 2018. ACM.

\bibitem{jockey:eurosys2012}
Andrew~D. Ferguson, Peter Bodik, Srikanth Kandula, Eric Boutin, and Rodrigo
  Fonseca.
\newblock Jockey: Guaranteed job latency in data parallel clusters.
\newblock In {\em Proceedings of the 7th ACM European Conference on Computer
  Systems}, EuroSys '12, pages 99--112, New York, NY, USA, 2012. ACM.

\bibitem{drf:nsdi11}
Ali Ghodsi, Matei Zaharia, Benjamin Hindman, Andy Konwinski, Scott Shenker, and
  Ion Stoica.
\newblock Dominant resource fairness: Fair allocation of multiple resource
  types.
\newblock In {\em Proceedings of the 8th USENIX Conference on Networked Systems
  Design and Implementation}, NSDI'11, pages 323--336, Berkeley, CA, USA, 2011.
  USENIX Association.

\bibitem{MultiResourcePackingForClusterSchedulers}
Robert Grandl, Ganesh Ananthanarayanan, Srikanth Kandula, Sriram Rao, and
  Aditya Akella.
\newblock Multi-resource packing for cluster schedulers.
\newblock In {\em Proceedings of the 2014 ACM Conference on SIGCOMM}, SIGCOMM
  '14, pages 455--466, New York, NY, USA, 2014. ACM.

\bibitem{AltruisticScheduling}
Robert Grandl, Mosharaf Chowdhury, Aditya Akella, and Ganesh Ananthanarayanan.
\newblock Altruistic scheduling in multi-resource clusters.
\newblock In {\em 12th {USENIX} Symposium on Operating Systems Design and
  Implementation ({OSDI} 16)}, pages 65--80, Savannah, GA, 2016. {USENIX}
  Association.

\bibitem{branchScheduling:IWQoS19}
Zhiyao Hu, Dongsheng Li, Yiming Zhang, Deke Guo, and Ziyang Li.
\newblock Branch scheduling: Dag-aware scheduling for speeding up data-parallel
  jobs.
\newblock In {\em Proceedings of the International Symposium on Quality of
  Service}, IWQoS '19, New York, NY, USA, 2019. Association for Computing
  Machinery.

\bibitem{cora:infocom2015}
Zhe Huang, Bharath Balasubramanian, Michael Wang, Tian Lan, Mung Chiang, and
  Danny~HK Tsang.
\newblock Need for speed: Cora scheduler for optimizing completion-times in the
  cloud.
\newblock In {\em 2015 IEEE Conference on Computer Communications (INFOCOM)},
  pages 891--899. IEEE, 2015.

\bibitem{DontCryOverSpilledRecords}
Calin Iorgulescu, Florin Dinu, Aunn Raza, Wajih~Ul Hassan, and Willy
  Zwaenepoel.
\newblock Don{\textquoteright}t cry over spilled records: Memory elasticity of
  data-parallel applications and its application to cluster scheduling.
\newblock In {\em 2017 {USENIX} Annual Technical Conference ({USENIX} {ATC}
  17)}, pages 97--109, Santa Clara, CA, 2017. {USENIX} Association.

\bibitem{dryad:eurosys2007}
Michael Isard, Mihai Budiu, Yuan Yu, Andrew Birrell, and Dennis Fetterly.
\newblock Dryad: Distributed data-parallel programs from sequential building
  blocks.
\newblock In {\em Proceedings of the 2Nd ACM SIGOPS/EuroSys European Conference
  on Computer Systems 2007}, EuroSys '07, pages 59--72, New York, NY, USA,
  2007. ACM.

\bibitem{jajoo2020exploiting}
Akshay Jajoo.
\newblock {\em EXPLOITING THE SPATIAL DIMENSION OF BIG DATA JOBS FOR EFFICIENT
  CLUSTER JOB SCHEDULING}.
\newblock PhD thesis, Purdue University Graduate School, 2020.

\bibitem{graviton:hotcloud16}
Akshay Jajoo, Rohan Gandhi, and Y.~Charlie Hu.
\newblock Graviton: Twisting space and time to speed-up coflows.
\newblock In {\em 8th {USENIX} Workshop on Hot Topics in Cloud Computing
  (HotCloud 16)}, Denver, CO, 2016. {USENIX} Association.

\bibitem{jajooSaath}
Akshay Jajoo, Rohan Gandhi, Y.~Charlie Hu, and Cheng-Kok Koh.
\newblock Saath: Speeding up coflows by exploiting the spatial dimension.
\newblock In {\em Proceedings of the 13th International Conference on Emerging
  Networking EXperiments and Technologies}, CoNEXT '17, pages 439--450, New
  York, NY, USA, 2017. ACM.

\bibitem{jajooPhilae}
Akshay Jajoo, Y.~Charlie Hu, and Xiaojun Lin.
\newblock Your coflow has many flows: Sampling them for fun and speed.
\newblock In {\em 2019 {USENIX} Annual Technical Conference ({USENIX} {ATC}
  19)}, pages 833--848, Renton, WA, 2019. {USENIX} Association.

\bibitem{philaeTechReport}
Akshay Jajoo, Y.~Charlie Hu, and Xiaojun Lin.
\newblock A case for sampling based learning techniques in coflow scheduling.
\newblock {\em CoRR}, abs/2108.11255, 2021.
\newblock \url{http://arxiv.org/abs/2108.11255}.

\bibitem{jajooSLearn}
Akshay Jajoo, Y.~Charlie~Hu Hu, Xiaojun Lin, and Nan Deng.
\newblock A case for task sampling based learning for cluster job scheduling.
\newblock In {\em Proceedings of the 19th USENIX Conference on Networked
  Systems Design and Implementation}, NSDI'22, Berkeley, CA, USA, 2022. USENIX
  Association.

\bibitem{corral}
Virajith Jalaparti, Peter Bodik, Ishai Menache, Sriram Rao, Konstantin
  Makarychev, and Matthew Caesar.
\newblock Network-aware scheduling for data-parallel jobs: Plan when you can.
\newblock In {\em Proceedings of the 2015 ACM Conference on Special Interest
  Group on Data Communication}, SIGCOMM '15, pages 407--420, New York, NY, USA,
  2015. ACM.

\bibitem{morpheus}
Sangeetha~Abdu Jyothi, Carlo Curino, Ishai Menache, Shravan~Matthur
  Narayanamurthy, Alexey Tumanov, Jonathan Yaniv, Ruslan Mavlyutov, Inigo
  Goiri, Subru Krishnan, Janardhan Kulkarni, and Sriram Rao.
\newblock Morpheus: Towards automated slos for enterprise clusters.
\newblock In {\em 12th {USENIX} Symposium on Operating Systems Design and
  Implementation ({OSDI} 16)}, pages 117--134, Savannah, GA, 2016. {USENIX}
  Association.

\bibitem{roughSetEstimation:IEEE:Shonali}
Shonali Krishnaswamy, {Seng Wai} Loke, and Arkady Zaslavsky.
\newblock Estimating computation times of data-intensive applications.
\newblock {\em IEEE Distributed Systems Online}, 5(4):1 -- 12, 2004.

\bibitem{nuyens:survey2008}
Misja Nuyens and Adam Wierman.
\newblock The foreground--background queue: a survey.
\newblock {\em Performance evaluation}, 65(3-4):286--307, 2008.

\bibitem{3Sigma}
Jun~Woo Park, Alexey Tumanov, Angela Jiang, Michael~A. Kozuch, and Gregory~R.
  Ganger.
\newblock 3sigma: Distribution-based cluster scheduling for runtime
  uncertainty.
\newblock In {\em Proceedings of the Thirteenth EuroSys Conference}, EuroSys
  '18, pages 2:1--2:17, New York, NY, USA, 2018. ACM.

\bibitem{raiLAS:sigmetrics2003}
Idris~A. Rai, Guillaume Urvoy-Keller, and Ernst~W. Biersack.
\newblock Analysis of las scheduling for job size distributions with high
  variance.
\newblock In {\em Proceedings of the 2003 ACM SIGMETRICS International
  Conference on Measurement and Modeling of Computer Systems}, SIGMETRICS '03,
  pages 218--228, New York, NY, USA, 2003. ACM.

\bibitem{perforator:socc2016}
Kaushik Rajan, Dharmesh Kakadia, Carlo Curino, and Subru Krishnan.
\newblock Perforator: Eloquent performance models for resource optimization.
\newblock In {\em Proceedings of the Seventh ACM Symposium on Cloud Computing},
  SoCC '16, pages 415--427, New York, NY, USA, 2016. ACM.

\bibitem{wsmith:IEEE98}
Warren Smith, Ian Foster, and Valerie Taylor.
\newblock Predicting application run times using historical information.
\newblock In Dror~G. Feitelson and Larry Rudolph, editors, {\em Job Scheduling
  Strategies for Parallel Processing}, pages 122--142, Berlin, Heidelberg,
  1998. Springer Berlin Heidelberg.

\bibitem{jamiasvu}
Alexey Tumanov, Angela Jiang, Jun~Woo Park, Michael~A. Kozuch, and Gregory~R.
  Ganger.
\newblock Jamaisvu: Robust scheduling with auto-estimated job runtimes.
\newblock In {\em Technical Report CMU-PDL-16-104}. Carnegie Mellon University,
  2016.

\bibitem{tetrisched}
Alexey Tumanov, Timothy Zhu, Jun~Woo Park, Michael~A. Kozuch, Mor
  Harchol-Balter, and Gregory~R. Ganger.
\newblock Tetrisched: Global rescheduling with adaptive plan-ahead in dynamic
  heterogeneous clusters.
\newblock In {\em Proceedings of the Eleventh European Conference on Computer
  Systems}, EuroSys '16, pages 35:1--35:16, New York, NY, USA, 2016. ACM.

\bibitem{borg}
Abhishek Verma, Luis Pedrosa, Madhukar~R. Korupolu, David Oppenheimer, Eric
  Tune, and John Wilkes.
\newblock Large-scale cluster management at {Google} with {Borg}.
\newblock In {\em Proceedings of the European Conference on Computer Systems
  (EuroSys)}, Bordeaux, France, 2015.

\bibitem{gandiva:osdi18}
Wencong Xiao, Romil Bhardwaj, Ramachandran Ramjee, Muthian Sivathanu, Nipun
  Kwatra, Zhenhua Han, Pratyush Patel, Xuan Peng, Hanyu Zhao, Quanlu Zhang, Fan
  Yang, and Lidong Zhou.
\newblock Gandiva: Introspective cluster scheduling for deep learning.
\newblock In {\em 13th {USENIX} Symposium on Operating Systems Design and
  Implementation ({OSDI} 18)}, pages 595--610, Carlsbad, CA, October 2018.
  {USENIX} Association.

\bibitem{cdef:atc18}
Yong Xu, Kaixin Sui, Randolph Yao, Hongyu Zhang, Qingwei Lin, Yingnong Dang,
  Peng Li, Keceng Jiang, Wenchi Zhang, Jian-Guang Lou, Murali Chintalapati, and
  Dongmei Zhang.
\newblock Improving service availability of cloud systems by predicting disk
  error.
\newblock In {\em 2018 {USENIX} Annual Technical Conference ({USENIX} {ATC}
  18)}, pages 481--494, Boston, MA, 2018. {USENIX} Association.

\bibitem{delay:eurosys10}
Matei Zaharia, Dhruba Borthakur, Joydeep Sen~Sarma, Khaled Elmeleegy, Scott
  Shenker, and Ion Stoica.
\newblock Delay scheduling: A simple technique for achieving locality and
  fairness in cluster scheduling.
\newblock In {\em Proceedings of the 5th European Conference on Computer
  Systems}, EuroSys '10, pages 265--278, New York, NY, USA, 2010. ACM.

\end{thebibliography}
